\documentclass[aps,twocolumn,prd,showpacs,showkeys,preprintnumbers,superscriptaddress,bibnotes,floatfix,longbibliography,notitlepage,nofootinbib]{revtex4-2}
\usepackage[colorlinks=true,breaklinks=true]{hyperref}
\usepackage[utf8]{inputenc}
\usepackage{url}
\usepackage{gensymb}
\usepackage{mathtools}
\usepackage[dvipsnames]{xcolor}
\definecolor{darkred}{rgb}{0.5,0,0}
\definecolor{darkblue}{rgb}{0,0,0.5}
\definecolor{firebrick}{rgb}{0.75,0.125,0.125}
\definecolor{darkgreen}{rgb}{0,0.5,0}
\hypersetup{urlcolor=darkblue,
	    citecolor=darkgreen,
	    linkcolor=firebrick}

\usepackage{orcidlink}
\usepackage{lipsum}

\long\def\exclude#1{}

\newcommand{\ie}{{i.e.}}

\newcommand{\eg}{{e.g.}}

\newcommand{\eq}{Eq.}

\newcommand{\fig}{Fig.}

\newcommand{\Refe}{Ref.}
\newcommand{\Refes}{Refs.}
\newcommand{\equ}[1]{\eq~(\ref{equ:#1})}
\newcommand{\figu}[1]{\fig~\ref{fig:#1}}
\newcommand{\orcid}[1]{\href{https://orcid.org/#1}{\includegraphics[width=10pt]{orcid.pdf}}}
\bibpunct{[}{]}{,}{n}{}{,}

\begin{document}

\title{Two-detector flavor sensitivity to ultra-high-energy cosmic neutrinos}

\author{Federico Testagrossa
\orcidlink{0009-0000-9401-1971}} 
\email{federico.testagrossa@studenti.unipd.it}
\affiliation{Dipartimento di Fisica e Astronomia, Università di Padova, Via Marzolo 8, 35131 Padova, Italy}

\author{Damiano F.~G.~Fiorillo 
\orcidlink{0000-0003-4927-9850}} 
\email{damiano.fiorillo@nbi.ku.dk}
\affiliation{Niels Bohr International Academy, Niels Bohr Institute, University of Copenhagen, 2100 Copenhagen, Denmark}

\author{Mauricio Bustamante
\orcidlink{0000-0001-6923-0865}} 
\email{mbustamante@nbi.ku.dk}
\affiliation{Niels Bohr International Academy, Niels Bohr Institute, University of Copenhagen, 2100 Copenhagen, Denmark}

\date{October 18, 2023}

\begin{abstract}
 Ultra-high-energy (UHE) cosmic neutrinos, with energies above 100~PeV, could be finally discovered in the near future.  Measuring their flavor composition would reveal information about their production and propagation, but new techniques are needed for UHE neutrino telescopes to have the capabilities to do it.  We address this by proposing a new way to measure the UHE neutrino flavor composition that does not rely on individual telescopes having flavor-identification capabilities.  We manufacture flavor sensitivity from the joint detection by one telescope sensitive to all flavors---the radio array of IceCube-Gen2---and one mostly sensitive to $\nu_\tau$---GRAND.  From this limited flavor sensitivity, predominantly to $\nu_\tau$, and even under conservative choices of neutrino flux and detector size, we extract important insight.  For astrophysics, we forecast meaningful constraints on the neutrino production mechanism; for fundamental physics, vastly improved constraints on Lorentz-invariance violation.  These are the first measurement forecasts of the UHE $\nu_\tau$ content.
\end{abstract}

\maketitle

\section{Introduction}
\label{sec:introduction}

High-energy cosmic neutrinos are incisive probes of extreme astrophysics and fundamental physics.  For astrophysics, they address the long-standing question of the origin of ultra-high-energy cosmic rays (UHECRs)~\cite{Anchordoqui:2013dnh, Ahlers:2018fkn, Anchordoqui:2018qom, Ackermann:2019ows, Meszaros:2019xej, Halzen:2019qkf, Palladino:2020jol, AlvesBatista:2021eeu, Ackermann:2022rqc, Guepin:2022qpl}.  For fundamental physics, they probe the highest neutrino energy scales, otherwise unreachable, where new physics may manifest~\cite{Gaisser:1994yf, Ahlers:2018mkf, Arguelles:2019rbn, Ackermann:2019cxh, AlvesBatista:2021eeu, Ackermann:2022rqc}.  On both fronts, their flavor composition, \ie, the proportion of neutrinos of different flavor, $\nu_e$, $\nu_\mu$, and $\nu_\tau$, in their flux is a particularly versatile observable~\cite{Rachen:1998fd, Athar:2000yw, Crocker:2001zs, Beacom:2002vi, Barenboim:2003jm, Beacom:2003nh, Beacom:2003eu, Beacom:2003zg, Beacom:2004jb, Serpico:2005bs, Kashti:2005qa, Mena:2006eq, Kachelriess:2006ksy, Lipari:2007su, Pakvasa:2007dc, Esmaili:2009dz, Choubey:2009jq, Esmaili:2009fk, Bhattacharya:2009tx, Hummer:2010ai, Bhattacharya:2010xj, Bustamante:2010nq, Mehta:2011qb, Baerwald:2012kc, Fu:2012zr, Pakvasa:2012db, Chatterjee:2013tza, Winter:2013cla, Xu:2014via, Aeikens:2014yga, Palladino:2015zua, Arguelles:2015dca, Bustamante:2015waa, Pagliaroli:2015rca, Shoemaker:2015qul, deSalas:2016svi, Gonzalez-Garcia:2016gpq, Bustamante:2016ciw, Biehl:2016psj, Rasmussen:2017ert, Dey:2017ede, Bustamante:2018mzu, Farzan:2018pnk, Ahlers:2018yom, Brdar:2018tce, Palladino:2019pid, Bustamante:2019sdb, Ahlers:2020miq, Bustamante:2020bxp, Karmakar:2020yzn, Fiorillo:2020gsb, Song:2020nfh, Telalovic:2023tcb}.  

Today, the IceCube neutrino telescope regularly detects high-energy astrophysical neutrinos of 10~TeV--10~PeV, the most energetic ones seen so far~\cite{IceCube:2013cdw, IceCube:2013low, IceCube:2014stg, IceCube:2015gsk, IceCube:2015qii, IceCube:2016umi, IceCube:2020wum, IceCube:2021uhz}.  They have already delivered valuable new insight, including via their flavor composition~\cite{Mena:2014sja, Palomares-Ruiz:2015mka, IceCube:2015rro, Palladino:2015vna, IceCube:2015gsk, Vincent:2016nut, IceCube:2018pgc, IceCube:2020fpi}.  Beyond them, long-foretold ultra-high-energy (UHE) neutrinos~\cite{Berezinsky:1969erk}, with energies larger than 100~PeV, hold the potential to deliver even more.  Yet, so far, they remain undiscovered~\cite{IceCube:2018fhm, PierreAuger:2019ens}: their flux, although unknown, is likely too low to be seen by existing detectors~\cite{Ackermann:2019ows, Ackermann:2022rqc}.  Fortunately, in the next decade, a host of new planned detectors~\cite{Ackermann:2019ows, MammenAbraham:2022xoc, Ackermann:2022rqc, Guepin:2022qpl} will have a realistic chance of discovering UHE neutrinos even if their flux is tiny~\cite{Valera:2022ylt, Fiorillo:2022ijt, Valera:2022wmu, Fiorillo:2023clw, Valera:2023ayh}.


However, the measurement of the flavor composition of UHE neutrinos has received little attention, the focus being instead on first establishing the discovery potential of the planned detectors; see, \eg, \Refes~\cite{Valera:2022ylt, Fiorillo:2022ijt, Valera:2022wmu, Valera:2023ayh}.  Further, because they use detection strategies different~\cite{MammenAbraham:2022xoc, Ackermann:2022rqc} from optical-light detection at IceCube~\cite{IceCube:2016zyt}---\ie, radio in-ice~\cite{IceCube-Gen2:2020qha, RNO-G:2020rmc} and in-air~\cite{Adams:2017fjh, GRAND:2018iaj, Deaconu:2019rdx, Prohira:2019glh, Nam:2020hng, Wissel:2020sec}, Cherenkov light~\cite{Otte:2018uxj, POEMMA:2020ykm}---new techniques are needed to distinguish between similar signals made by neutrinos of different flavor.  For in-ice radio-based detection, there are promising prospects~\cite{Wang:2013njo, Stjarnholm:2021xpj, Glaser:2021hfi, Coleman:2024scd}.  

We sideline this challenge by proposing a new way to measure the UHE neutrino flavor composition that bypasses the need for individual detectors to have flavor-identification capabilities.  Instead, we exploit the synergy between different detection strategies.

\begin{figure}[t!]
 \centering
 \includegraphics[trim={0.34cm 0 0 0},clip,width=1.05\columnwidth]{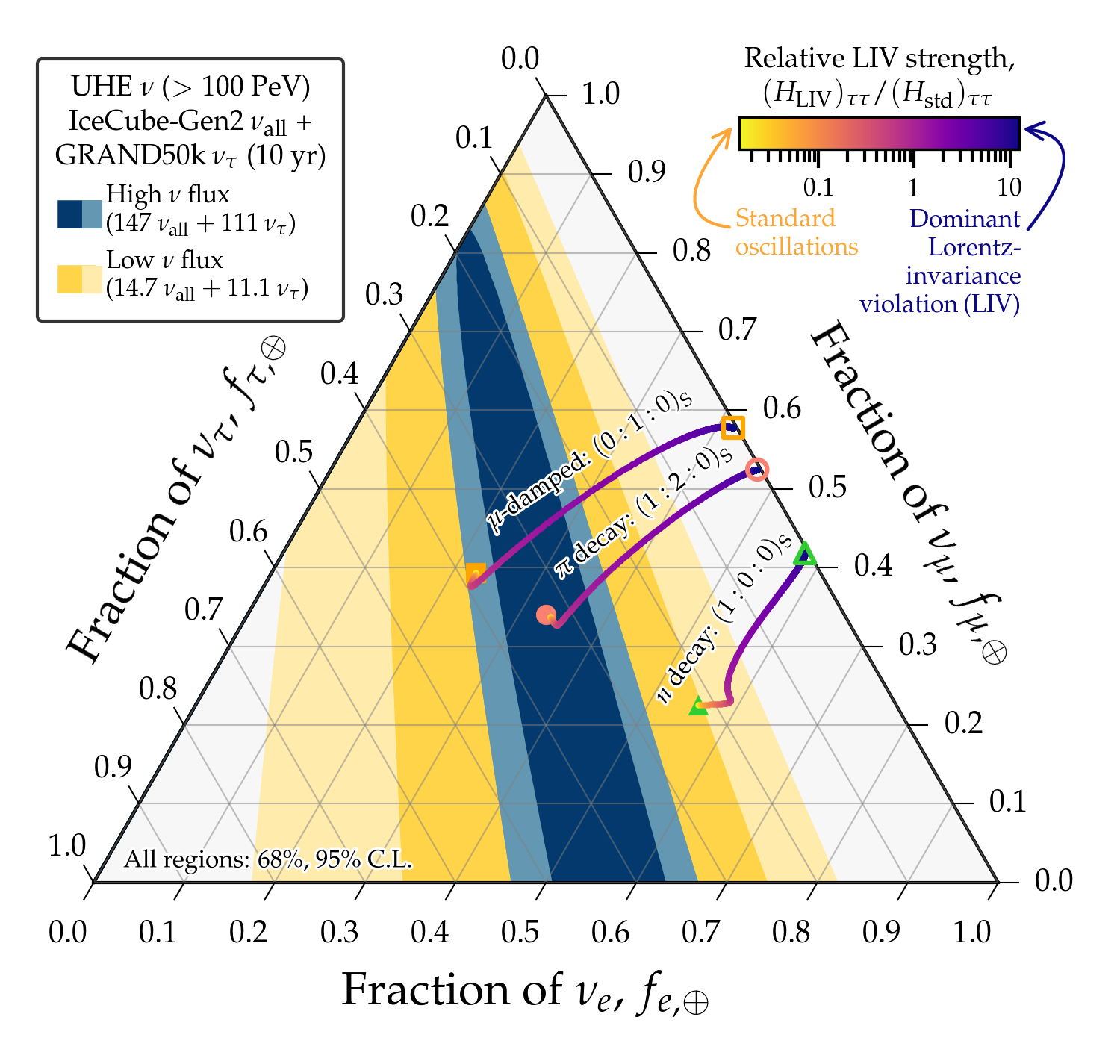}
 \caption{\textbf{\textit{Projected measurement of the flavor composition of ultra-high-energy neutrinos.}} The regions are of allowed flavor composition, obtained by combining measurements of neutrinos of all flavors in the radio array of IceCube-Gen2~\cite{IceCube-Gen2:2021rkf} and of $\nu_\tau$ in GRAND50k~\cite{GRAND:2018iaj}, assuming high and low benchmark neutrino flux models.  Overlaid is the flavor composition at the Earth under standard oscillations and in the presence of Lorentz-invariance violation (LIV) of varying strength, for three benchmarks of the flavor composition at the neutrino sources. }
 \label{fig:ternary_grand50k}
\end{figure}

\smallskip


We manufacture flavor sensitivity from the combined action of two planned detectors: one that detects all neutrino flavors roughly indistinctly---\eg, the planned radio array of IceCube-Gen2---and one that detects predominantly $\nu_\tau$---\eg, GRAND.  This allows us to measure the fraction of $\nu_\tau$ in the diffuse UHE flux.

Figure~\ref{fig:ternary_grand50k} shows, for the first time, that this measurement is possible, even under conservative assumptions: a relatively low UHE neutrino flux and an intermediate-size configuration of GRAND.  Even this limited measurement, of a single flavor fraction, affords important new insight into neutrino astrophysics and fundamental physics.  For astrophysics, it allows us to infer the flavor composition with which the neutrinos were produced, which helps identify their production process and sources (\figu{likelihood_sources}).  For fundamental physics, it allows us to improve constraints of physics beyond the Standard Model by many orders of magnitude (\figu{limits_tt_vs_dim}).

The rest of the paper is structured as follows. In Sec.~\ref{sec:flux_char} we introduce the main features of UHE neutrinos and their flavor composition. In Sec.~\ref{sec:detector_char} we characterize the detector response and show how the event rates in each experiment are computed. In Sec.~\ref{sec:measurement} we introduce the statistical analysis used to measure the flavor composition.  In Sec.~\ref{sec:astro}, we show our results on astrophysics.  In Sec.~\ref{sec:LIV}, we show our results on fundamental physics.  In Sec.~\ref{sec:summary}, we summarize and conclude.

\smallskip


\section{UHE neutrino fluxes}
\label{sec:flux_char}

In the 1960s, it was realized that the existence of UHE neutrinos, above 100~PeV, could be expected on general grounds~\cite{Berezinsky:1969erk}.  UHECR protons with energies above $10^{12}$~GeV propagating in extragalactic space should interact with the cosmic microwave background, producing UHE {\it cosmogenic} neutrinos of comparable energies~\cite{Greisen:1966jv, Zatsepin:1966jv}.  UHECRs could also interact with matter or radiation inside their sources, yielding {\it astrophysical} UHE neutrinos~\cite{Mannheim:1995mm, Atoyan:2001ey, Atoyan:2002gu, Alvarez-Muniz:2004xlu, Murase:2014foa, Kimura:2014jba, Murase:2014foa, Padovani:2015mba, Palladino:2018lov, Rodrigues:2020pli, Righi:2020ufi, Kimura:2020thg, Neronov:2020fww,Paczynski:1994uv, Waxman:1997ti, Murase:2006mm,    Bustamante:2014oka, Senno:2015tsn, Pitik:2021xhb, Rudolph:2022dky, Rudolph:2022ppp,Fang:2013vla, Fang:2014qva,Farrar:2008ex, Wang:2011ip, Dai:2016gtz, Senno:2016bso, Lunardini:2016xwi, Zhang:2017hom, Guepin:2017abw, Winter:2020ptf, Winter:2022fpf}.  UHE neutrinos may reach the Earth
after traveling Gpc-scale distances undeterred by matter or radiation, their energies dampened only mildly by the adiabatic cosmological expansion.  Thus, they are direct probes of UHECRs and of processes at ultra-high energies.

Today, UHE neutrinos remain undetected.  Predictions of the size and shape of their flux are uncertain~\cite{Aloisio:2009sj, Kotera:2010yn, Ahlers:2012rz, Fang:2013vla, Padovani:2015mba, Heinze:2015hhp, Fang:2017zjf, Romero-Wolf:2017xqe, AlvesBatista:2018zui, Heinze:2019jou, Muzio:2019leu, Rodrigues:2020pli, Anker:2020lre, IceCube:2020wum, Muzio:2021zud, IceCube:2021uhz} because they rely on uncertainly known properties of UHECRs---their mass composition, maximum energies, and shape of their energy spectrum---and of their unknown sources---mainly, their abundance at different redshifts~\cite{Anchordoqui:2018qom, AlvesBatista:2021eeu}.  Broadly stated, light UHECR mass composition, low maximum energies, a spectrum that falls slowly with energy, and an abundance of distant sources yield higher UHE neutrino fluxes; changes in each of these factors in the opposite direction yield lower fluxes; see, \eg, \Refes~\cite{Kotera:2010yn, Romero-Wolf:2017xqe, AlvesBatista:2018zui, Heinze:2019jou}. Figure 5 in \Refe~\cite{Valera:2022ylt} and Fig.~2 in \Refe~\cite{Valera:2022wmu} illustrate the present-day breadth of predictions of the diffuse UHE neutrino flux. Our choice of benchmark for the diffuse flux introduced below, in \equ{flux_def}, is motivated by them.

\subsection{Energy spectrum}

Since our purpose is to measure the flavor composition of UHE neutrinos, what matters is their energy distribution, which determines how many we can expect to detect. Thus, in lieu of detailed flux modeling, we adopt for the neutrino energy spectrum a heuristic shape that captures the essential features of many  predictions.  For the diffuse flux of UHE $\nu_\alpha + \bar{\nu}_\alpha$ ($\alpha = e, \mu, \tau$), this is~\cite{Fiorillo:2022rft}
\begin{equation}
 \label{equ:flux_def}
 E_\nu^2 \Phi_\alpha
 =
 \Phi_0 
 f_{\alpha, \oplus} 
 \exp\left[
 -w \log^2\left(\frac{E_\nu}{E_\mathrm{bump}}\right)
 \right] \;,
\end{equation}
\ie, a log-parabola with height $\Phi_0$, width $w$, and centered at energy $E_\mathrm{bump}$, that captures the bump-like feature in the neutrino spectrum expected in production via proton-photon interactions~\cite{Stecker:1991vm, Waxman:1997ti, Learned:2000sw, Winter:2012xq, Fiorillo:2021hty}.  For the flavor composition at Earth, $f_{\alpha, \oplus}$, we adopt the expectation from pion decay (see Sec.~\ref{sec:flavor}).  

As benchmark, we adopt for $\Phi_0$, $w$, and $E_\mathrm{bump}$ values that yields an all-flavor flux, $\sum_\alpha \Phi_\alpha$ from \equ{flux_def}, that approximates the predicted neutrino flux from newborn pulsars, from \Refe~\cite{Fang:2013vla}, which peaks at $\sim$$3 \cdot 10^8$~GeV, and is representative of the breadth in predictions.  We compare predictions made with this flux (``high'') {\it vs.}~predictions made with a flux ten times smaller (``low'').  They yield, respectively, $\sim$$100$ and $\sim$$10$ events in 10 years in IceCube-Gen2 and GRAND50k (see Sec.~\ref{sec:event_rate}).

\subsection{Flavor composition}
\label{sec:flavor}
UHECR interactions produce high-energy pions that, upon decaying, make high-energy neutrinos, \ie, $\pi^+ \to \mu^+ + \nu_\mu$ followed by $\mu^+ \to \bar{\nu}_\mu + e^+ + \nu_e$, and their charge-conjugated processes.  Thus, the full pion decay chain yields a flavor composition at the sources (S) of $(f_{e, {\rm S}}, f_{\mu, {\rm S}}, f_{\tau, {\rm S}}) = \left( \frac{1}{3}, \frac{2}{3}, 0 \right)$, where $f_{\alpha, {\rm S}}$ is the ratio of the flux of $\nu_\alpha + \bar{\nu}_\alpha$ to the total flux emitted.  This is our nominal expectation.  If the neutrinos are produced in regions that harbor an intense magnetic field, the intermediate muons may cool by synchrotron radiation before decaying, yielding instead $(0,1,0)_{\rm S}$.  And, if neutrinos are made in the beta decay of neutrons, we expect only $\bar{\nu}_e$, \ie, $(1,0,0)_{\rm S}$.  We consider the above three production scenarios as benchmarks, but pick pion decay to produce our main results.  

En route to Earth, neutrinos oscillate; as a result, their flavor composition at the Earth, $f_{\alpha, \oplus}$, is different from that at the sources. 
For high-energy cosmic neutrinos, because oscillations are rapid and the energy resolution of neutrino telescopes is limited, there is sensitivity only to the average flavor-transition probability.  For $\nu_\alpha \to \nu_\beta$ ($\alpha, \beta = e, \mu, \tau$), this is $P_{\nu_\alpha \to \nu_\beta} = \sum_i \lvert U_{\alpha i} \rvert^2 \lvert U_{\beta i} \rvert^2$, where the sum is over the three neutrino mass eigenstates, and $U$ is the Pontecorvo-Maki-Nakagawa-Sakata (PMNS) lepton mixing matrix~\cite{Pontecorvo:1957qd,Maki:1962mu}, which depends on parameters whose values are known experimentally~\cite{Esteban:2020cvm, NuFit5.2}.  For a given flavor composition at the sources, $f_{\alpha, \oplus} = \sum_\beta P_{\nu_\beta \to \nu_\alpha} f_{\beta, {\rm S}}$.  Figure~\ref{fig:ternary_grand50k} shows the expectation for our three benchmarks production scenarios; the nominal expectation, from pion decay, is about $(0.33, 0.34, 0.33)_\oplus$ .

\subsection{Determining the diffuse neutrino flux}\label{sec:determining_diffuse}

We assume that the diffuse neutrino flux is due to a population of identical, nondescript astrophysical sources distributed in redshift, each of which injects the same neutrino spectrum. The rate of $\nu_\alpha + \bar{\nu}_\alpha$ emitted per unit energy emitted by an individual source is
\begin{equation}
 E_\nu^2 \frac{dN_\alpha}{dE_\nu dt}=C_\nu f_{\alpha,\mathrm{S}}\exp\left[-w \log^2\left(\frac{E_\nu}{E_{\rm bump}}\right)\right] \;,
\end{equation}
which has the same shape as our benchmark diffuse flux, \equ{flux_def}, with $C_\nu$ a normalization constant (more on it later in this section).  The diffuse flux at Earth is the sum of contributions across all redshifts, \ie,
\begin{eqnarray}
 \label{equ:diffuse_flux}
 &&
 \Phi_\alpha(E_\nu, \boldsymbol{\tilde{\theta}}, \boldsymbol{f}_{\rm S})
 =
 \int \frac{dz}{H(z)} 
 \rho_{\rm src}(z) 
 \\
 && ~\times
 \sum_{\beta} 
 \frac{dN_\alpha}{dE_\nu dt}[E_\nu(1+z), \boldsymbol{\tilde{\theta}}, \boldsymbol{f}_{\rm S}]
 P_{\nu_\beta \to \nu_\alpha} 
 \nonumber \;,
\end{eqnarray}
where $\boldsymbol{\tilde{\theta}} \equiv (C_\nu, w, E_{\rm bump})$ are the flux shape parameters, $\boldsymbol{f}_{\rm S} \equiv (f_{e,{\rm S}}, f_{\mu,{\rm S}}, f_{\tau,{\rm S}})$ are the flavor fractions at the sources, $H(z) \equiv H_0 [\Omega_\Lambda + \Omega_m (1+z)^3]^{1/2}$ is the Hubble parameter, and $\Omega_\Lambda = 0.68$ and $\Omega_m = 0.32$ are, respectively, the adimensional energy densities of vacuum and matter~\cite{Planck:2018vyg}. The right-hand side of \equ{diffuse_flux} is evaluated at an energy $E_\nu (1+z)$ to account for the redshifting of the neutrino energy due to the cosmological expansion.  

We assume that the number density of the sources, $\rho_{\rm src}$, follows the star-formation rate, parametrized as in \Refe~\cite{2dFGRS:2000opy}:
$\rho_{\rm src}(z) = \rho_0 (a + b z)h / [1 + \left(z/c\right)^d]$, where $a = 0.017$, $b = 0.13$, $h = 0.7$, $c = 3.3$, $d = 5.3$ for the modified Salpeter initial mass function~\cite{Hopkins:2006bw}; the local source density, $\rho_0$, is a parameter whose value we do not need to specify explicitly, as we discuss below.  This places most of the sources at $z \approx 2$, a few Gpc away; sources at higher redshifts are rarer and contribute little to the diffuse flux. 

In \equ{diffuse_flux}, the unknown per-source normalization parameter, $C_\nu$, is degenerate with the unknown local density of the sources, $\rho_0$.  Thus, it is convenient to define an effective normalization parameter, $\phi_0$, with units of flux, via $ E_\nu^2\phi_\alpha \equiv (\rho_0/H_0) (dN_{\alpha}/dE_\nu dt)$, \ie,
\begin{equation}
 \label{equ:flux_source}
 E_\nu^2\phi_\alpha
 =
 \phi_0 f_{\alpha,\rm S} \exp\left[-w \log^2\left(\frac{E_\nu}{E_{\rm bump}}\right)\right] \;.
\end{equation}
With this, the diffuse flux at Earth becomes
\begin{eqnarray}
 \label{equ:diffuse_flux_v2}
 &&
 \Phi_\alpha(E_\nu, \boldsymbol{\theta}, \boldsymbol{f}_{\rm S})
 =
 \int \frac{dz}{\tilde{H}(z)} 
 \tilde{\rho}_{\rm src}(z) 
 \\
 && ~\times
 \sum_{\beta} 
 \phi_\beta[E_\nu(1+z), \boldsymbol{\theta}, \boldsymbol{f}_{\rm S}]
 P_{\nu_\beta \to \nu_\alpha} 
 \nonumber \;,
\end{eqnarray}
where now the flux parameters are $\boldsymbol{\theta}=(\phi_0,w,E_{\rm bump})$, $\tilde{\rho}_{\rm src} = \rho_{\rm src}/\rho_0$ and $\tilde{H} = H/H_0$.  Equations~(\ref{equ:flux_source}) and (\ref{equ:diffuse_flux_v2}) are the expressions we use in our calculations, and $\boldsymbol{\theta}$ are the free parameters that we vary.

\begin{figure}[t!]
 \centering
 \includegraphics[width=\columnwidth]{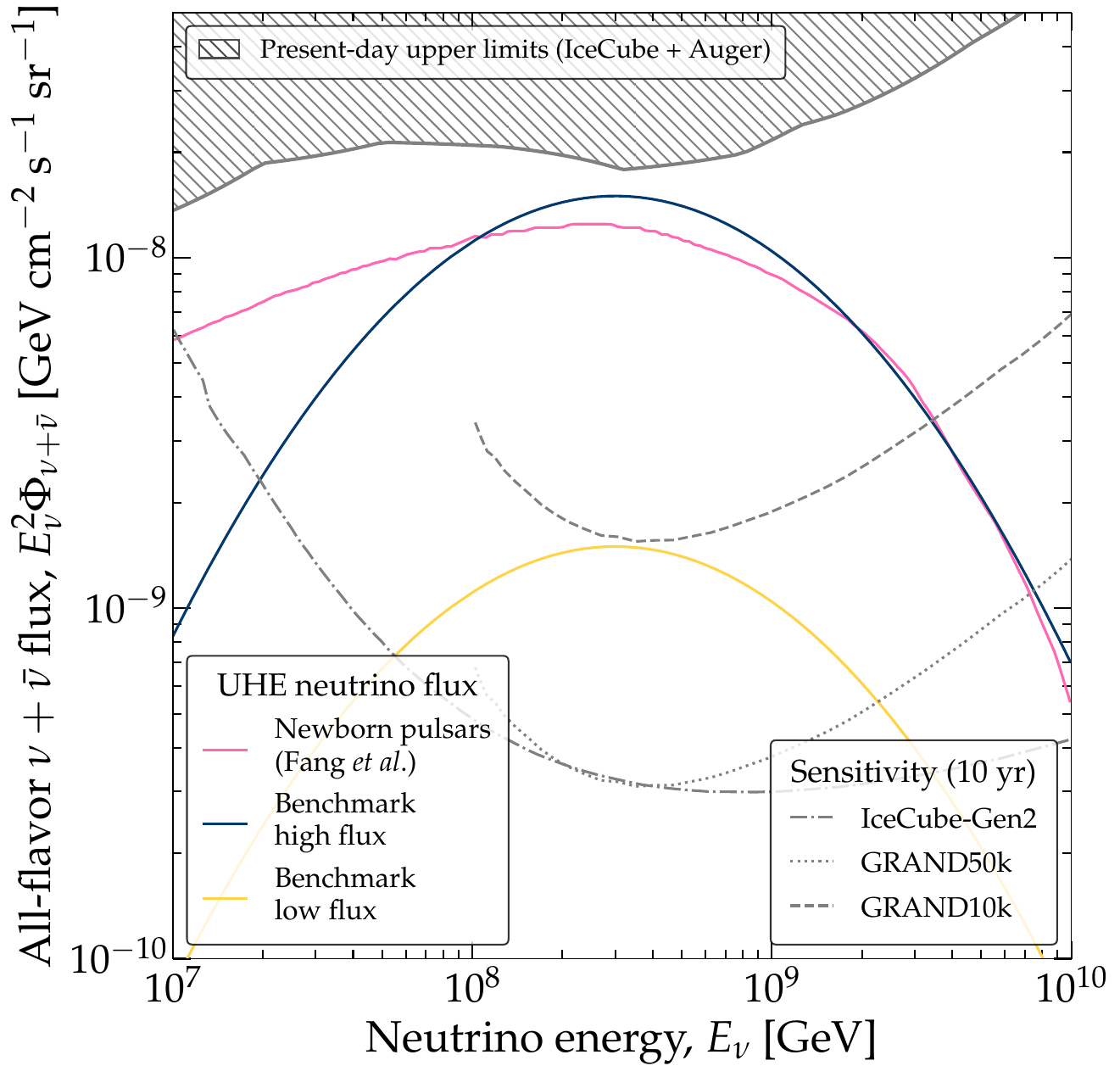}
 \caption{\textbf{\textit{UHE neutrino flux benchmark models used in our analysis.}}Our benchmark high and low flux models are built from \equ{flux_def}, based off of the predicted flux from newborn pulsars from \Refe~\cite{Fang:2013vla}. 
 The sensitivity of the radio array of IceCube-Gen2 is from \Refe~\cite{IceCube-Gen2:2021rkf}.  The sensitivities of GRAND50k and GRAND10k are computed by inverting \equ{aeff_gen2}, using the effective areas of GRAND50k and GRAND10.  The upper limits are from IceCube~\cite{IceCube:2018fhm} and Auger~\cite{PierreAuger:2019ens}.  }
 \label{fig:nu_flux}
\end{figure}

We assume neutrino production via pion decay, \ie, $\boldsymbol{f}_{\rm S} = \left( \frac{1}{3}, \frac{2}{3}, 0 \right)$.  We fix the normalization of the diffuse flux in \equ{diffuse_flux} by demanding that the all-flavor flux, $\sum_\alpha \Phi_\alpha$, approximates our high benchmark flux (\figu{nu_flux}).  This fixes the shape parameters to $\phi_0 = 2.2 \cdot 10^{-7}~ \mathrm{GeV}~ \mathrm{cm}^{-2}~ \mathrm{s}^{-1} ~\mathrm{sr}^{-1}$, $w = 0.18$, and $E_\textrm{bump} = 5 \cdot 10^8~\mathrm{GeV}$.  To reproduce our low benchmark flux, we use a normalization constant, $\phi_0$, ten times smaller.

Figure~\ref{fig:nu_flux} shows the all-flavor benchmark UHE  diffuse neutrino flux models that we use in our forecasts.  They are built using \equ{flux_def}, summed over all flavors.  Our high benchmark flux approximates the predicted flux emitted by newborn pulsars from \Refe~\cite{Fang:2013vla} by computing \equ{flux_def} with $\Phi_0 = 1.5 \cdot 10^{-8}~\mathrm{GeV}~\mathrm{cm}^{-2}~\mathrm{s}^{-1}~\mathrm{sr}^{-1}$, $w=0.25$, and $E_{\mathrm{bump}} = 3 \cdot 10^8~\mathrm{GeV}$.  The low flux has a normalization, $\Phi_0$, ten times smaller.  We use the flux shape given by \equ{flux_def} rather than directly the prediction of \Refe~\cite{Fang:2013vla} because doing so allows us to account for uncertainties in the flux shape by varying $\alpha$ and $E_{\rm bump}$.  For details about our choice of spectrum in \equ{flux_def}, see \Refe~\cite{Fiorillo:2022rft}.

\begin{figure}[t!]
 \centering
 \includegraphics[width=0.48\textwidth]{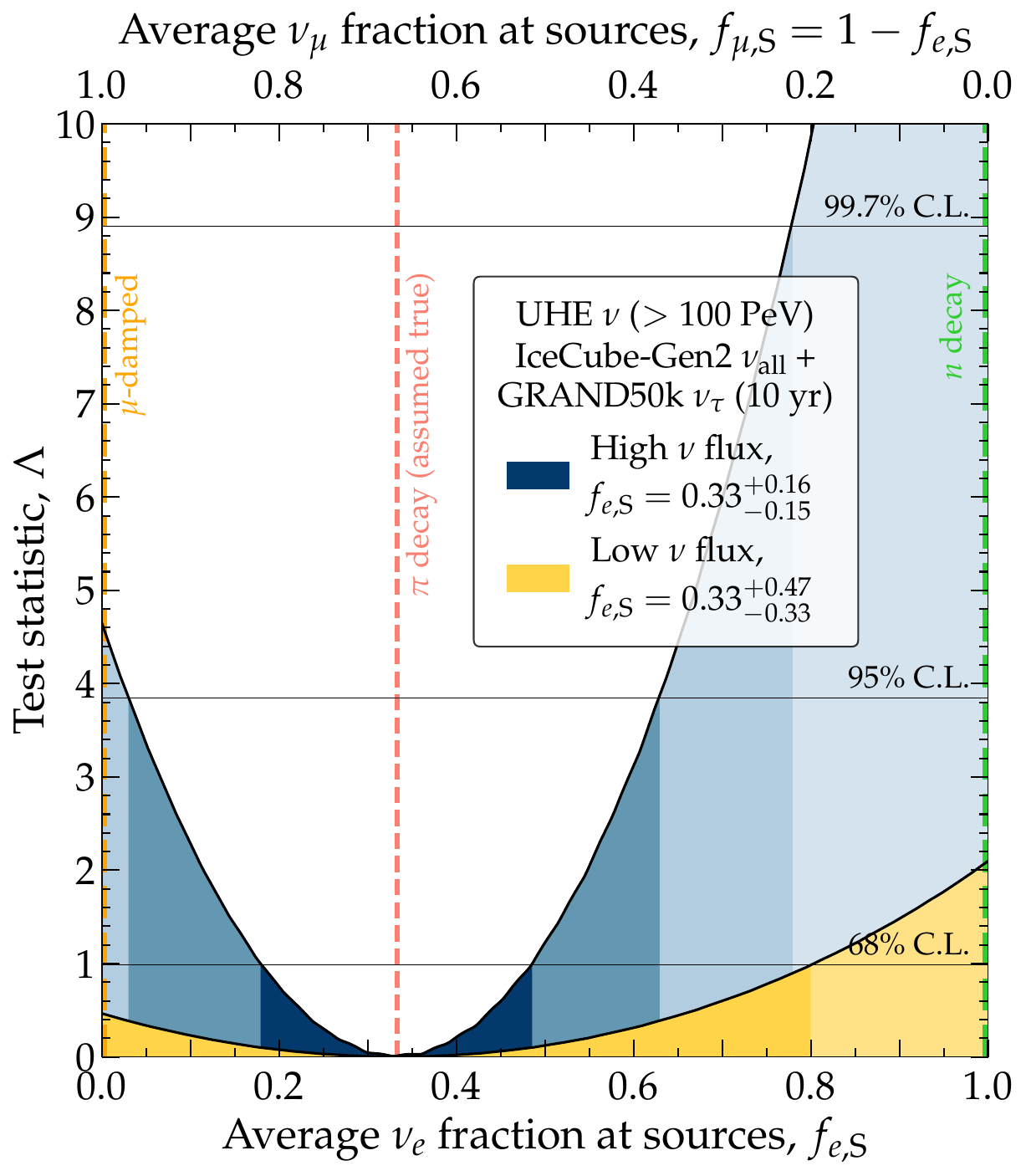}
 \caption{\textbf{\textit{Inferred flavor composition of UHE cosmic neutrinos at their sources.}}  The results are projections obtained from measuring the fraction $f_{\tau, \oplus}$ of $\nu_\tau$ at Earth by combining the detection by the radio array of IceCube-Gen2 and GRAND50k (\figu{ternary_grand50k}), using methods from \Refes~\cite{Bustamante:2019sdb, Song:2020nfh}.  We assume neutrino production via pion decay, no $\nu_\tau$ production (\ie, $f_{\tau, {\rm S}} = 0$), and two benchmark UHE neutrino fluxes, high and low. }
 \label{fig:likelihood_sources}
\end{figure}

\begin{figure*}[t!]
 \includegraphics[width=\textwidth]{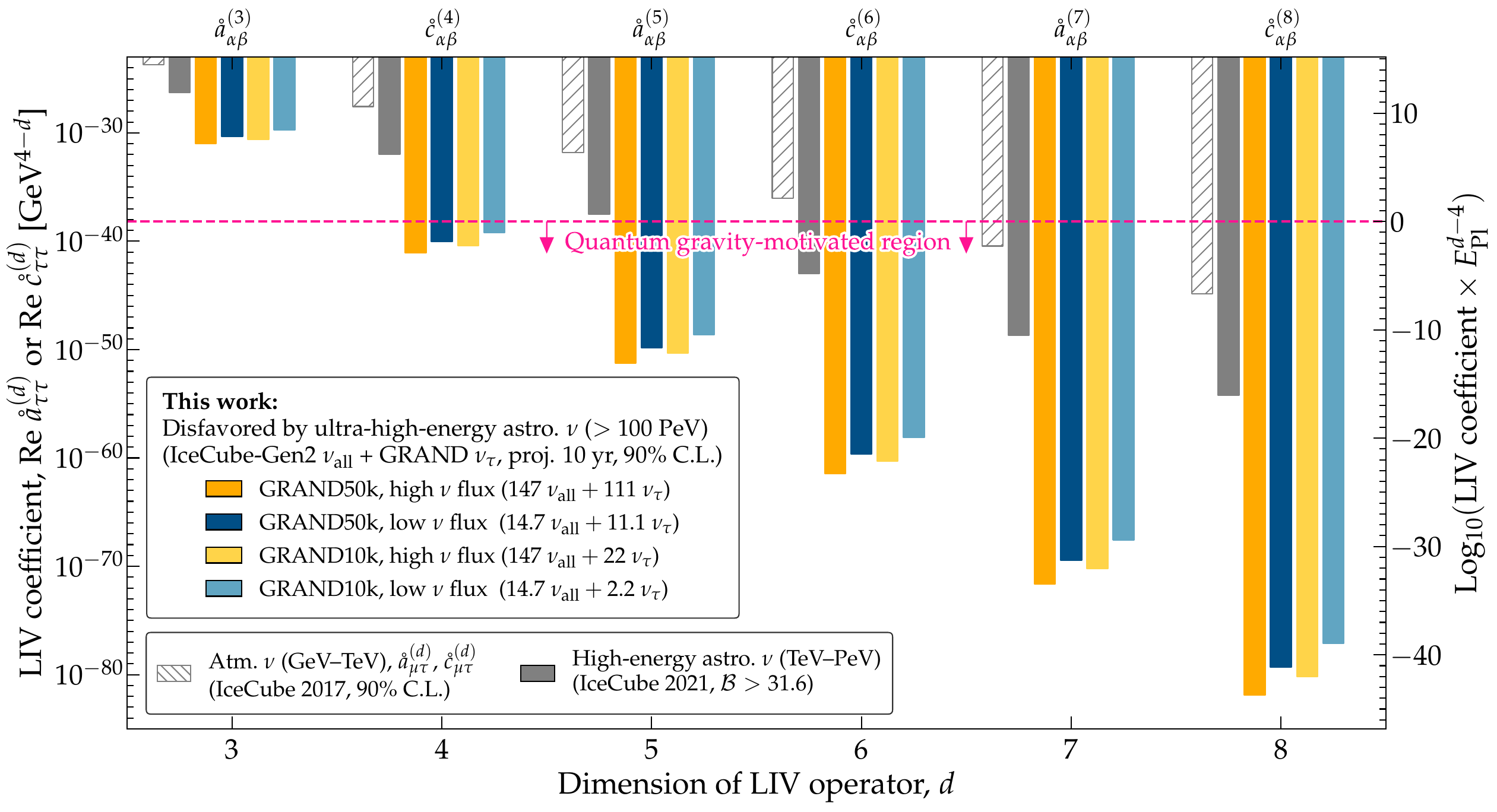}
 \caption{\textbf{\textit{Projected limits on LIV from UHE neutrinos.}}  Our limits come from combining the detection of neutrinos of all flavors by the radio array of IceCube-Gen2 and of $\nu_\tau$ by GRAND.  Limits are on the isotropic LIV coefficients $\mathring{a}_{\tau\tau}^{(d)}$ and $\mathring{c}_{\tau\tau}^{(d)}$ of the Standard Model Extension, which are CPT-odd and CPT-even, respectively. 
 Existing limits come IceCube observations of atmospheric neutrinos~\cite{IceCube:2017qyp} (on $\mathring{a}_{\mu\tau}^{(d)}$ and $\mathring{c}_{\mu\tau}^{(d)}$ instead) and TeV--PeV astrophysical neutrinos~\cite{IceCube:2021tdn}.  For the latter and our projections, we show limits assuming the canonical flavor composition at the sources, $\left(\frac{1}{3}:\frac{2}{3}:0\right)_{\rm S}$.  Our limits are profiled over the flavor composition at the sources, and the size and shape of the neutrino spectrum.}
 \label{fig:limits_tt_vs_dim}
\end{figure*}

\smallskip


\section{Detecting UHE neutrinos}
\label{sec:detector_char}
IceCube-Gen2~\cite{IceCube-Gen2:2020qha} and GRAND~\cite{GRAND:2018iaj} target the radio-detection of UHE neutrinos~\cite{Schroder:2016hrv}.  While they share common features, \ie, they are both affected by the severe attenuation of the UHE neutrino flux inside Earth and both exploit the long attenuation length of radio in ice and air, they adopt different, complementary strategies.  Below, we outline them. 

The radio array of IceCube-Gen2 looks for radio signals in-ice, emitted by particle showers triggered by UHE neutrinos scattering off nucleons~\cite{IceCube:2017roe, Bustamante:2017xuy, IceCube:2018pgc, IceCube:2020rnc}.  The showers accrue an excess of electrons that is discharged as an Askaryan~\cite{Askaryan:1961pfb} radio pulse~\cite{Halzen:1990vt, Zas:1991jv} detectable by antennas in the ice.  
The detection sensitivity is comparable for all neutrino flavors, albeit slightly better for $\nu_e$. 

GRAND looks for radio signals above ground, emitted by showers triggered by UHE neutrinos scattering with nucleons just below the surface, including in nearby mountains.  This strategy averts a large background of showers triggered instead by UHECRs.  GRAND targets predominantly Earth-skimming $\nu_\tau$~\cite{Fargion:1999se}, more resilient than $\nu_e$ and $\nu_\mu$ to in-Earth attenuation, whose interactions generate tauons that, upon exiting into the air, decay and trigger extensive air showers.  The negatively and positively charged particles in them are separated by the geomagnetic field, and the imbalance is discharged as a radio pulse~\cite{geomagnetic} detectable by antennas on the ground.  

\subsection{Detector response and event rate computation}
\label{sec:event_rate}
We compute the expected rate of detected events after 10~years of exposure, in bins of a quarter of a decade in energy from $10^7$ to $10^{10}$~GeV.  For IceCube-Gen2, we sum the contributions from all flavors, based on \Refes~\cite{IceCube-Gen2:2021rkf, vanSanten:2022wss}; for GRAND, we consider $\nu_\tau$ only, based on \Refe~\cite{GRAND:2018iaj}, and ignore potential subdominant contributions from other flavors.  We make our predictions conservative by considering only the partial configurations GRAND50k and GRAND10k, which represent one fourth and one tenth of the final size of GRAND, respectively.  

In an UHE neutrino detector (det), the differential number of events induced by $\nu_\alpha$ of energy $E_\nu$ is
\begin{equation}
 \label{equ:eventrate}
 \frac{dN_\alpha^\textrm{det}}{dE_\nu}=
 \Omega \,T \,\Phi_\alpha \,A_\alpha^\textrm{det} \;,
\end{equation}
where $\Omega$ is the solid angle of the sky to which the detector is sensitive, $T$ is the detector exposure time, and $A_\alpha^\textrm{det}$ is the effective area of the detector for $\nu_\alpha$.  We consider detection in the radio array of IceCube-Gen2 and in GRAND, for an exposure time of 10 years, representative of their design runtimes.

For the radio array of IceCube-Gen2, even though we do not attribute to it flavor-identification capabilities, we compute the contributions of neutrinos of different flavors separately.  To this end, we use flavor-specific effective areas, which we build in three steps.  First, we infer the flavor-averaged effective area from the differential IceCube-Gen2 sensitivity, $S(E_\nu)$, recently reported by the Collaboration in \Refe~\cite{IceCube-Gen2:2021rkf}, \ie,
\begin{equation}
 \label{equ:aeff_gen2}
 A_\text{avg}^\textrm{IC-Gen2}
 =
 \frac{2.44 E_\nu}{\Omega~S(E_\nu)~T \ln 10} \;,
\end{equation}
where the 2.44 events come from using the background-free prescription in the Feldman-Cousins approach~\cite{Feldman_1998}, and, in this case, $\Omega = 4 \pi$ because the sensitivity is reported as all-sky.  Second, we pick the flavor-specific effective areas from the $\texttt{toise}$ framework~\cite{vanSanten:2022wss} (see also \Refe~\cite{Glaser:2019cws}), $A_e^\textrm{IC-Gen2}$, $A_\mu^\textrm{IC-Gen2}$, and $A_\tau^\textrm{IC-Gen2}$ for a benchmark in-ice radio-based neutrino telescope; see Fig.~16 in \Refe~\cite{vanSanten:2022wss}.  From them, we compute the flavor-averaged effective area, $(\sum_\alpha A_\alpha^\textrm{IC-Gen2})/3$.  Third, and finally, we equate this to $A_\text{avg}^\textrm{IC-Gen2}$ from \equ{aeff_gen2} to reweigh the flavor-specific effective areas.  Thus, the resulting areas, which we use in our forecasts, reflect both the flavor sensitivity of the detector and its most recent flux-discovery potential.  Because we do not ascribe flavor-identification capabilities to the radio array of IceCube-Gen2, we use the event rate summed over all flavors, \ie,
\begin{equation}
 \frac{dN^\textrm{IC-Gen2}}{dE_\nu}
 =
 \sum_\alpha
 \frac{dN_\alpha^\textrm{IC-Gen2}}{dE_\nu} \;,
\end{equation}
where the contribution of each flavor is computed using \equ{aeff_gen2} with the flavor-specific effective area.

For GRAND, we start with the effective area of the full-sized array, GRAND200k, reported by the Collaboration in Fig.~25 of \Refe~\cite{GRAND:2018iaj}.  The effective area is defined for $\nu_\tau$ only and for detection of neutrinos from $3^\circ$--$4^\circ$ around the horizontal direction, including the contribution of a nearby mountain where neutrinos can interact, \ie, $\Omega = 2\pi(\cos{86^\circ} - \cos{93^\circ})$, where the angles are zenith angles measured from the South Pole.  For GRAND50k and GRAND10k, we divide the GRAND200k effective area by 4 and 20, respectively.  We ignore the potential subdominant contribution of neutrinos of other flavors interacting with the air near the detector~\cite{GRAND:2018iaj}, since it has not been estimated in detail yet.  Thus, for GRAND50k, we use the event rate due to $\nu_\tau$ only, \ie,
\begin{equation}
 \frac{dN^\textrm{GRAND50k}}{dE_\nu}
 =
 \frac{dN_\tau^\textrm{GRAND50k}}{dE_\nu} \;,
\end{equation}
computed using \equ{aeff_gen2} with the effective area of GRAND50k estimated as outlined above, and similarly for GRAND10k.

\begin{figure}[t!]
 \centering
 \includegraphics[width=\columnwidth]{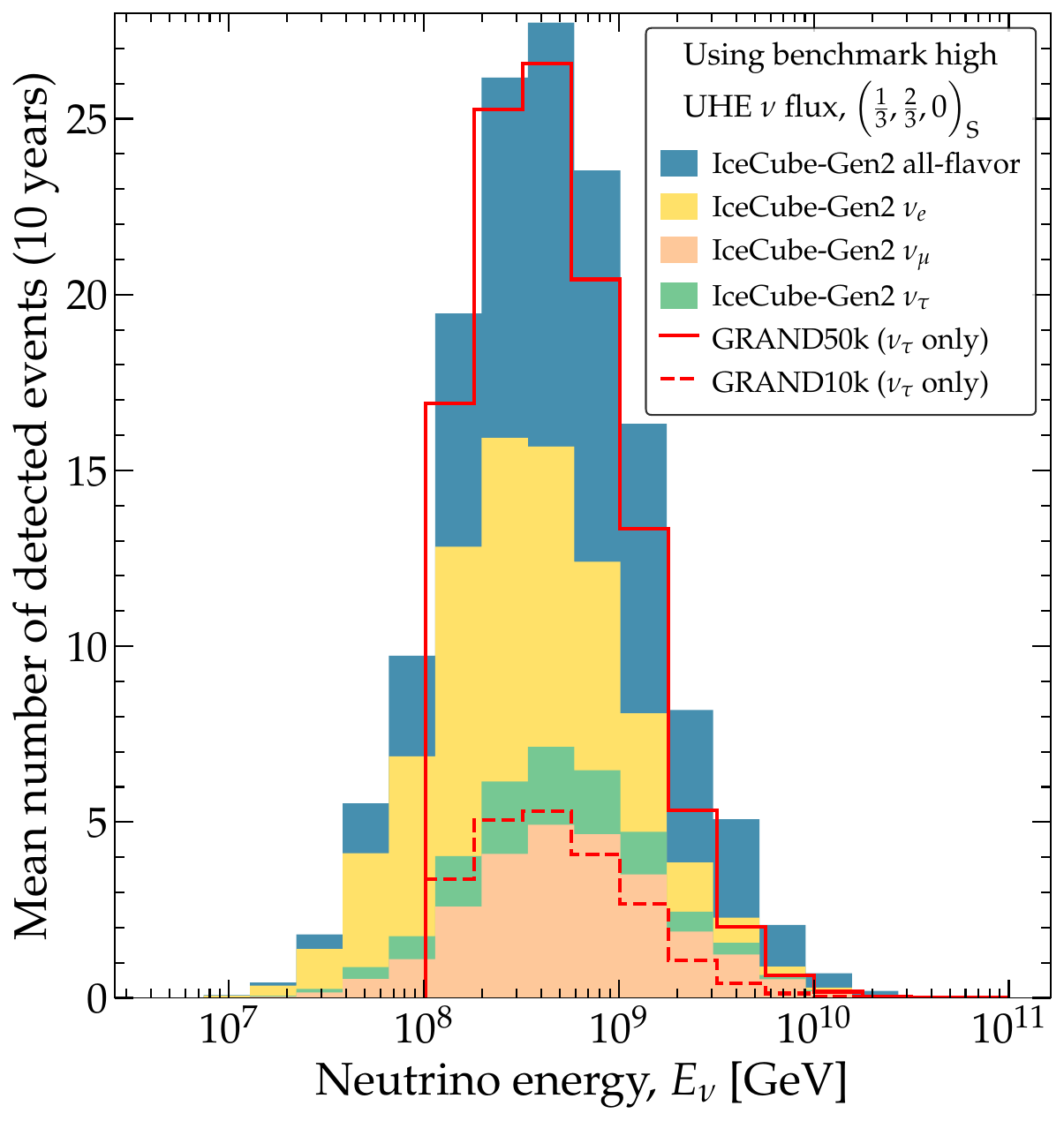}
 \caption{\textbf{\textit{Mean expected number of detected events in the radio array of IceCube-Gen2 and GRAND.}} In this plot, event rates are computed assuming our benchmark high UHE neutrino flux introduced in the main text.  For our benchmark low flux, the event rates (not shown) are ten times smaller.  The bin sizes are the same for both detectors, but their positions are slightly different because the energy ranges of IceCube-Gen2 and GRAND that we use are different.}
 \label{fig:event_rate}
\end{figure}

Figure~\ref{fig:event_rate} shows the mean expected energy distributions of events computed assuming our high benchmark UHE neutrino flux.  For each detector, we compute the event rate in the $i$-th energy bin by integrating the differential event rate over the width of the bin, $\Delta E_{\nu, i}$, \ie,
\begin{equation}
 N_i^\textrm{det}
 =
 \int_{\Delta E_{\nu, i}} \frac{dN^\textrm{det}}{dE_\nu} dE_\nu \;.
\end{equation}
We use four energy bins per energy decade, evenly spaced in logarithmic scale, from $5 \cdot 10^6$ to $5 \cdot 10^{10}$~GeV for IceCube-Gen2 and from $10^8$ to $10^{11}$~GeV for GRAND, the energy ranges reported in \Refes~\cite{IceCube-Gen2:2021rkf, GRAND:2018iaj}.

\smallskip


\section{Measuring flavor composition}
\label{sec:measurement}
For our high and low benchmark UHE neutrino flux separately, we compute the {\it true} mean expected number of detected events in IceCube-Gen2 and GRAND, $\bar{\mu}_\textrm{IC-Gen2}$ and $\bar{\mu}_{\rm GRAND}$.  Then, for {\it test} fluxes, \ie, for test values of the shape parameters $\boldsymbol{\theta} \equiv ( \Phi_0, w, E_{\rm bump} )$ and the flavor composition $\boldsymbol{f}_\oplus \equiv ( f_{e, \oplus}, f_{\mu, \oplus}, f_{\tau, \oplus})$ in \equ{flux_def}, we compute the mean expected numbers, $\mu_\textrm{IC-Gen2}(\boldsymbol{f}_\oplus, \boldsymbol{\theta})$ and $\mu_{\rm GRAND}(\boldsymbol{f}_\oplus, \boldsymbol{\theta})$.  We compare Asimov event samples---the mean true {\it vs.}~test rates---in each bin in each experiment via Poisson likelihood functions, and combine them into a total likelihood, \ie,
\begin{eqnarray}
 \label{equ:likelihood}
 \ln\mathcal{L}(\boldsymbol{f}_\oplus, \boldsymbol{\theta}) 
 &=& 
 \sum_{\textrm{exp}}\sum_i^{\rm bins}
 \left[\nonumber 
 \bar{\mu}_{i, \textrm{exp}} 
 \ln  \mu_{i, \textrm{exp}}(\boldsymbol{f}_\oplus, \boldsymbol{\theta})
 \right.
 \nonumber \\ 
 && \qquad
 \left.
 -~
 \mu_{i, \textrm{exp}}(\boldsymbol{f}_\oplus, \boldsymbol{\theta})
 \right] \;,
\end{eqnarray}
where $\textrm{exp} \in \{\textrm{IC-Gen2}, \textrm{GRAND}\}$.
We build our test statistic by minimizing the likelihood over the flux shape parameters, \ie, $\Lambda(\boldsymbol{f}_\oplus) = 2 \left[\mathrm{min}_{\boldsymbol{\theta}} \ln \mathcal{L}(\boldsymbol{f}_\oplus, \boldsymbol{\theta}) - \ln \bar{\mathcal{L}} \right]$, where $\bar{\mathcal{L}}$ is \equ{likelihood} computed with $\mu_{i, \textrm{IC-Gen2}}(\boldsymbol{f}_\oplus, \boldsymbol{\theta}) \to \bar{\mu}_{i, \textrm{IC-Gen2}}$ and $\mu_{i, \textrm{GRAND}}(\boldsymbol{f}_\oplus, \boldsymbol{\theta}) \to \bar{\mu}_{i, \textrm{GRAND}}$.  Wilks' theorem ensures that $\Lambda$ follows a $\chi^2$ distribution with two degrees of freedom, which we choose to be $f_{e, \oplus}$ and $f_{\mu, \oplus}$ (because $f_{\tau, \oplus} \equiv 1-f_{e, \oplus}-f_{\mu, \oplus}$, it is not independent).  We constrain $f_{e, \oplus}$ and $f_{\mu, \oplus}$ at the 68\% and 95\% confidence level (C.L.) by demanding $\Lambda = 2.28$ and 6, respectively.

Figure~\ref{fig:ternary_grand50k} shows the resulting allowed regions of flavor composition at Earth.  The allowed regions are roughly aligned with the $f_{\tau,\oplus}$ axis, since this is the predominant flavor fraction extracted from combining IceCube-Gen2 and GRAND.  The misalignment comes from IceCube-Gen2 being slightly more sensitive to $\nu_e$ than to the other flavors~\cite{vanSanten:2022wss} because they are more likely to trigger radio-emitting electromagnetic showers; this is what allows us to disfavor values of $f_{e, \oplus}$ that differ significantly from its true value of about $\frac{1}{3}$, especially those higher.  

If the flux is high, the true flavor composition, from pion decay, can be distinguished from the muon-damped one at nearly 95\%~C.L.~and from neutron decay at more than that.  If the flux is low, the three benchmarks become indistinguishable at 68\%~C.L.  Prima facie, this flavor sensitivity is scant---but it is only deceptively so.  Below, we show that, combined with a sensible assumption on $\nu_\tau$ production, it yields ample insight into astrophysics and fundamental physics.

\smallskip


\subsection{Flavor composition at the sources}
\label{sec:astro}
Knowledge of the flavor composition at Earth allows us to infer the flavor composition with which neutrinos are produced at their sources (S), $f_{\alpha, {\rm S}}$, averaged over all contributing sources, by undoing the effects of neutrino oscillations.  This could allow us to distinguish between competing neutrino production mechanisms~\cite{Mena:2014sja, Bustamante:2019sdb, Song:2020nfh}.

Following \Refe~\cite{Bustamante:2019sdb} (see also \Refes~\cite{Mena:2014sja, Song:2020nfh}), we infer the flavor composition at the sources by computing the same test statistic as above, but as a function of $\boldsymbol{f}_{\alpha, {\rm S}} \equiv (f_{e, {\rm S}}, f_{\mu, {\rm S}}, f_{\tau, {\rm S}})$, \ie, $\mathcal{L}(\boldsymbol{f}_{\rm S}, \boldsymbol{\theta}) = \mathcal{L}[\boldsymbol{f}_\oplus(\boldsymbol{f}_{\rm S}), \boldsymbol{\theta}]$, where $\boldsymbol{f}_\oplus$ is obtained from $\boldsymbol{f}_{\rm S}$ as indicated earlier.  We make two reasonable assumptions.  First, we ignore uncertainties in the mixing parameters since by the year 2040 they should be known precisely enough to render the effect of their uncertainty on $\boldsymbol{f}_\oplus$ negligible~\cite{Song:2020nfh}.  Second, we assume that there is no UHE $\nu_\tau$ production, so that we need only infer $f_{e, {\rm S}}$, since in that case $f_{\mu, {\rm S}} \equiv 1 - f_{e, {\rm S}}$; this reduces the likelihood to a single degree of freedom.  This is justified because $\nu_\tau$ come from the decay of charmed mesons, whose production is suppressed~\cite{Enberg:2008jm, Carpio:2020wzg, Bhattacharya:2023mmp}. 

Figure~\ref{fig:likelihood_sources} shows that our projections are promising.  Using our high benchmark flux, the true value of $f_{e, {\rm S}} = 1/3$ is inferred with enough precision to separate it from the alternative muon-damped and neutron-decay compositions at more than 95\%~C.L.~and 99.7\%~C.L., respectively.  Using our low flux, the separation worsens, but remains significant against neutron decay.  Similar conclusions hold when using GRAND10k and when assuming $(0:1:0)_{\rm S}$ as true instead; see Appendix~\ref{app:additional_flavor_composition}.  

Distinguishing between the pion-decay and muon-damped flavor composition could constrain the magnetic field intensity in the neutrino sources~\cite{Hummer:2010vx, Winter:2013cla, Bustamante:2020bxp} and, indirectly, their identity.  Further, since extragalactic magnetic fields are believed to be weak, inferring a flavor composition compatible with muon-damped could hint at the diffuse UHE neutrino flux being of astrophysical rather than cosmogenic origin.

\smallskip


\section{Lorentz-invariance violation}
\label{sec:LIV}
Neutrino oscillations en route to Earth may undergo effects beyond the Standard Model that manifest as deviations to the flavor composition compared to the expectation from standard oscillations~\cite{Mehta:2011qb, Rasmussen:2017ert, Ahlers:2018mkf, Ackermann:2019cxh, Arguelles:2019rbn, Ackermann:2022rqc, Arguelles:2022tki}.  Lorentz-invariance violation (LIV), postulated in proposals of quantum gravity, is capable of inducing large deviations~\cite{Barenboim:2003jm, Bustamante:2010nq, Arguelles:2015dca, Bustamante:2015waa, Rasmussen:2017ert}, making it particularly amenable to testing in experiments.   So far, there is no evidence for LIV~\cite{Kostelecky:2008ts}, but the strongest limits on it for neutrinos come from measurements of the flavor composition of the IceCube TeV--PeV astrophysical neutrinos~\cite{IceCube:2021tdn}.  Because the intensity of LIV is expected to grow with neutrino energy---possibly much faster than linearly---using UHE neutrinos promises vast improvement in the discovery and limiting opportunities.

We adopt the LIV treatment from the Standard Model Extension~\cite{Colladay:1998fq, Kostelecky:2003cr, Kostelecky:2008ts, Diaz:2011ia}, an effective field theory that couples neutrinos to spacetime features.  Under LIV, neutrinos are affected by a series of new CPT-odd and CPT-even operators, $\mathring{a}^{(d)}$ and $\mathring{c}^{(d)}$ of dimension $d$, each a $3 \times 3$ matrix in the flavor basis.  They modify neutrino mixing via the Hamiltonian $H = H_{\rm std} + H_{\rm LIV}$, where $H_{\rm std} = M^2/(2E_\nu)$ is the standard term that drives oscillations due to the difference in neutrino masses, with $M^2 \equiv (m_1^2, m_2^2, m_3^2)$ and $m_i$ the mass of the $i$-th mass eigenstate, and
\begin{equation}
 \label{equ:hamiltonian}
  H_{\rm LIV} 
  =
  \mathring{a}^{(3)}
  -
  E_\nu \cdot \mathring{c}^{(4)}
  +
  E_\nu^2 \cdot \mathring{a}^{(5)}
  -
  E_\nu^3 \cdot \mathring{c}^{(6)} 
  +
  \ldots \;.
\end{equation}
the contribution from LIV.  We compute the average flavor-transition probability as before, but using instead the energy-dependent matrix that diagonalizes the total Hamiltonian, $U^\prime(E_\nu, \mathring{a}_{\alpha\beta}^{(d)}, \mathring{c}_{\alpha\beta}^{(d)})$.
Specifically, the mixing probabilities introduced in Sec.~\ref{sec:determining_diffuse} are changed to
\begin{eqnarray}
 &&
 P_{\nu_\beta \to \nu_\alpha}
 (E_\nu, z, \mathring{a}_{\kappa\lambda}^{(d)}, \mathring{c}_{\kappa\lambda}^{(d)})
 \nonumber \\
 && \qquad
 =
 \sum_i 
 \lvert 
 U_{\beta i}^\prime
 \left[E_\nu(1+z)\right] 
 \rvert^2 
 \lvert 
 U_{\alpha i}^\prime(E_\nu) 
 \rvert^2 \;.
\end{eqnarray}

The flavor composition at Earth depends on energy, the redshift at which the neutrino was produced, $z$, since cosmological expansion dampens the energy, and the LIV coefficients, \ie, $f_{\alpha, \oplus}^\prime(E_\nu, z) = \sum_{\beta, i} \lvert U_{\beta i}^\prime\left[E_\nu(1+z)\right] \rvert^2 \lvert U_{\alpha i}^\prime(E_\nu) \rvert^2 f_{\beta, {\rm S}}$.  The values of the LIV coefficients are a priori unknown; below, we constrain them.  Since we are mostly sensitive to the $\nu_\tau$ fraction and since from pion decay most of the emitted flux is in $\nu_\mu$, we have higher sensitivity to coefficients that inhibit $\nu_\mu \to \nu_\tau$ oscillations, \ie, $\mathring{a}_{e\mu}^{(d)}$, $\mathring{a}_{\tau\tau}^{(d)}$, $\mathring{a}_{e\tau}^{(d)}$, and $\mathring{a}_{\mu\mu}^{(d)}$, and equivalently for $\mathring{c}_{\alpha\beta}^{(d)}$.

Figure~\ref{fig:ternary_grand50k} illustrates the effect of pure-$\tau \tau$ LIV [$(H_{\rm LIV})_{\tau\tau}$], with varying strength relative to standard oscillations [$(H_{\rm std})_{\tau\tau}$], on our three flavor-composition benchmarks.  Under weak LIV, we recover standard oscillations.  Under dominant LIV, $\nu_\tau$ mixing is suppressed in our simplified scenario where $(H_{\rm LIV})_{\tau\tau}$ is the only nonzero LIV component.  In-between, interference between the standard and LIV contributions creates the wiggles seen in \figu{ternary_grand50k}.  

\smallskip


\subsection{Constraints on LIV}

We compute the diffuse flux of UHE neutrinos at Earth under LIV from a nondescript population of sources distributed in redshift.  The flux is normalized to match our low or high benchmark fluxes.  We assume neutrino production via pion decay, \ie, $\left( \frac{1}{3}, \frac{2}{3}, 0\right)_{\rm S}$ (but marginalize over $f_{\alpha, {\rm S}}$ later).  Since the number of LIV coefficients is large, we turn on a single one at a time; in the main text, this is either $\mathring{a}_{\tau\tau}^{(d)}$ or $\mathring{c}_{\tau\tau}^{(d)}$.  To forecast constraints on them, we compute the expected event rates at IceCube-Gen2 and GRAND, and use a test statistic similar to the one introduced earlier, modified to yield upper limits~\cite{Cowan:2010js}, with a likelihood like \equ{likelihood} but now also dependent on the LIV coefficient, \eg, $\mathcal{L}(\mathring{a}_{\tau\tau}^{(d)}, \boldsymbol{f}_{\rm S}, \boldsymbol{\theta})$.  Like before, we profile over $\boldsymbol{f}_{\rm S}$ and $\boldsymbol{\theta}$, and assume no $\nu_\tau$ production.  In Appendix~\ref{app:LIV_operators} we provide results for different coefficients.

Figure~\ref{fig:limits_tt_vs_dim} shows our resulting limits for operators of dimension 3--8.  Even in the most pessimistic scenario---low neutrino flux and using GRAND10k---all of our projected limits are better than the present ones~\cite{IceCube:2021tdn} by orders of magnitude, since $H_{\rm LIV}$ grows with energy.  The relative improvement grows with operator dimension, as the energy dependence grows increasingly faster than linearly [\equ{hamiltonian}].  This is particularly dramatic for dimension-4 and -5 operators, where our limits reach into the quantum gravity-motivated region, unlike existing limits.

\smallskip


\section{Summary and outlook}
\label{sec:summary}

The near-future discovery of UHE neutrinos, with energies larger than 100~PeV, would bring transformative progress to astrophysics and fundamental physics.  Their flavor composition is a particularly versatile probe of both, yet is notoriously difficult to measure, since different neutrino flavors make similar signals in the detectors.  

We propose a new strategy to measure the UHE flavor composition that does not rely on individual UHE detectors having flavor-identification capabilities.  Combining the indistinct detection of all flavors of one detector---in our work, the radio array of IceCube-Gen2---with the innate sensitivity to predominantly $\nu_\tau$ of another one---in our work, GRAND---begets access to the fraction of UHE $\nu_\tau$.  This could be achieved even under conservative conditions: even if the neutrino flux is  low, with 10--20 events detected in 10 years, and using intermediate versions of GRAND, one fourth or one tenth of its full planned size.

This flavor sensitivity, though limited, is ripe with insight.  For astrophysics, it could identify the production mechanism of UHE neutrinos, distinguishing between benchmark expectations at 95\%~C.L.~or more.  For fundamental physics, it could lead to vast improvement in the constraints on Lorentz-invariance violation, emblematic of the power to test other UHE new physics, \eg, secret neutrino interactions and active-sterile neutrino mixing.

Our methods can be applied to other planned detectors, like RNO-G~\cite{RNO-G:2020rmc}, sensitive to all flavors, and POEMMA~\cite{POEMMA:2020ykm}, sensitive to $\nu_\tau$.  Further, recent work~\cite{Coleman:2024scd} has shown that in-ice radio-based UHE neutrino telescopes may measure the $\nu_e$ and $\nu_\mu + \nu_\tau$ fractions.  Our methods would  break the degeneracy between $\nu_\mu$ and $\nu_\tau$, providing access to the full flavor composition to tap into its inherent physics potential.

\smallskip


\section{Acknowledgments}

DF and MB are supported by the {\sc Villum Fonden} under project no.~29388. This work used resources provided by the High Performance Computing Center at the University of Copenhagen. This project has received funding from the European Union’s Horizon 2020 research and innovation program under the Marie Sklodowska-Curie Grant Agreement No.~847523 ‘INTERACTIONS’.


\bibliography{refs.bib}

\begin{thebibliography}{187}%
\makeatletter
\providecommand \@ifxundefined [1]{%
 \@ifx{#1\undefined}
}%
\providecommand \@ifnum [1]{%
 \ifnum #1\expandafter \@firstoftwo
 \else \expandafter \@secondoftwo
 \fi
}%
\providecommand \@ifx [1]{%
 \ifx #1\expandafter \@firstoftwo
 \else \expandafter \@secondoftwo
 \fi
}%
\providecommand \natexlab [1]{#1}%
\providecommand \enquote  [1]{``#1''}%
\providecommand \bibnamefont  [1]{#1}%
\providecommand \bibfnamefont [1]{#1}%
\providecommand \citenamefont [1]{#1}%
\providecommand \href@noop [0]{\@secondoftwo}%
\providecommand \href [0]{\begingroup \@sanitize@url \@href}%
\providecommand \@href[1]{\@@startlink{#1}\@@href}%
\providecommand \@@href[1]{\endgroup#1\@@endlink}%
\providecommand \@sanitize@url [0]{\catcode `\\12\catcode `\$12\catcode
  `\&12\catcode `\#12\catcode `\^12\catcode `\_12\catcode `\%12\relax}%
\providecommand \@@startlink[1]{}%
\providecommand \@@endlink[0]{}%
\providecommand \url  [0]{\begingroup\@sanitize@url \@url }%
\providecommand \@url [1]{\endgroup\@href {#1}{\urlprefix }}%
\providecommand \urlprefix  [0]{URL }%
\providecommand \Eprint [0]{\href }%
\providecommand \doibase [0]{https://doi.org/}%
\providecommand \selectlanguage [0]{\@gobble}%
\providecommand \bibinfo  [0]{\@secondoftwo}%
\providecommand \bibfield  [0]{\@secondoftwo}%
\providecommand \translation [1]{[#1]}%
\providecommand \BibitemOpen [0]{}%
\providecommand \bibitemStop [0]{}%
\providecommand \bibitemNoStop [0]{.\EOS\space}%
\providecommand \EOS [0]{\spacefactor3000\relax}%
\providecommand \BibitemShut  [1]{\csname bibitem#1\endcsname}%
\let\auto@bib@innerbib\@empty
\bibitem [{\citenamefont {Anchordoqui}\ \emph {et~al.}(2014)\citenamefont
  {Anchordoqui} \emph {et~al.}}]{Anchordoqui:2013dnh}%
  \BibitemOpen
  \bibfield  {author} {\bibinfo {author} {\bibfnamefont {L.~A.}\ \bibnamefont
  {Anchordoqui}} \emph {et~al.},\ }\bibfield  {title} {\bibinfo {title}
  {{Cosmic Neutrino Pevatrons: A Brand New Pathway to Astronomy, Astrophysics,
  and Particle Physics}},\ }\href {https://doi.org/10.1016/j.jheap.2014.01.001}
  {\bibfield  {journal} {\bibinfo  {journal} {JHEAp}\ }\textbf {\bibinfo
  {volume} {1-2}},\ \bibinfo {pages} {1} (\bibinfo {year} {2014})},\ \Eprint
  {https://arxiv.org/abs/1312.6587} {arXiv:1312.6587 [astro-ph.HE]}
  \BibitemShut {NoStop}%
\bibitem [{\citenamefont {Ahlers}\ and\ \citenamefont
  {Halzen}(2018)}]{Ahlers:2018fkn}%
  \BibitemOpen
  \bibfield  {author} {\bibinfo {author} {\bibfnamefont {M.}~\bibnamefont
  {Ahlers}}\ and\ \bibinfo {author} {\bibfnamefont {F.}~\bibnamefont
  {Halzen}},\ }\bibfield  {title} {\bibinfo {title} {{Opening a New Window onto
  the Universe with IceCube}},\ }\href
  {https://doi.org/10.1016/j.ppnp.2018.05.001} {\bibfield  {journal} {\bibinfo
  {journal} {Prog. Part. Nucl. Phys.}\ }\textbf {\bibinfo {volume} {102}},\
  \bibinfo {pages} {73} (\bibinfo {year} {2018})},\ \Eprint
  {https://arxiv.org/abs/1805.11112} {arXiv:1805.11112 [astro-ph.HE]}
  \BibitemShut {NoStop}%
\bibitem [{\citenamefont {Anchordoqui}(2019)}]{Anchordoqui:2018qom}%
  \BibitemOpen
  \bibfield  {author} {\bibinfo {author} {\bibfnamefont {L.~A.}\ \bibnamefont
  {Anchordoqui}},\ }\bibfield  {title} {\bibinfo {title} {{Ultra-High-Energy
  Cosmic Rays}},\ }\href {https://doi.org/10.1016/j.physrep.2019.01.002}
  {\bibfield  {journal} {\bibinfo  {journal} {Phys. Rept.}\ }\textbf {\bibinfo
  {volume} {801}},\ \bibinfo {pages} {1} (\bibinfo {year} {2019})},\ \Eprint
  {https://arxiv.org/abs/1807.09645} {arXiv:1807.09645 [astro-ph.HE]}
  \BibitemShut {NoStop}%
\bibitem [{\citenamefont {Ackermann}\ \emph
  {et~al.}(2019{\natexlab{a}})\citenamefont {Ackermann} \emph
  {et~al.}}]{Ackermann:2019ows}%
  \BibitemOpen
  \bibfield  {author} {\bibinfo {author} {\bibfnamefont {M.}~\bibnamefont
  {Ackermann}} \emph {et~al.},\ }\bibfield  {title} {\bibinfo {title}
  {{Astrophysics Uniquely Enabled by Observations of High-Energy Cosmic
  Neutrinos}},\ }\href@noop {} {\bibfield  {journal} {\bibinfo  {journal}
  {Bull. Am. Astron. Soc.}\ }\textbf {\bibinfo {volume} {51}},\ \bibinfo
  {pages} {185} (\bibinfo {year} {2019}{\natexlab{a}})},\ \Eprint
  {https://arxiv.org/abs/1903.04334} {arXiv:1903.04334 [astro-ph.HE]}
  \BibitemShut {NoStop}%
\bibitem [{\citenamefont {M\'esz\'aros}\ \emph {et~al.}(2019)\citenamefont
  {M\'esz\'aros}, \citenamefont {Fox}, \citenamefont {Hanna},\ and\
  \citenamefont {Murase}}]{Meszaros:2019xej}%
  \BibitemOpen
  \bibfield  {author} {\bibinfo {author} {\bibfnamefont {P.}~\bibnamefont
  {M\'esz\'aros}}, \bibinfo {author} {\bibfnamefont {D.~B.}\ \bibnamefont
  {Fox}}, \bibinfo {author} {\bibfnamefont {C.}~\bibnamefont {Hanna}},\ and\
  \bibinfo {author} {\bibfnamefont {K.}~\bibnamefont {Murase}},\ }\bibfield
  {title} {\bibinfo {title} {{Multi-Messenger Astrophysics}},\ }\href
  {https://doi.org/10.1038/s42254-019-0101-z} {\bibfield  {journal} {\bibinfo
  {journal} {Nature Rev. Phys.}\ }\textbf {\bibinfo {volume} {1}},\ \bibinfo
  {pages} {585} (\bibinfo {year} {2019})},\ \Eprint
  {https://arxiv.org/abs/1906.10212} {arXiv:1906.10212 [astro-ph.HE]}
  \BibitemShut {NoStop}%
\bibitem [{\citenamefont {Halzen}\ and\ \citenamefont
  {Kheirandish}(2019)}]{Halzen:2019qkf}%
  \BibitemOpen
  \bibfield  {author} {\bibinfo {author} {\bibfnamefont {F.}~\bibnamefont
  {Halzen}}\ and\ \bibinfo {author} {\bibfnamefont {A.}~\bibnamefont
  {Kheirandish}},\ }\bibfield  {title} {\bibinfo {title} {{Multimessenger
  Search for the Sources of Cosmic Rays Using Cosmic Neutrinos}},\ }\href
  {https://doi.org/10.3389/fspas.2019.00032} {\bibfield  {journal} {\bibinfo
  {journal} {Front. Astron. Space Sci.}\ }\textbf {\bibinfo {volume} {6}},\
  \bibinfo {pages} {32} (\bibinfo {year} {2019})}\BibitemShut {NoStop}%
\bibitem [{\citenamefont {Palladino}\ \emph {et~al.}(2020)\citenamefont
  {Palladino}, \citenamefont {Spurio},\ and\ \citenamefont
  {Vissani}}]{Palladino:2020jol}%
  \BibitemOpen
  \bibfield  {author} {\bibinfo {author} {\bibfnamefont {A.}~\bibnamefont
  {Palladino}}, \bibinfo {author} {\bibfnamefont {M.}~\bibnamefont {Spurio}},\
  and\ \bibinfo {author} {\bibfnamefont {F.}~\bibnamefont {Vissani}},\
  }\bibfield  {title} {\bibinfo {title} {{Neutrino Telescopes and High-Energy
  Cosmic Neutrinos}},\ }\href {https://doi.org/10.3390/universe6020030}
  {\bibfield  {journal} {\bibinfo  {journal} {Universe}\ }\textbf {\bibinfo
  {volume} {6}},\ \bibinfo {pages} {30} (\bibinfo {year} {2020})},\ \Eprint
  {https://arxiv.org/abs/2009.01919} {arXiv:2009.01919 [astro-ph.HE]}
  \BibitemShut {NoStop}%
\bibitem [{\citenamefont {Alves~Batista}\ \emph {et~al.}(2021)\citenamefont
  {Alves~Batista} \emph {et~al.}}]{AlvesBatista:2021eeu}%
  \BibitemOpen
  \bibfield  {author} {\bibinfo {author} {\bibfnamefont {R.}~\bibnamefont
  {Alves~Batista}} \emph {et~al.},\ }\bibfield  {title} {\bibinfo {title}
  {{EuCAPT White Paper: Opportunities and Challenges for Theoretical
  Astroparticle Physics in the Next Decade}},\ }\href@noop {} {\  (\bibinfo
  {year} {2021})},\ \Eprint {https://arxiv.org/abs/2110.10074}
  {arXiv:2110.10074 [astro-ph.HE]} \BibitemShut {NoStop}%
\bibitem [{\citenamefont {Ackermann}\ \emph {et~al.}(2022)\citenamefont
  {Ackermann} \emph {et~al.}}]{Ackermann:2022rqc}%
  \BibitemOpen
  \bibfield  {author} {\bibinfo {author} {\bibfnamefont {M.}~\bibnamefont
  {Ackermann}} \emph {et~al.},\ }\bibfield  {title} {\bibinfo {title}
  {{High-energy and ultra-high-energy neutrinos: A Snowmass white paper}},\
  }\href {https://doi.org/10.1016/j.jheap.2022.08.001} {\bibfield  {journal}
  {\bibinfo  {journal} {JHEAp}\ }\textbf {\bibinfo {volume} {36}},\ \bibinfo
  {pages} {55} (\bibinfo {year} {2022})},\ \Eprint
  {https://arxiv.org/abs/2203.08096} {arXiv:2203.08096 [hep-ph]} \BibitemShut
  {NoStop}%
\bibitem [{\citenamefont {Gu\'epin}\ \emph {et~al.}(2022)\citenamefont
  {Gu\'epin}, \citenamefont {Kotera},\ and\ \citenamefont
  {Oikonomou}}]{Guepin:2022qpl}%
  \BibitemOpen
  \bibfield  {author} {\bibinfo {author} {\bibfnamefont {C.}~\bibnamefont
  {Gu\'epin}}, \bibinfo {author} {\bibfnamefont {K.}~\bibnamefont {Kotera}},\
  and\ \bibinfo {author} {\bibfnamefont {F.}~\bibnamefont {Oikonomou}},\
  }\bibfield  {title} {\bibinfo {title} {{High-energy neutrino transients and
  the future of multi-messenger astronomy}},\ }\href
  {https://doi.org/10.1038/s42254-022-00504-9} {\bibfield  {journal} {\bibinfo
  {journal} {Nature Rev. Phys.}\ }\textbf {\bibinfo {volume} {4}},\ \bibinfo
  {pages} {697} (\bibinfo {year} {2022})},\ \Eprint
  {https://arxiv.org/abs/2207.12205} {arXiv:2207.12205 [astro-ph.HE]}
  \BibitemShut {NoStop}%
\bibitem [{\citenamefont {Gaisser}\ \emph {et~al.}(1995)\citenamefont
  {Gaisser}, \citenamefont {Halzen},\ and\ \citenamefont
  {Stanev}}]{Gaisser:1994yf}%
  \BibitemOpen
  \bibfield  {author} {\bibinfo {author} {\bibfnamefont {T.~K.}\ \bibnamefont
  {Gaisser}}, \bibinfo {author} {\bibfnamefont {F.}~\bibnamefont {Halzen}},\
  and\ \bibinfo {author} {\bibfnamefont {T.}~\bibnamefont {Stanev}},\
  }\bibfield  {title} {\bibinfo {title} {{Particle astrophysics with
  high-energy neutrinos}},\ }\href
  {https://doi.org/10.1016/0370-1573(95)00003-Y} {\bibfield  {journal}
  {\bibinfo  {journal} {Phys. Rept.}\ }\textbf {\bibinfo {volume} {258}},\
  \bibinfo {pages} {173} (\bibinfo {year} {1995})},\ \bibinfo {note} {[Erratum:
  Phys.Rept. 271, 355--356 (1996)]},\ \Eprint
  {https://arxiv.org/abs/hep-ph/9410384} {arXiv:hep-ph/9410384} \BibitemShut
  {NoStop}%
\bibitem [{\citenamefont {Ahlers}\ \emph
  {et~al.}(2018{\natexlab{a}})\citenamefont {Ahlers}, \citenamefont {Helbing},\
  and\ \citenamefont {P\'erez de~los Heros}}]{Ahlers:2018mkf}%
  \BibitemOpen
  \bibfield  {author} {\bibinfo {author} {\bibfnamefont {M.}~\bibnamefont
  {Ahlers}}, \bibinfo {author} {\bibfnamefont {K.}~\bibnamefont {Helbing}},\
  and\ \bibinfo {author} {\bibfnamefont {C.}~\bibnamefont {P\'erez de~los
  Heros}},\ }\bibfield  {title} {\bibinfo {title} {{Probing Particle Physics
  with IceCube}},\ }\href {https://doi.org/10.1140/epjc/s10052-018-6369-9}
  {\bibfield  {journal} {\bibinfo  {journal} {Eur. Phys. J. C}\ }\textbf
  {\bibinfo {volume} {78}},\ \bibinfo {pages} {924} (\bibinfo {year}
  {2018}{\natexlab{a}})},\ \Eprint {https://arxiv.org/abs/1806.05696}
  {arXiv:1806.05696 [astro-ph.HE]} \BibitemShut {NoStop}%
\bibitem [{\citenamefont {Arg\"uelles}\ \emph {et~al.}(2020)\citenamefont
  {Arg\"uelles}, \citenamefont {Bustamante}, \citenamefont {Kheirandish},
  \citenamefont {Palomares-Ruiz}, \citenamefont {Salvad\'o},\ and\
  \citenamefont {Vincent}}]{Arguelles:2019rbn}%
  \BibitemOpen
  \bibfield  {author} {\bibinfo {author} {\bibfnamefont {C.~A.}\ \bibnamefont
  {Arg\"uelles}}, \bibinfo {author} {\bibfnamefont {M.}~\bibnamefont
  {Bustamante}}, \bibinfo {author} {\bibfnamefont {A.}~\bibnamefont
  {Kheirandish}}, \bibinfo {author} {\bibfnamefont {S.}~\bibnamefont
  {Palomares-Ruiz}}, \bibinfo {author} {\bibfnamefont {J.}~\bibnamefont
  {Salvad\'o}},\ and\ \bibinfo {author} {\bibfnamefont {A.~C.}\ \bibnamefont
  {Vincent}},\ }\bibfield  {title} {\bibinfo {title} {{Fundamental physics with
  high-energy cosmic neutrinos today and in the future}},\ }\href
  {https://doi.org/10.22323/1.358.0849} {\bibfield  {journal} {\bibinfo
  {journal} {PoS}\ }\textbf {\bibinfo {volume} {ICRC2019}},\ \bibinfo {pages}
  {849} (\bibinfo {year} {2020})},\ \Eprint {https://arxiv.org/abs/1907.08690}
  {arXiv:1907.08690 [astro-ph.HE]} \BibitemShut {NoStop}%
\bibitem [{\citenamefont {Ackermann}\ \emph
  {et~al.}(2019{\natexlab{b}})\citenamefont {Ackermann} \emph
  {et~al.}}]{Ackermann:2019cxh}%
  \BibitemOpen
  \bibfield  {author} {\bibinfo {author} {\bibfnamefont {M.}~\bibnamefont
  {Ackermann}} \emph {et~al.},\ }\bibfield  {title} {\bibinfo {title}
  {{Fundamental Physics with High-Energy Cosmic Neutrinos}},\ }\href@noop {}
  {\bibfield  {journal} {\bibinfo  {journal} {Bull. Am. Astron. Soc.}\ }\textbf
  {\bibinfo {volume} {51}},\ \bibinfo {pages} {215} (\bibinfo {year}
  {2019}{\natexlab{b}})},\ \Eprint {https://arxiv.org/abs/1903.04333}
  {arXiv:1903.04333 [astro-ph.HE]} \BibitemShut {NoStop}%
\bibitem [{\citenamefont {Rachen}\ and\ \citenamefont
  {Meszaros}(1998)}]{Rachen:1998fd}%
  \BibitemOpen
  \bibfield  {author} {\bibinfo {author} {\bibfnamefont {J.~P.}\ \bibnamefont
  {Rachen}}\ and\ \bibinfo {author} {\bibfnamefont {P.}~\bibnamefont
  {Meszaros}},\ }\bibfield  {title} {\bibinfo {title} {{Photohadronic neutrinos
  from transients in astrophysical sources}},\ }\href
  {https://doi.org/10.1103/PhysRevD.58.123005} {\bibfield  {journal} {\bibinfo
  {journal} {Phys. Rev. D}\ }\textbf {\bibinfo {volume} {58}},\ \bibinfo
  {pages} {123005} (\bibinfo {year} {1998})},\ \Eprint
  {https://arxiv.org/abs/astro-ph/9802280} {arXiv:astro-ph/9802280}
  \BibitemShut {NoStop}%
\bibitem [{\citenamefont {Athar}\ \emph {et~al.}(2000)\citenamefont {Athar},
  \citenamefont {Jezabek},\ and\ \citenamefont {Yasuda}}]{Athar:2000yw}%
  \BibitemOpen
  \bibfield  {author} {\bibinfo {author} {\bibfnamefont {H.}~\bibnamefont
  {Athar}}, \bibinfo {author} {\bibfnamefont {M.}~\bibnamefont {Jezabek}},\
  and\ \bibinfo {author} {\bibfnamefont {O.}~\bibnamefont {Yasuda}},\
  }\bibfield  {title} {\bibinfo {title} {{Effects of neutrino mixing on
  high-energy cosmic neutrino flux}},\ }\href
  {https://doi.org/10.1103/PhysRevD.62.103007} {\bibfield  {journal} {\bibinfo
  {journal} {Phys. Rev. D}\ }\textbf {\bibinfo {volume} {62}},\ \bibinfo
  {pages} {103007} (\bibinfo {year} {2000})},\ \Eprint
  {https://arxiv.org/abs/hep-ph/0005104} {arXiv:hep-ph/0005104} \BibitemShut
  {NoStop}%
\bibitem [{\citenamefont {Crocker}\ \emph {et~al.}(2002)\citenamefont
  {Crocker}, \citenamefont {Melia},\ and\ \citenamefont
  {Volkas}}]{Crocker:2001zs}%
  \BibitemOpen
  \bibfield  {author} {\bibinfo {author} {\bibfnamefont {R.~M.}\ \bibnamefont
  {Crocker}}, \bibinfo {author} {\bibfnamefont {F.}~\bibnamefont {Melia}},\
  and\ \bibinfo {author} {\bibfnamefont {R.~R.}\ \bibnamefont {Volkas}},\
  }\bibfield  {title} {\bibinfo {title} {{Searching for long wavelength
  neutrino oscillations in the distorted neutrino spectrum of galactic
  supernova remnants}},\ }\href {https://doi.org/10.1086/340278} {\bibfield
  {journal} {\bibinfo  {journal} {Astrophys. J. Suppl.}\ }\textbf {\bibinfo
  {volume} {141}},\ \bibinfo {pages} {147} (\bibinfo {year} {2002})},\ \Eprint
  {https://arxiv.org/abs/astro-ph/0106090} {arXiv:astro-ph/0106090}
  \BibitemShut {NoStop}%
\bibitem [{\citenamefont {Beacom}\ \emph
  {et~al.}(2003{\natexlab{a}})\citenamefont {Beacom}, \citenamefont {Bell},
  \citenamefont {Hooper}, \citenamefont {Pakvasa},\ and\ \citenamefont
  {Weiler}}]{Beacom:2002vi}%
  \BibitemOpen
  \bibfield  {author} {\bibinfo {author} {\bibfnamefont {J.~F.}\ \bibnamefont
  {Beacom}}, \bibinfo {author} {\bibfnamefont {N.~F.}\ \bibnamefont {Bell}},
  \bibinfo {author} {\bibfnamefont {D.}~\bibnamefont {Hooper}}, \bibinfo
  {author} {\bibfnamefont {S.}~\bibnamefont {Pakvasa}},\ and\ \bibinfo {author}
  {\bibfnamefont {T.~J.}\ \bibnamefont {Weiler}},\ }\bibfield  {title}
  {\bibinfo {title} {{Decay of High-Energy Astrophysical Neutrinos}},\ }\href
  {https://doi.org/10.1103/PhysRevLett.90.181301} {\bibfield  {journal}
  {\bibinfo  {journal} {Phys. Rev. Lett.}\ }\textbf {\bibinfo {volume} {90}},\
  \bibinfo {pages} {181301} (\bibinfo {year} {2003}{\natexlab{a}})},\ \Eprint
  {https://arxiv.org/abs/hep-ph/0211305} {arXiv:hep-ph/0211305} \BibitemShut
  {NoStop}%
\bibitem [{\citenamefont {Barenboim}\ and\ \citenamefont
  {Quigg}(2003)}]{Barenboim:2003jm}%
  \BibitemOpen
  \bibfield  {author} {\bibinfo {author} {\bibfnamefont {G.}~\bibnamefont
  {Barenboim}}\ and\ \bibinfo {author} {\bibfnamefont {C.}~\bibnamefont
  {Quigg}},\ }\bibfield  {title} {\bibinfo {title} {{Neutrino observatories can
  characterize cosmic sources and neutrino properties}},\ }\href
  {https://doi.org/10.1103/PhysRevD.67.073024} {\bibfield  {journal} {\bibinfo
  {journal} {Phys. Rev. D}\ }\textbf {\bibinfo {volume} {67}},\ \bibinfo
  {pages} {073024} (\bibinfo {year} {2003})},\ \Eprint
  {https://arxiv.org/abs/hep-ph/0301220} {arXiv:hep-ph/0301220} \BibitemShut
  {NoStop}%
\bibitem [{\citenamefont {Beacom}\ \emph
  {et~al.}(2003{\natexlab{b}})\citenamefont {Beacom}, \citenamefont {Bell},
  \citenamefont {Hooper}, \citenamefont {Pakvasa},\ and\ \citenamefont
  {Weiler}}]{Beacom:2003nh}%
  \BibitemOpen
  \bibfield  {author} {\bibinfo {author} {\bibfnamefont {J.~F.}\ \bibnamefont
  {Beacom}}, \bibinfo {author} {\bibfnamefont {N.~F.}\ \bibnamefont {Bell}},
  \bibinfo {author} {\bibfnamefont {D.}~\bibnamefont {Hooper}}, \bibinfo
  {author} {\bibfnamefont {S.}~\bibnamefont {Pakvasa}},\ and\ \bibinfo {author}
  {\bibfnamefont {T.~J.}\ \bibnamefont {Weiler}},\ }\bibfield  {title}
  {\bibinfo {title} {{Measuring flavor ratios of high-energy astrophysical
  neutrinos}},\ }\href {https://doi.org/10.1103/PhysRevD.68.093005} {\bibfield
  {journal} {\bibinfo  {journal} {Phys. Rev. D}\ }\textbf {\bibinfo {volume}
  {68}},\ \bibinfo {pages} {093005} (\bibinfo {year} {2003}{\natexlab{b}})},\
  \bibinfo {note} {[Erratum: Phys.~Rev.~D 72, 019901 (2005)]},\ \Eprint
  {https://arxiv.org/abs/hep-ph/0307025} {arXiv:hep-ph/0307025} \BibitemShut
  {NoStop}%
\bibitem [{\citenamefont {Beacom}\ \emph
  {et~al.}(2004{\natexlab{a}})\citenamefont {Beacom}, \citenamefont {Bell},
  \citenamefont {Hooper}, \citenamefont {Learned}, \citenamefont {Pakvasa},\
  and\ \citenamefont {Weiler}}]{Beacom:2003eu}%
  \BibitemOpen
  \bibfield  {author} {\bibinfo {author} {\bibfnamefont {J.~F.}\ \bibnamefont
  {Beacom}}, \bibinfo {author} {\bibfnamefont {N.~F.}\ \bibnamefont {Bell}},
  \bibinfo {author} {\bibfnamefont {D.}~\bibnamefont {Hooper}}, \bibinfo
  {author} {\bibfnamefont {J.~G.}\ \bibnamefont {Learned}}, \bibinfo {author}
  {\bibfnamefont {S.}~\bibnamefont {Pakvasa}},\ and\ \bibinfo {author}
  {\bibfnamefont {T.~J.}\ \bibnamefont {Weiler}},\ }\bibfield  {title}
  {\bibinfo {title} {{PseudoDirac neutrinos: A Challenge for neutrino
  telescopes}},\ }\href {https://doi.org/10.1103/PhysRevLett.92.011101}
  {\bibfield  {journal} {\bibinfo  {journal} {Phys. Rev. Lett.}\ }\textbf
  {\bibinfo {volume} {92}},\ \bibinfo {pages} {011101} (\bibinfo {year}
  {2004}{\natexlab{a}})},\ \Eprint {https://arxiv.org/abs/hep-ph/0307151}
  {arXiv:hep-ph/0307151} \BibitemShut {NoStop}%
\bibitem [{\citenamefont {Beacom}\ \emph
  {et~al.}(2004{\natexlab{b}})\citenamefont {Beacom}, \citenamefont {Bell},
  \citenamefont {Hooper}, \citenamefont {Pakvasa},\ and\ \citenamefont
  {Weiler}}]{Beacom:2003zg}%
  \BibitemOpen
  \bibfield  {author} {\bibinfo {author} {\bibfnamefont {J.~F.}\ \bibnamefont
  {Beacom}}, \bibinfo {author} {\bibfnamefont {N.~F.}\ \bibnamefont {Bell}},
  \bibinfo {author} {\bibfnamefont {D.}~\bibnamefont {Hooper}}, \bibinfo
  {author} {\bibfnamefont {S.}~\bibnamefont {Pakvasa}},\ and\ \bibinfo {author}
  {\bibfnamefont {T.~J.}\ \bibnamefont {Weiler}},\ }\bibfield  {title}
  {\bibinfo {title} {{Sensitivity to $\theta_{13}$ and $\delta$ in the decaying
  astrophysical neutrino scenario}},\ }\href
  {https://doi.org/10.1103/PhysRevD.69.017303} {\bibfield  {journal} {\bibinfo
  {journal} {Phys. Rev. D}\ }\textbf {\bibinfo {volume} {69}},\ \bibinfo
  {pages} {017303} (\bibinfo {year} {2004}{\natexlab{b}})},\ \Eprint
  {https://arxiv.org/abs/hep-ph/0309267} {arXiv:hep-ph/0309267} \BibitemShut
  {NoStop}%
\bibitem [{\citenamefont {Beacom}\ and\ \citenamefont
  {Candia}(2004)}]{Beacom:2004jb}%
  \BibitemOpen
  \bibfield  {author} {\bibinfo {author} {\bibfnamefont {J.~F.}\ \bibnamefont
  {Beacom}}\ and\ \bibinfo {author} {\bibfnamefont {J.}~\bibnamefont
  {Candia}},\ }\bibfield  {title} {\bibinfo {title} {{Shower power: Isolating
  the prompt atmospheric neutrino flux using electron neutrinos}},\ }\href
  {https://doi.org/10.1088/1475-7516/2004/11/009} {\bibfield  {journal}
  {\bibinfo  {journal} {JCAP}\ }\textbf {\bibinfo {volume} {11}},\ \bibinfo
  {pages} {009}},\ \Eprint {https://arxiv.org/abs/hep-ph/0409046}
  {arXiv:hep-ph/0409046} \BibitemShut {NoStop}%
\bibitem [{\citenamefont {Serpico}(2006)}]{Serpico:2005bs}%
  \BibitemOpen
  \bibfield  {author} {\bibinfo {author} {\bibfnamefont {P.~D.}\ \bibnamefont
  {Serpico}},\ }\bibfield  {title} {\bibinfo {title} {{Probing the 2-3 leptonic
  mixing at high-energy neutrino telescopes}},\ }\href
  {https://doi.org/10.1103/PhysRevD.73.047301} {\bibfield  {journal} {\bibinfo
  {journal} {Phys. Rev. D}\ }\textbf {\bibinfo {volume} {73}},\ \bibinfo
  {pages} {047301} (\bibinfo {year} {2006})},\ \Eprint
  {https://arxiv.org/abs/hep-ph/0511313} {arXiv:hep-ph/0511313} \BibitemShut
  {NoStop}%
\bibitem [{\citenamefont {Kashti}\ and\ \citenamefont
  {Waxman}(2005)}]{Kashti:2005qa}%
  \BibitemOpen
  \bibfield  {author} {\bibinfo {author} {\bibfnamefont {T.}~\bibnamefont
  {Kashti}}\ and\ \bibinfo {author} {\bibfnamefont {E.}~\bibnamefont
  {Waxman}},\ }\bibfield  {title} {\bibinfo {title} {{Flavoring astrophysical
  neutrinos: Flavor ratios depend on energy}},\ }\href
  {https://doi.org/10.1103/PhysRevLett.95.181101} {\bibfield  {journal}
  {\bibinfo  {journal} {Phys. Rev. Lett.}\ }\textbf {\bibinfo {volume} {95}},\
  \bibinfo {pages} {181101} (\bibinfo {year} {2005})},\ \Eprint
  {https://arxiv.org/abs/astro-ph/0507599} {arXiv:astro-ph/0507599}
  \BibitemShut {NoStop}%
\bibitem [{\citenamefont {Mena}\ \emph {et~al.}(2007)\citenamefont {Mena},
  \citenamefont {Mocioiu},\ and\ \citenamefont {Razzaque}}]{Mena:2006eq}%
  \BibitemOpen
  \bibfield  {author} {\bibinfo {author} {\bibfnamefont {O.}~\bibnamefont
  {Mena}}, \bibinfo {author} {\bibfnamefont {I.}~\bibnamefont {Mocioiu}},\ and\
  \bibinfo {author} {\bibfnamefont {S.}~\bibnamefont {Razzaque}},\ }\bibfield
  {title} {\bibinfo {title} {{Oscillation effects on high-energy neutrino
  fluxes from astrophysical hidden sources}},\ }\href
  {https://doi.org/10.1103/PhysRevD.75.063003} {\bibfield  {journal} {\bibinfo
  {journal} {Phys. Rev. D}\ }\textbf {\bibinfo {volume} {75}},\ \bibinfo
  {pages} {063003} (\bibinfo {year} {2007})},\ \Eprint
  {https://arxiv.org/abs/astro-ph/0612325} {arXiv:astro-ph/0612325}
  \BibitemShut {NoStop}%
\bibitem [{\citenamefont {Kachelrie{\ss}}\ and\ \citenamefont
  {Tom{\`a}s}(2006)}]{Kachelriess:2006ksy}%
  \BibitemOpen
  \bibfield  {author} {\bibinfo {author} {\bibfnamefont {M.}~\bibnamefont
  {Kachelrie{\ss}}}\ and\ \bibinfo {author} {\bibfnamefont {R.}~\bibnamefont
  {Tom{\`a}s}},\ }\bibfield  {title} {\bibinfo {title} {{High energy neutrino
  yields from astrophysical sources I: Weakly magnetized sources}},\ }\href
  {https://doi.org/10.1103/PhysRevD.74.063009} {\bibfield  {journal} {\bibinfo
  {journal} {Phys. Rev. D}\ }\textbf {\bibinfo {volume} {74}},\ \bibinfo
  {pages} {063009} (\bibinfo {year} {2006})},\ \Eprint
  {https://arxiv.org/abs/astro-ph/0606406} {arXiv:astro-ph/0606406}
  \BibitemShut {NoStop}%
\bibitem [{\citenamefont {Lipari}\ \emph {et~al.}(2007)\citenamefont {Lipari},
  \citenamefont {Lusignoli},\ and\ \citenamefont {Meloni}}]{Lipari:2007su}%
  \BibitemOpen
  \bibfield  {author} {\bibinfo {author} {\bibfnamefont {P.}~\bibnamefont
  {Lipari}}, \bibinfo {author} {\bibfnamefont {M.}~\bibnamefont {Lusignoli}},\
  and\ \bibinfo {author} {\bibfnamefont {D.}~\bibnamefont {Meloni}},\
  }\bibfield  {title} {\bibinfo {title} {{Flavor Composition and Energy
  Spectrum of Astrophysical Neutrinos}},\ }\href
  {https://doi.org/10.1103/PhysRevD.75.123005} {\bibfield  {journal} {\bibinfo
  {journal} {Phys. Rev. D}\ }\textbf {\bibinfo {volume} {75}},\ \bibinfo
  {pages} {123005} (\bibinfo {year} {2007})},\ \Eprint
  {https://arxiv.org/abs/0704.0718} {arXiv:0704.0718 [astro-ph]} \BibitemShut
  {NoStop}%
\bibitem [{\citenamefont {Pakvasa}\ \emph {et~al.}(2008)\citenamefont
  {Pakvasa}, \citenamefont {Rodejohann},\ and\ \citenamefont
  {Weiler}}]{Pakvasa:2007dc}%
  \BibitemOpen
  \bibfield  {author} {\bibinfo {author} {\bibfnamefont {S.}~\bibnamefont
  {Pakvasa}}, \bibinfo {author} {\bibfnamefont {W.}~\bibnamefont
  {Rodejohann}},\ and\ \bibinfo {author} {\bibfnamefont {T.~J.}\ \bibnamefont
  {Weiler}},\ }\bibfield  {title} {\bibinfo {title} {{Flavor Ratios of
  Astrophysical Neutrinos: Implications for Precision Measurements}},\ }\href
  {https://doi.org/10.1088/1126-6708/2008/02/005} {\bibfield  {journal}
  {\bibinfo  {journal} {JHEP}\ }\textbf {\bibinfo {volume} {02}},\ \bibinfo
  {pages} {005}},\ \Eprint {https://arxiv.org/abs/0711.4517} {arXiv:0711.4517
  [hep-ph]} \BibitemShut {NoStop}%
\bibitem [{\citenamefont {Esmaili}\ and\ \citenamefont
  {Farzan}(2009)}]{Esmaili:2009dz}%
  \BibitemOpen
  \bibfield  {author} {\bibinfo {author} {\bibfnamefont {A.}~\bibnamefont
  {Esmaili}}\ and\ \bibinfo {author} {\bibfnamefont {Y.}~\bibnamefont
  {Farzan}},\ }\bibfield  {title} {\bibinfo {title} {{An Analysis of Cosmic
  Neutrinos: Flavor Composition at Source and Neutrino Mixing Parameters}},\
  }\href {https://doi.org/10.1016/j.nuclphysb.2009.06.017} {\bibfield
  {journal} {\bibinfo  {journal} {Nucl. Phys. B}\ }\textbf {\bibinfo {volume}
  {821}},\ \bibinfo {pages} {197} (\bibinfo {year} {2009})},\ \Eprint
  {https://arxiv.org/abs/0905.0259} {arXiv:0905.0259 [hep-ph]} \BibitemShut
  {NoStop}%
\bibitem [{\citenamefont {Choubey}\ and\ \citenamefont
  {Rodejohann}(2009)}]{Choubey:2009jq}%
  \BibitemOpen
  \bibfield  {author} {\bibinfo {author} {\bibfnamefont {S.}~\bibnamefont
  {Choubey}}\ and\ \bibinfo {author} {\bibfnamefont {W.}~\bibnamefont
  {Rodejohann}},\ }\bibfield  {title} {\bibinfo {title} {{Flavor Composition of
  UHE Neutrinos at Source and at Neutrino Telescopes}},\ }\href
  {https://doi.org/10.1103/PhysRevD.80.113006} {\bibfield  {journal} {\bibinfo
  {journal} {Phys. Rev. D}\ }\textbf {\bibinfo {volume} {80}},\ \bibinfo
  {pages} {113006} (\bibinfo {year} {2009})},\ \Eprint
  {https://arxiv.org/abs/0909.1219} {arXiv:0909.1219 [hep-ph]} \BibitemShut
  {NoStop}%
\bibitem [{\citenamefont {Esmaili}(2010)}]{Esmaili:2009fk}%
  \BibitemOpen
  \bibfield  {author} {\bibinfo {author} {\bibfnamefont {A.}~\bibnamefont
  {Esmaili}},\ }\bibfield  {title} {\bibinfo {title} {{Pseudo-Dirac Neutrino
  Scenario: Cosmic Neutrinos at Neutrino Telescopes}},\ }\href
  {https://doi.org/10.1103/PhysRevD.81.013006} {\bibfield  {journal} {\bibinfo
  {journal} {Phys. Rev. D}\ }\textbf {\bibinfo {volume} {81}},\ \bibinfo
  {pages} {013006} (\bibinfo {year} {2010})},\ \Eprint
  {https://arxiv.org/abs/0909.5410} {arXiv:0909.5410 [hep-ph]} \BibitemShut
  {NoStop}%
\bibitem [{\citenamefont {Bhattacharya}\ \emph
  {et~al.}(2010{\natexlab{a}})\citenamefont {Bhattacharya}, \citenamefont
  {Choubey}, \citenamefont {Gandhi},\ and\ \citenamefont
  {Watanabe}}]{Bhattacharya:2009tx}%
  \BibitemOpen
  \bibfield  {author} {\bibinfo {author} {\bibfnamefont {A.}~\bibnamefont
  {Bhattacharya}}, \bibinfo {author} {\bibfnamefont {S.}~\bibnamefont
  {Choubey}}, \bibinfo {author} {\bibfnamefont {R.}~\bibnamefont {Gandhi}},\
  and\ \bibinfo {author} {\bibfnamefont {A.}~\bibnamefont {Watanabe}},\
  }\bibfield  {title} {\bibinfo {title} {{Diffuse Ultra-High Energy Neutrino
  Fluxes and Physics Beyond the Standard Model}},\ }\href
  {https://doi.org/10.1016/j.physletb.2010.04.078} {\bibfield  {journal}
  {\bibinfo  {journal} {Phys. Lett. B}\ }\textbf {\bibinfo {volume} {690}},\
  \bibinfo {pages} {42} (\bibinfo {year} {2010}{\natexlab{a}})},\ \Eprint
  {https://arxiv.org/abs/0910.4396} {arXiv:0910.4396 [hep-ph]} \BibitemShut
  {NoStop}%
\bibitem [{\citenamefont {H{\"u}mmer}\ \emph
  {et~al.}(2010{\natexlab{a}})\citenamefont {H{\"u}mmer}, \citenamefont
  {Maltoni}, \citenamefont {Winter},\ and\ \citenamefont
  {Yaguna}}]{Hummer:2010ai}%
  \BibitemOpen
  \bibfield  {author} {\bibinfo {author} {\bibfnamefont {S.}~\bibnamefont
  {H{\"u}mmer}}, \bibinfo {author} {\bibfnamefont {M.}~\bibnamefont {Maltoni}},
  \bibinfo {author} {\bibfnamefont {W.}~\bibnamefont {Winter}},\ and\ \bibinfo
  {author} {\bibfnamefont {C.}~\bibnamefont {Yaguna}},\ }\bibfield  {title}
  {\bibinfo {title} {{Energy dependent neutrino flavor ratios from cosmic
  accelerators on the Hillas plot}},\ }\href
  {https://doi.org/10.1016/j.astropartphys.2010.07.003} {\bibfield  {journal}
  {\bibinfo  {journal} {Astropart. Phys.}\ }\textbf {\bibinfo {volume} {34}},\
  \bibinfo {pages} {205} (\bibinfo {year} {2010}{\natexlab{a}})},\ \Eprint
  {https://arxiv.org/abs/1007.0006} {arXiv:1007.0006 [astro-ph.HE]}
  \BibitemShut {NoStop}%
\bibitem [{\citenamefont {Bhattacharya}\ \emph
  {et~al.}(2010{\natexlab{b}})\citenamefont {Bhattacharya}, \citenamefont
  {Choubey}, \citenamefont {Gandhi},\ and\ \citenamefont
  {Watanabe}}]{Bhattacharya:2010xj}%
  \BibitemOpen
  \bibfield  {author} {\bibinfo {author} {\bibfnamefont {A.}~\bibnamefont
  {Bhattacharya}}, \bibinfo {author} {\bibfnamefont {S.}~\bibnamefont
  {Choubey}}, \bibinfo {author} {\bibfnamefont {R.}~\bibnamefont {Gandhi}},\
  and\ \bibinfo {author} {\bibfnamefont {A.}~\bibnamefont {Watanabe}},\
  }\bibfield  {title} {\bibinfo {title} {{Ultra-high neutrino fluxes as a probe
  for non-standard physics}},\ }\href
  {https://doi.org/10.1088/1475-7516/2010/09/009} {\bibfield  {journal}
  {\bibinfo  {journal} {JCAP}\ }\textbf {\bibinfo {volume} {09}},\ \bibinfo
  {pages} {009}},\ \Eprint {https://arxiv.org/abs/1006.3082} {arXiv:1006.3082
  [hep-ph]} \BibitemShut {NoStop}%
\bibitem [{\citenamefont {Bustamante}\ \emph {et~al.}(2010)\citenamefont
  {Bustamante}, \citenamefont {Gago},\ and\ \citenamefont
  {Pe{\~n}a-Garay}}]{Bustamante:2010nq}%
  \BibitemOpen
  \bibfield  {author} {\bibinfo {author} {\bibfnamefont {M.}~\bibnamefont
  {Bustamante}}, \bibinfo {author} {\bibfnamefont {A.~M.}\ \bibnamefont
  {Gago}},\ and\ \bibinfo {author} {\bibfnamefont {C.}~\bibnamefont
  {Pe{\~n}a-Garay}},\ }\bibfield  {title} {\bibinfo {title}
  {{Energy-Independent New Physics in the Flavour Ratios of High-Energy
  Astrophysical Neutrinos}},\ }\href {https://doi.org/10.1007/JHEP04(2010)066}
  {\bibfield  {journal} {\bibinfo  {journal} {JHEP}\ }\textbf {\bibinfo
  {volume} {04}},\ \bibinfo {pages} {066}},\ \Eprint
  {https://arxiv.org/abs/1001.4878} {arXiv:1001.4878 [hep-ph]} \BibitemShut
  {NoStop}%
\bibitem [{\citenamefont {Mehta}\ and\ \citenamefont
  {Winter}(2011)}]{Mehta:2011qb}%
  \BibitemOpen
  \bibfield  {author} {\bibinfo {author} {\bibfnamefont {P.}~\bibnamefont
  {Mehta}}\ and\ \bibinfo {author} {\bibfnamefont {W.}~\bibnamefont {Winter}},\
  }\bibfield  {title} {\bibinfo {title} {{Interplay of energy dependent
  astrophysical neutrino flavor ratios and new physics effects}},\ }\href
  {https://doi.org/10.1088/1475-7516/2011/03/041} {\bibfield  {journal}
  {\bibinfo  {journal} {JCAP}\ }\textbf {\bibinfo {volume} {03}},\ \bibinfo
  {pages} {041}},\ \Eprint {https://arxiv.org/abs/1101.2673} {arXiv:1101.2673
  [hep-ph]} \BibitemShut {NoStop}%
\bibitem [{\citenamefont {Baerwald}\ \emph {et~al.}(2012)\citenamefont
  {Baerwald}, \citenamefont {Bustamante},\ and\ \citenamefont
  {Winter}}]{Baerwald:2012kc}%
  \BibitemOpen
  \bibfield  {author} {\bibinfo {author} {\bibfnamefont {P.}~\bibnamefont
  {Baerwald}}, \bibinfo {author} {\bibfnamefont {M.}~\bibnamefont
  {Bustamante}},\ and\ \bibinfo {author} {\bibfnamefont {W.}~\bibnamefont
  {Winter}},\ }\bibfield  {title} {\bibinfo {title} {{Neutrino Decays over
  Cosmological Distances and the Implications for Neutrino Telescopes}},\
  }\href {https://doi.org/10.1088/1475-7516/2012/10/020} {\bibfield  {journal}
  {\bibinfo  {journal} {JCAP}\ }\textbf {\bibinfo {volume} {10}},\ \bibinfo
  {pages} {020}},\ \Eprint {https://arxiv.org/abs/1208.4600} {arXiv:1208.4600
  [astro-ph.CO]} \BibitemShut {NoStop}%
\bibitem [{\citenamefont {Fu}\ \emph {et~al.}(2012)\citenamefont {Fu},
  \citenamefont {Ho},\ and\ \citenamefont {Weiler}}]{Fu:2012zr}%
  \BibitemOpen
  \bibfield  {author} {\bibinfo {author} {\bibfnamefont {L.}~\bibnamefont
  {Fu}}, \bibinfo {author} {\bibfnamefont {C.~M.}\ \bibnamefont {Ho}},\ and\
  \bibinfo {author} {\bibfnamefont {T.~J.}\ \bibnamefont {Weiler}},\ }\bibfield
   {title} {\bibinfo {title} {{Cosmic Neutrino Flavor Ratios with Broken
  $\nu_\mu$-$\nu_\tau$ Symmetry}},\ }\href
  {https://doi.org/10.1016/j.physletb.2012.11.011} {\bibfield  {journal}
  {\bibinfo  {journal} {Phys. Lett. B}\ }\textbf {\bibinfo {volume} {718}},\
  \bibinfo {pages} {558} (\bibinfo {year} {2012})},\ \Eprint
  {https://arxiv.org/abs/1209.5382} {arXiv:1209.5382 [hep-ph]} \BibitemShut
  {NoStop}%
\bibitem [{\citenamefont {Pakvasa}\ \emph {et~al.}(2013)\citenamefont
  {Pakvasa}, \citenamefont {Joshipura},\ and\ \citenamefont
  {Mohanty}}]{Pakvasa:2012db}%
  \BibitemOpen
  \bibfield  {author} {\bibinfo {author} {\bibfnamefont {S.}~\bibnamefont
  {Pakvasa}}, \bibinfo {author} {\bibfnamefont {A.}~\bibnamefont {Joshipura}},\
  and\ \bibinfo {author} {\bibfnamefont {S.}~\bibnamefont {Mohanty}},\
  }\bibfield  {title} {\bibinfo {title} {{Explanation for the low flux of high
  energy astrophysical muon-neutrinos}},\ }\href
  {https://doi.org/10.1103/PhysRevLett.110.171802} {\bibfield  {journal}
  {\bibinfo  {journal} {Phys. Rev. Lett.}\ }\textbf {\bibinfo {volume} {110}},\
  \bibinfo {pages} {171802} (\bibinfo {year} {2013})},\ \Eprint
  {https://arxiv.org/abs/1209.5630} {arXiv:1209.5630 [hep-ph]} \BibitemShut
  {NoStop}%
\bibitem [{\citenamefont {Chatterjee}\ \emph {et~al.}(2014)\citenamefont
  {Chatterjee}, \citenamefont {Devi}, \citenamefont {Ghosh}, \citenamefont
  {Moharana},\ and\ \citenamefont {Raut}}]{Chatterjee:2013tza}%
  \BibitemOpen
  \bibfield  {author} {\bibinfo {author} {\bibfnamefont {A.}~\bibnamefont
  {Chatterjee}}, \bibinfo {author} {\bibfnamefont {M.~M.}\ \bibnamefont
  {Devi}}, \bibinfo {author} {\bibfnamefont {M.}~\bibnamefont {Ghosh}},
  \bibinfo {author} {\bibfnamefont {R.}~\bibnamefont {Moharana}},\ and\
  \bibinfo {author} {\bibfnamefont {S.~K.}\ \bibnamefont {Raut}},\ }\bibfield
  {title} {\bibinfo {title} {{Probing CP violation with the first three years
  of ultrahigh energy neutrinos from IceCube}},\ }\href
  {https://doi.org/10.1103/PhysRevD.90.073003} {\bibfield  {journal} {\bibinfo
  {journal} {Phys. Rev. D}\ }\textbf {\bibinfo {volume} {90}},\ \bibinfo
  {pages} {073003} (\bibinfo {year} {2014})},\ \Eprint
  {https://arxiv.org/abs/1312.6593} {arXiv:1312.6593 [hep-ph]} \BibitemShut
  {NoStop}%
\bibitem [{\citenamefont {Winter}(2013)}]{Winter:2013cla}%
  \BibitemOpen
  \bibfield  {author} {\bibinfo {author} {\bibfnamefont {W.}~\bibnamefont
  {Winter}},\ }\bibfield  {title} {\bibinfo {title} {{Photohadronic Origin of
  the TeV-PeV Neutrinos Observed in IceCube}},\ }\href
  {https://doi.org/10.1103/PhysRevD.88.083007} {\bibfield  {journal} {\bibinfo
  {journal} {Phys. Rev. D}\ }\textbf {\bibinfo {volume} {88}},\ \bibinfo
  {pages} {083007} (\bibinfo {year} {2013})},\ \Eprint
  {https://arxiv.org/abs/1307.2793} {arXiv:1307.2793 [astro-ph.HE]}
  \BibitemShut {NoStop}%
\bibitem [{\citenamefont {Xu}\ \emph {et~al.}(2014)\citenamefont {Xu},
  \citenamefont {He},\ and\ \citenamefont {Rodejohann}}]{Xu:2014via}%
  \BibitemOpen
  \bibfield  {author} {\bibinfo {author} {\bibfnamefont {X.-J.}\ \bibnamefont
  {Xu}}, \bibinfo {author} {\bibfnamefont {H.-J.}\ \bibnamefont {He}},\ and\
  \bibinfo {author} {\bibfnamefont {W.}~\bibnamefont {Rodejohann}},\ }\bibfield
   {title} {\bibinfo {title} {{Constraining Astrophysical Neutrino Flavor
  Composition from Leptonic Unitarity}},\ }\href
  {https://doi.org/10.1088/1475-7516/2014/12/039} {\bibfield  {journal}
  {\bibinfo  {journal} {JCAP}\ }\textbf {\bibinfo {volume} {12}},\ \bibinfo
  {pages} {039}},\ \Eprint {https://arxiv.org/abs/1407.3736} {arXiv:1407.3736
  [hep-ph]} \BibitemShut {NoStop}%
\bibitem [{\citenamefont {Aeikens}\ \emph {et~al.}(2015)\citenamefont
  {Aeikens}, \citenamefont {P\"as}, \citenamefont {Pakvasa},\ and\
  \citenamefont {Sicking}}]{Aeikens:2014yga}%
  \BibitemOpen
  \bibfield  {author} {\bibinfo {author} {\bibfnamefont {E.}~\bibnamefont
  {Aeikens}}, \bibinfo {author} {\bibfnamefont {H.}~\bibnamefont {P\"as}},
  \bibinfo {author} {\bibfnamefont {S.}~\bibnamefont {Pakvasa}},\ and\ \bibinfo
  {author} {\bibfnamefont {P.}~\bibnamefont {Sicking}},\ }\bibfield  {title}
  {\bibinfo {title} {{Flavor ratios of extragalactic neutrinos and neutrino
  shortcuts in extra dimensions}},\ }\href
  {https://doi.org/10.1088/1475-7516/2015/10/005} {\bibfield  {journal}
  {\bibinfo  {journal} {JCAP}\ }\textbf {\bibinfo {volume} {10}},\ \bibinfo
  {pages} {005}},\ \Eprint {https://arxiv.org/abs/1410.0408} {arXiv:1410.0408
  [hep-ph]} \BibitemShut {NoStop}%
\bibitem [{\citenamefont {Palladino}\ \emph {et~al.}(2015)\citenamefont
  {Palladino}, \citenamefont {Pagliaroli}, \citenamefont {Villante},\ and\
  \citenamefont {Vissani}}]{Palladino:2015zua}%
  \BibitemOpen
  \bibfield  {author} {\bibinfo {author} {\bibfnamefont {A.}~\bibnamefont
  {Palladino}}, \bibinfo {author} {\bibfnamefont {G.}~\bibnamefont
  {Pagliaroli}}, \bibinfo {author} {\bibfnamefont {F.~L.}\ \bibnamefont
  {Villante}},\ and\ \bibinfo {author} {\bibfnamefont {F.}~\bibnamefont
  {Vissani}},\ }\bibfield  {title} {\bibinfo {title} {{What is the Flavor of
  the Cosmic Neutrinos Seen by IceCube?}},\ }\href
  {https://doi.org/10.1103/PhysRevLett.114.171101} {\bibfield  {journal}
  {\bibinfo  {journal} {Phys. Rev. Lett.}\ }\textbf {\bibinfo {volume} {114}},\
  \bibinfo {pages} {171101} (\bibinfo {year} {2015})},\ \Eprint
  {https://arxiv.org/abs/1502.02923} {arXiv:1502.02923 [astro-ph.HE]}
  \BibitemShut {NoStop}%
\bibitem [{\citenamefont {Arg\"uelles}\ \emph {et~al.}(2015)\citenamefont
  {Arg\"uelles}, \citenamefont {Katori},\ and\ \citenamefont
  {Salvad\'o}}]{Arguelles:2015dca}%
  \BibitemOpen
  \bibfield  {author} {\bibinfo {author} {\bibfnamefont {C.~A.}\ \bibnamefont
  {Arg\"uelles}}, \bibinfo {author} {\bibfnamefont {T.}~\bibnamefont
  {Katori}},\ and\ \bibinfo {author} {\bibfnamefont {J.}~\bibnamefont
  {Salvad\'o}},\ }\bibfield  {title} {\bibinfo {title} {{New Physics in
  Astrophysical Neutrino Flavor}},\ }\href
  {https://doi.org/10.1103/PhysRevLett.115.161303} {\bibfield  {journal}
  {\bibinfo  {journal} {Phys. Rev. Lett.}\ }\textbf {\bibinfo {volume} {115}},\
  \bibinfo {pages} {161303} (\bibinfo {year} {2015})},\ \Eprint
  {https://arxiv.org/abs/1506.02043} {arXiv:1506.02043 [hep-ph]} \BibitemShut
  {NoStop}%
\bibitem [{\citenamefont {Bustamante}\ \emph
  {et~al.}(2015{\natexlab{a}})\citenamefont {Bustamante}, \citenamefont
  {Beacom},\ and\ \citenamefont {Winter}}]{Bustamante:2015waa}%
  \BibitemOpen
  \bibfield  {author} {\bibinfo {author} {\bibfnamefont {M.}~\bibnamefont
  {Bustamante}}, \bibinfo {author} {\bibfnamefont {J.~F.}\ \bibnamefont
  {Beacom}},\ and\ \bibinfo {author} {\bibfnamefont {W.}~\bibnamefont
  {Winter}},\ }\bibfield  {title} {\bibinfo {title} {{Theoretically palatable
  flavor combinations of astrophysical neutrinos}},\ }\href
  {https://doi.org/10.1103/PhysRevLett.115.161302} {\bibfield  {journal}
  {\bibinfo  {journal} {Phys. Rev. Lett.}\ }\textbf {\bibinfo {volume} {115}},\
  \bibinfo {pages} {161302} (\bibinfo {year} {2015}{\natexlab{a}})},\ \Eprint
  {https://arxiv.org/abs/1506.02645} {arXiv:1506.02645 [astro-ph.HE]}
  \BibitemShut {NoStop}%
\bibitem [{\citenamefont {Pagliaroli}\ \emph {et~al.}(2015)\citenamefont
  {Pagliaroli}, \citenamefont {Palladino}, \citenamefont {Villante},\ and\
  \citenamefont {Vissani}}]{Pagliaroli:2015rca}%
  \BibitemOpen
  \bibfield  {author} {\bibinfo {author} {\bibfnamefont {G.}~\bibnamefont
  {Pagliaroli}}, \bibinfo {author} {\bibfnamefont {A.}~\bibnamefont
  {Palladino}}, \bibinfo {author} {\bibfnamefont {F.~L.}\ \bibnamefont
  {Villante}},\ and\ \bibinfo {author} {\bibfnamefont {F.}~\bibnamefont
  {Vissani}},\ }\bibfield  {title} {\bibinfo {title} {{Testing nonradiative
  neutrino decay scenarios with IceCube data}},\ }\href
  {https://doi.org/10.1103/PhysRevD.92.113008} {\bibfield  {journal} {\bibinfo
  {journal} {Phys. Rev. D}\ }\textbf {\bibinfo {volume} {92}},\ \bibinfo
  {pages} {113008} (\bibinfo {year} {2015})},\ \Eprint
  {https://arxiv.org/abs/1506.02624} {arXiv:1506.02624 [hep-ph]} \BibitemShut
  {NoStop}%
\bibitem [{\citenamefont {Shoemaker}\ and\ \citenamefont
  {Murase}(2016)}]{Shoemaker:2015qul}%
  \BibitemOpen
  \bibfield  {author} {\bibinfo {author} {\bibfnamefont {I.~M.}\ \bibnamefont
  {Shoemaker}}\ and\ \bibinfo {author} {\bibfnamefont {K.}~\bibnamefont
  {Murase}},\ }\bibfield  {title} {\bibinfo {title} {{Probing BSM Neutrino
  Physics with Flavor and Spectral Distortions: Prospects for Future
  High-Energy Neutrino Telescopes}},\ }\href
  {https://doi.org/10.1103/PhysRevD.93.085004} {\bibfield  {journal} {\bibinfo
  {journal} {Phys. Rev. D}\ }\textbf {\bibinfo {volume} {93}},\ \bibinfo
  {pages} {085004} (\bibinfo {year} {2016})},\ \Eprint
  {https://arxiv.org/abs/1512.07228} {arXiv:1512.07228 [astro-ph.HE]}
  \BibitemShut {NoStop}%
\bibitem [{\citenamefont {de~Salas}\ \emph {et~al.}(2016)\citenamefont
  {de~Salas}, \citenamefont {Lineros},\ and\ \citenamefont
  {T\'ortola}}]{deSalas:2016svi}%
  \BibitemOpen
  \bibfield  {author} {\bibinfo {author} {\bibfnamefont {P.~F.}\ \bibnamefont
  {de~Salas}}, \bibinfo {author} {\bibfnamefont {R.~A.}\ \bibnamefont
  {Lineros}},\ and\ \bibinfo {author} {\bibfnamefont {M.}~\bibnamefont
  {T\'ortola}},\ }\bibfield  {title} {\bibinfo {title} {{Neutrino propagation
  in the galactic dark matter halo}},\ }\href
  {https://doi.org/10.1103/PhysRevD.94.123001} {\bibfield  {journal} {\bibinfo
  {journal} {Phys. Rev. D}\ }\textbf {\bibinfo {volume} {94}},\ \bibinfo
  {pages} {123001} (\bibinfo {year} {2016})},\ \Eprint
  {https://arxiv.org/abs/1601.05798} {arXiv:1601.05798 [astro-ph.HE]}
  \BibitemShut {NoStop}%
\bibitem [{\citenamefont {Gonz\'alez-Garc\'ia}\ \emph
  {et~al.}(2016)\citenamefont {Gonz\'alez-Garc\'ia}, \citenamefont {Maltoni},
  \citenamefont {Mart\'inez-Soler},\ and\ \citenamefont
  {Song}}]{Gonzalez-Garcia:2016gpq}%
  \BibitemOpen
  \bibfield  {author} {\bibinfo {author} {\bibfnamefont {M.~C.}\ \bibnamefont
  {Gonz\'alez-Garc\'ia}}, \bibinfo {author} {\bibfnamefont {M.}~\bibnamefont
  {Maltoni}}, \bibinfo {author} {\bibfnamefont {I.}~\bibnamefont
  {Mart\'inez-Soler}},\ and\ \bibinfo {author} {\bibfnamefont {N.}~\bibnamefont
  {Song}},\ }\bibfield  {title} {\bibinfo {title} {{Non-standard neutrino
  interactions in the Earth and the flavor of astrophysical neutrinos}},\
  }\href {https://doi.org/10.1016/j.astropartphys.2016.07.001} {\bibfield
  {journal} {\bibinfo  {journal} {Astropart. Phys.}\ }\textbf {\bibinfo
  {volume} {84}},\ \bibinfo {pages} {15} (\bibinfo {year} {2016})},\ \Eprint
  {https://arxiv.org/abs/1605.08055} {arXiv:1605.08055 [hep-ph]} \BibitemShut
  {NoStop}%
\bibitem [{\citenamefont {Bustamante}\ \emph {et~al.}(2017)\citenamefont
  {Bustamante}, \citenamefont {Beacom},\ and\ \citenamefont
  {Murase}}]{Bustamante:2016ciw}%
  \BibitemOpen
  \bibfield  {author} {\bibinfo {author} {\bibfnamefont {M.}~\bibnamefont
  {Bustamante}}, \bibinfo {author} {\bibfnamefont {J.~F.}\ \bibnamefont
  {Beacom}},\ and\ \bibinfo {author} {\bibfnamefont {K.}~\bibnamefont
  {Murase}},\ }\bibfield  {title} {\bibinfo {title} {{Testing decay of
  astrophysical neutrinos with incomplete information}},\ }\href
  {https://doi.org/10.1103/PhysRevD.95.063013} {\bibfield  {journal} {\bibinfo
  {journal} {Phys. Rev. D}\ }\textbf {\bibinfo {volume} {95}},\ \bibinfo
  {pages} {063013} (\bibinfo {year} {2017})},\ \Eprint
  {https://arxiv.org/abs/1610.02096} {arXiv:1610.02096 [astro-ph.HE]}
  \BibitemShut {NoStop}%
\bibitem [{\citenamefont {Biehl}\ \emph {et~al.}(2017)\citenamefont {Biehl},
  \citenamefont {Fedynitch}, \citenamefont {Palladino}, \citenamefont
  {Weiler},\ and\ \citenamefont {Winter}}]{Biehl:2016psj}%
  \BibitemOpen
  \bibfield  {author} {\bibinfo {author} {\bibfnamefont {D.}~\bibnamefont
  {Biehl}}, \bibinfo {author} {\bibfnamefont {A.}~\bibnamefont {Fedynitch}},
  \bibinfo {author} {\bibfnamefont {A.}~\bibnamefont {Palladino}}, \bibinfo
  {author} {\bibfnamefont {T.~J.}\ \bibnamefont {Weiler}},\ and\ \bibinfo
  {author} {\bibfnamefont {W.}~\bibnamefont {Winter}},\ }\bibfield  {title}
  {\bibinfo {title} {{Astrophysical Neutrino Production Diagnostics with the
  Glashow Resonance}},\ }\href {https://doi.org/10.1088/1475-7516/2017/01/033}
  {\bibfield  {journal} {\bibinfo  {journal} {JCAP}\ }\textbf {\bibinfo
  {volume} {01}},\ \bibinfo {pages} {033}},\ \Eprint
  {https://arxiv.org/abs/1611.07983} {arXiv:1611.07983 [astro-ph.HE]}
  \BibitemShut {NoStop}%
\bibitem [{\citenamefont {Rasmussen}\ \emph {et~al.}(2017)\citenamefont
  {Rasmussen}, \citenamefont {Lechner}, \citenamefont {Ackermann},
  \citenamefont {Kowalski},\ and\ \citenamefont {Winter}}]{Rasmussen:2017ert}%
  \BibitemOpen
  \bibfield  {author} {\bibinfo {author} {\bibfnamefont {R.~W.}\ \bibnamefont
  {Rasmussen}}, \bibinfo {author} {\bibfnamefont {L.}~\bibnamefont {Lechner}},
  \bibinfo {author} {\bibfnamefont {M.}~\bibnamefont {Ackermann}}, \bibinfo
  {author} {\bibfnamefont {M.}~\bibnamefont {Kowalski}},\ and\ \bibinfo
  {author} {\bibfnamefont {W.}~\bibnamefont {Winter}},\ }\bibfield  {title}
  {\bibinfo {title} {{Astrophysical neutrinos flavored with Beyond the Standard
  Model physics}},\ }\href {https://doi.org/10.1103/PhysRevD.96.083018}
  {\bibfield  {journal} {\bibinfo  {journal} {Phys. Rev. D}\ }\textbf {\bibinfo
  {volume} {96}},\ \bibinfo {pages} {083018} (\bibinfo {year} {2017})},\
  \Eprint {https://arxiv.org/abs/1707.07684} {arXiv:1707.07684 [hep-ph]}
  \BibitemShut {NoStop}%
\bibitem [{\citenamefont {Dey}\ \emph {et~al.}(2018)\citenamefont {Dey},
  \citenamefont {Kar}, \citenamefont {Mitra}, \citenamefont {Spannowsky},\ and\
  \citenamefont {Vincent}}]{Dey:2017ede}%
  \BibitemOpen
  \bibfield  {author} {\bibinfo {author} {\bibfnamefont {U.~K.}\ \bibnamefont
  {Dey}}, \bibinfo {author} {\bibfnamefont {D.}~\bibnamefont {Kar}}, \bibinfo
  {author} {\bibfnamefont {M.}~\bibnamefont {Mitra}}, \bibinfo {author}
  {\bibfnamefont {M.}~\bibnamefont {Spannowsky}},\ and\ \bibinfo {author}
  {\bibfnamefont {A.~C.}\ \bibnamefont {Vincent}},\ }\bibfield  {title}
  {\bibinfo {title} {{Searching for Leptoquarks at IceCube and the LHC}},\
  }\href {https://doi.org/10.1103/PhysRevD.98.035014} {\bibfield  {journal}
  {\bibinfo  {journal} {Phys. Rev. D}\ }\textbf {\bibinfo {volume} {98}},\
  \bibinfo {pages} {035014} (\bibinfo {year} {2018})},\ \Eprint
  {https://arxiv.org/abs/1709.02009} {arXiv:1709.02009 [hep-ph]} \BibitemShut
  {NoStop}%
\bibitem [{\citenamefont {Bustamante}\ and\ \citenamefont
  {Agarwalla}(2019)}]{Bustamante:2018mzu}%
  \BibitemOpen
  \bibfield  {author} {\bibinfo {author} {\bibfnamefont {M.}~\bibnamefont
  {Bustamante}}\ and\ \bibinfo {author} {\bibfnamefont {S.~K.}\ \bibnamefont
  {Agarwalla}},\ }\bibfield  {title} {\bibinfo {title} {{Universe's Worth of
  Electrons to Probe Long-Range Interactions of High-Energy Astrophysical
  Neutrinos}},\ }\href {https://doi.org/10.1103/PhysRevLett.122.061103}
  {\bibfield  {journal} {\bibinfo  {journal} {Phys. Rev. Lett.}\ }\textbf
  {\bibinfo {volume} {122}},\ \bibinfo {pages} {061103} (\bibinfo {year}
  {2019})},\ \Eprint {https://arxiv.org/abs/1808.02042} {arXiv:1808.02042
  [astro-ph.HE]} \BibitemShut {NoStop}%
\bibitem [{\citenamefont {Farzan}\ and\ \citenamefont
  {Palomares-Ruiz}(2019)}]{Farzan:2018pnk}%
  \BibitemOpen
  \bibfield  {author} {\bibinfo {author} {\bibfnamefont {Y.}~\bibnamefont
  {Farzan}}\ and\ \bibinfo {author} {\bibfnamefont {S.}~\bibnamefont
  {Palomares-Ruiz}},\ }\bibfield  {title} {\bibinfo {title} {{Flavor of cosmic
  neutrinos preserved by ultralight dark matter}},\ }\href
  {https://doi.org/10.1103/PhysRevD.99.051702} {\bibfield  {journal} {\bibinfo
  {journal} {Phys. Rev. D}\ }\textbf {\bibinfo {volume} {99}},\ \bibinfo
  {pages} {051702} (\bibinfo {year} {2019})},\ \Eprint
  {https://arxiv.org/abs/1810.00892} {arXiv:1810.00892 [hep-ph]} \BibitemShut
  {NoStop}%
\bibitem [{\citenamefont {Ahlers}\ \emph
  {et~al.}(2018{\natexlab{b}})\citenamefont {Ahlers}, \citenamefont
  {Bustamante},\ and\ \citenamefont {Mu}}]{Ahlers:2018yom}%
  \BibitemOpen
  \bibfield  {author} {\bibinfo {author} {\bibfnamefont {M.}~\bibnamefont
  {Ahlers}}, \bibinfo {author} {\bibfnamefont {M.}~\bibnamefont {Bustamante}},\
  and\ \bibinfo {author} {\bibfnamefont {S.}~\bibnamefont {Mu}},\ }\bibfield
  {title} {\bibinfo {title} {{Unitarity Bounds of Astrophysical Neutrinos}},\
  }\href {https://doi.org/10.1103/PhysRevD.98.123023} {\bibfield  {journal}
  {\bibinfo  {journal} {Phys. Rev. D}\ }\textbf {\bibinfo {volume} {98}},\
  \bibinfo {pages} {123023} (\bibinfo {year} {2018}{\natexlab{b}})},\ \Eprint
  {https://arxiv.org/abs/1810.00893} {arXiv:1810.00893 [astro-ph.HE]}
  \BibitemShut {NoStop}%
\bibitem [{\citenamefont {Brdar}\ and\ \citenamefont
  {Hansen}(2019)}]{Brdar:2018tce}%
  \BibitemOpen
  \bibfield  {author} {\bibinfo {author} {\bibfnamefont {V.}~\bibnamefont
  {Brdar}}\ and\ \bibinfo {author} {\bibfnamefont {R.~S.~L.}\ \bibnamefont
  {Hansen}},\ }\bibfield  {title} {\bibinfo {title} {{IceCube Flavor Ratios
  with Identified Astrophysical Sources: Towards Improving New Physics
  Testability}},\ }\href {https://doi.org/10.1088/1475-7516/2019/02/023}
  {\bibfield  {journal} {\bibinfo  {journal} {JCAP}\ }\textbf {\bibinfo
  {volume} {02}},\ \bibinfo {pages} {023}},\ \Eprint
  {https://arxiv.org/abs/1812.05541} {arXiv:1812.05541 [hep-ph]} \BibitemShut
  {NoStop}%
\bibitem [{\citenamefont {Palladino}(2019)}]{Palladino:2019pid}%
  \BibitemOpen
  \bibfield  {author} {\bibinfo {author} {\bibfnamefont {A.}~\bibnamefont
  {Palladino}},\ }\bibfield  {title} {\bibinfo {title} {{The flavor composition
  of astrophysical neutrinos after 8 years of IceCube: an indication of neutron
  decay scenario?}},\ }\href {https://doi.org/10.1140/epjc/s10052-019-7018-7}
  {\bibfield  {journal} {\bibinfo  {journal} {Eur. Phys. J. C}\ }\textbf
  {\bibinfo {volume} {79}},\ \bibinfo {pages} {500} (\bibinfo {year} {2019})},\
  \Eprint {https://arxiv.org/abs/1902.08630} {arXiv:1902.08630 [astro-ph.HE]}
  \BibitemShut {NoStop}%
\bibitem [{\citenamefont {Bustamante}\ and\ \citenamefont
  {Ahlers}(2019)}]{Bustamante:2019sdb}%
  \BibitemOpen
  \bibfield  {author} {\bibinfo {author} {\bibfnamefont {M.}~\bibnamefont
  {Bustamante}}\ and\ \bibinfo {author} {\bibfnamefont {M.}~\bibnamefont
  {Ahlers}},\ }\bibfield  {title} {\bibinfo {title} {{Inferring the flavor of
  high-energy astrophysical neutrinos at their sources}},\ }\href
  {https://doi.org/10.1103/PhysRevLett.122.241101} {\bibfield  {journal}
  {\bibinfo  {journal} {Phys. Rev. Lett.}\ }\textbf {\bibinfo {volume} {122}},\
  \bibinfo {pages} {241101} (\bibinfo {year} {2019})},\ \Eprint
  {https://arxiv.org/abs/1901.10087} {arXiv:1901.10087 [astro-ph.HE]}
  \BibitemShut {NoStop}%
\bibitem [{\citenamefont {Ahlers}\ \emph {et~al.}(2021)\citenamefont {Ahlers},
  \citenamefont {Bustamante},\ and\ \citenamefont {Willesen}}]{Ahlers:2020miq}%
  \BibitemOpen
  \bibfield  {author} {\bibinfo {author} {\bibfnamefont {M.}~\bibnamefont
  {Ahlers}}, \bibinfo {author} {\bibfnamefont {M.}~\bibnamefont {Bustamante}},\
  and\ \bibinfo {author} {\bibfnamefont {N.~G.~N.}\ \bibnamefont {Willesen}},\
  }\bibfield  {title} {\bibinfo {title} {{Flavors of astrophysical neutrinos
  with active-sterile mixing}},\ }\href
  {https://doi.org/10.1088/1475-7516/2021/07/029} {\bibfield  {journal}
  {\bibinfo  {journal} {JCAP}\ }\textbf {\bibinfo {volume} {07}},\ \bibinfo
  {pages} {029}},\ \Eprint {https://arxiv.org/abs/2009.01253} {arXiv:2009.01253
  [hep-ph]} \BibitemShut {NoStop}%
\bibitem [{\citenamefont {Bustamante}\ and\ \citenamefont
  {Tamborra}(2020)}]{Bustamante:2020bxp}%
  \BibitemOpen
  \bibfield  {author} {\bibinfo {author} {\bibfnamefont {M.}~\bibnamefont
  {Bustamante}}\ and\ \bibinfo {author} {\bibfnamefont {I.}~\bibnamefont
  {Tamborra}},\ }\bibfield  {title} {\bibinfo {title} {{Using high-energy
  neutrinos as cosmic magnetometers}},\ }\href
  {https://doi.org/10.1103/PhysRevD.102.123008} {\bibfield  {journal} {\bibinfo
   {journal} {Phys. Rev. D}\ }\textbf {\bibinfo {volume} {102}},\ \bibinfo
  {pages} {123008} (\bibinfo {year} {2020})},\ \Eprint
  {https://arxiv.org/abs/2009.01306} {arXiv:2009.01306 [astro-ph.HE]}
  \BibitemShut {NoStop}%
\bibitem [{\citenamefont {Karmakar}\ \emph {et~al.}(2021)\citenamefont
  {Karmakar}, \citenamefont {Pandey},\ and\ \citenamefont
  {Rakshit}}]{Karmakar:2020yzn}%
  \BibitemOpen
  \bibfield  {author} {\bibinfo {author} {\bibfnamefont {S.}~\bibnamefont
  {Karmakar}}, \bibinfo {author} {\bibfnamefont {S.}~\bibnamefont {Pandey}},\
  and\ \bibinfo {author} {\bibfnamefont {S.}~\bibnamefont {Rakshit}},\
  }\bibfield  {title} {\bibinfo {title} {{Astronomy with energy dependent
  flavour ratios of extragalactic neutrinos}},\ }\href
  {https://doi.org/10.1007/JHEP10(2021)004} {\bibfield  {journal} {\bibinfo
  {journal} {JHEP}\ }\textbf {\bibinfo {volume} {10}},\ \bibinfo {pages}
  {004}},\ \Eprint {https://arxiv.org/abs/2010.07336} {arXiv:2010.07336
  [hep-ph]} \BibitemShut {NoStop}%
\bibitem [{\citenamefont {Fiorillo}\ \emph
  {et~al.}(2021{\natexlab{a}})\citenamefont {Fiorillo}, \citenamefont
  {Mangano}, \citenamefont {Morisi},\ and\ \citenamefont
  {Pisanti}}]{Fiorillo:2020gsb}%
  \BibitemOpen
  \bibfield  {author} {\bibinfo {author} {\bibfnamefont {D.~F.~G.}\
  \bibnamefont {Fiorillo}}, \bibinfo {author} {\bibfnamefont {G.}~\bibnamefont
  {Mangano}}, \bibinfo {author} {\bibfnamefont {S.}~\bibnamefont {Morisi}},\
  and\ \bibinfo {author} {\bibfnamefont {O.}~\bibnamefont {Pisanti}},\
  }\bibfield  {title} {\bibinfo {title} {{IceCube constraints on violation of
  equivalence principle}},\ }\href
  {https://doi.org/10.1088/1475-7516/2021/04/079} {\bibfield  {journal}
  {\bibinfo  {journal} {JCAP}\ }\textbf {\bibinfo {volume} {04}},\ \bibinfo
  {pages} {079}},\ \Eprint {https://arxiv.org/abs/2012.07867} {arXiv:2012.07867
  [hep-ph]} \BibitemShut {NoStop}%
\bibitem [{\citenamefont {Song}\ \emph {et~al.}(2021)\citenamefont {Song},
  \citenamefont {Li}, \citenamefont {Arg\"uelles}, \citenamefont {Bustamante},\
  and\ \citenamefont {Vincent}}]{Song:2020nfh}%
  \BibitemOpen
  \bibfield  {author} {\bibinfo {author} {\bibfnamefont {N.}~\bibnamefont
  {Song}}, \bibinfo {author} {\bibfnamefont {S.~W.}\ \bibnamefont {Li}},
  \bibinfo {author} {\bibfnamefont {C.~A.}\ \bibnamefont {Arg\"uelles}},
  \bibinfo {author} {\bibfnamefont {M.}~\bibnamefont {Bustamante}},\ and\
  \bibinfo {author} {\bibfnamefont {A.~C.}\ \bibnamefont {Vincent}},\
  }\bibfield  {title} {\bibinfo {title} {{The Future of High-Energy
  Astrophysical Neutrino Flavor Measurements}},\ }\href
  {https://doi.org/10.1088/1475-7516/2021/04/054} {\bibfield  {journal}
  {\bibinfo  {journal} {JCAP}\ }\textbf {\bibinfo {volume} {04}},\ \bibinfo
  {pages} {054}},\ \Eprint {https://arxiv.org/abs/2012.12893} {arXiv:2012.12893
  [hep-ph]} \BibitemShut {NoStop}%
\bibitem [{\citenamefont {Telalovic}\ and\ \citenamefont
  {Bustamante}(2023)}]{Telalovic:2023tcb}%
  \BibitemOpen
  \bibfield  {author} {\bibinfo {author} {\bibfnamefont {B.}~\bibnamefont
  {Telalovic}}\ and\ \bibinfo {author} {\bibfnamefont {M.}~\bibnamefont
  {Bustamante}},\ }\bibfield  {title} {\bibinfo {title} {{Flavor Anisotropy in
  the High-Energy Astrophysical Neutrino Sky}},\ }\href@noop {} {\  (\bibinfo
  {year} {2023})},\ \Eprint {https://arxiv.org/abs/2310.15224}
  {arXiv:2310.15224 [astro-ph.HE]} \BibitemShut {NoStop}%
\bibitem [{\citenamefont {Aartsen}\ \emph
  {et~al.}(2013{\natexlab{a}})\citenamefont {Aartsen} \emph
  {et~al.}}]{IceCube:2013cdw}%
  \BibitemOpen
  \bibfield  {author} {\bibinfo {author} {\bibfnamefont {M.~G.}\ \bibnamefont
  {Aartsen}} \emph {et~al.} (\bibinfo {collaboration} {IceCube}),\ }\bibfield
  {title} {\bibinfo {title} {{First observation of PeV-energy neutrinos with
  IceCube}},\ }\href {https://doi.org/10.1103/PhysRevLett.111.021103}
  {\bibfield  {journal} {\bibinfo  {journal} {Phys. Rev. Lett.}\ }\textbf
  {\bibinfo {volume} {111}},\ \bibinfo {pages} {021103} (\bibinfo {year}
  {2013}{\natexlab{a}})},\ \Eprint {https://arxiv.org/abs/1304.5356}
  {arXiv:1304.5356 [astro-ph.HE]} \BibitemShut {NoStop}%
\bibitem [{\citenamefont {Aartsen}\ \emph
  {et~al.}(2013{\natexlab{b}})\citenamefont {Aartsen} \emph
  {et~al.}}]{IceCube:2013low}%
  \BibitemOpen
  \bibfield  {author} {\bibinfo {author} {\bibfnamefont {M.~G.}\ \bibnamefont
  {Aartsen}} \emph {et~al.} (\bibinfo {collaboration} {IceCube}),\ }\bibfield
  {title} {\bibinfo {title} {{Evidence for High-Energy Extraterrestrial
  Neutrinos at the IceCube Detector}},\ }\href
  {https://doi.org/10.1126/science.1242856} {\bibfield  {journal} {\bibinfo
  {journal} {Science}\ }\textbf {\bibinfo {volume} {342}},\ \bibinfo {pages}
  {1242856} (\bibinfo {year} {2013}{\natexlab{b}})},\ \Eprint
  {https://arxiv.org/abs/1311.5238} {arXiv:1311.5238 [astro-ph.HE]}
  \BibitemShut {NoStop}%
\bibitem [{\citenamefont {Aartsen}\ \emph {et~al.}(2014)\citenamefont {Aartsen}
  \emph {et~al.}}]{IceCube:2014stg}%
  \BibitemOpen
  \bibfield  {author} {\bibinfo {author} {\bibfnamefont {M.~G.}\ \bibnamefont
  {Aartsen}} \emph {et~al.} (\bibinfo {collaboration} {IceCube}),\ }\bibfield
  {title} {\bibinfo {title} {{Observation of High-Energy Astrophysical
  Neutrinos in Three Years of IceCube Data}},\ }\href
  {https://doi.org/10.1103/PhysRevLett.113.101101} {\bibfield  {journal}
  {\bibinfo  {journal} {Phys. Rev. Lett.}\ }\textbf {\bibinfo {volume} {113}},\
  \bibinfo {pages} {101101} (\bibinfo {year} {2014})},\ \Eprint
  {https://arxiv.org/abs/1405.5303} {arXiv:1405.5303 [astro-ph.HE]}
  \BibitemShut {NoStop}%
\bibitem [{\citenamefont {Aartsen}\ \emph
  {et~al.}(2015{\natexlab{a}})\citenamefont {Aartsen} \emph
  {et~al.}}]{IceCube:2015gsk}%
  \BibitemOpen
  \bibfield  {author} {\bibinfo {author} {\bibfnamefont {M.~G.}\ \bibnamefont
  {Aartsen}} \emph {et~al.} (\bibinfo {collaboration} {IceCube}),\ }\bibfield
  {title} {\bibinfo {title} {{A combined maximum-likelihood analysis of the
  high-energy astrophysical neutrino flux measured with IceCube}},\ }\href
  {https://doi.org/10.1088/0004-637X/809/1/98} {\bibfield  {journal} {\bibinfo
  {journal} {Astrophys. J.}\ }\textbf {\bibinfo {volume} {809}},\ \bibinfo
  {pages} {98} (\bibinfo {year} {2015}{\natexlab{a}})},\ \Eprint
  {https://arxiv.org/abs/1507.03991} {arXiv:1507.03991 [astro-ph.HE]}
  \BibitemShut {NoStop}%
\bibitem [{\citenamefont {Aartsen}\ \emph
  {et~al.}(2015{\natexlab{b}})\citenamefont {Aartsen} \emph
  {et~al.}}]{IceCube:2015qii}%
  \BibitemOpen
  \bibfield  {author} {\bibinfo {author} {\bibfnamefont {M.~G.}\ \bibnamefont
  {Aartsen}} \emph {et~al.} (\bibinfo {collaboration} {IceCube}),\ }\bibfield
  {title} {\bibinfo {title} {{Evidence for Astrophysical Muon Neutrinos from
  the Northern Sky with IceCube}},\ }\href
  {https://doi.org/10.1103/PhysRevLett.115.081102} {\bibfield  {journal}
  {\bibinfo  {journal} {Phys. Rev. Lett.}\ }\textbf {\bibinfo {volume} {115}},\
  \bibinfo {pages} {081102} (\bibinfo {year} {2015}{\natexlab{b}})},\ \Eprint
  {https://arxiv.org/abs/1507.04005} {arXiv:1507.04005 [astro-ph.HE]}
  \BibitemShut {NoStop}%
\bibitem [{\citenamefont {Aartsen}\ \emph {et~al.}(2016)\citenamefont {Aartsen}
  \emph {et~al.}}]{IceCube:2016umi}%
  \BibitemOpen
  \bibfield  {author} {\bibinfo {author} {\bibfnamefont {M.~G.}\ \bibnamefont
  {Aartsen}} \emph {et~al.} (\bibinfo {collaboration} {IceCube}),\ }\bibfield
  {title} {\bibinfo {title} {{Observation and Characterization of a Cosmic Muon
  Neutrino Flux from the Northern Hemisphere using six years of IceCube
  data}},\ }\href {https://doi.org/10.3847/0004-637X/833/1/3} {\bibfield
  {journal} {\bibinfo  {journal} {Astrophys. J.}\ }\textbf {\bibinfo {volume}
  {833}},\ \bibinfo {pages} {3} (\bibinfo {year} {2016})},\ \Eprint
  {https://arxiv.org/abs/1607.08006} {arXiv:1607.08006 [astro-ph.HE]}
  \BibitemShut {NoStop}%
\bibitem [{\citenamefont {Abbasi}\ \emph
  {et~al.}(2021{\natexlab{a}})\citenamefont {Abbasi} \emph
  {et~al.}}]{IceCube:2020wum}%
  \BibitemOpen
  \bibfield  {author} {\bibinfo {author} {\bibfnamefont {R.~U.}\ \bibnamefont
  {Abbasi}} \emph {et~al.} (\bibinfo {collaboration} {IceCube}),\ }\bibfield
  {title} {\bibinfo {title} {{The IceCube high-energy starting event sample:
  Description and flux characterization with 7.5 years of data}},\ }\href
  {https://doi.org/10.1103/PhysRevD.104.022002} {\bibfield  {journal} {\bibinfo
   {journal} {Phys. Rev. D}\ }\textbf {\bibinfo {volume} {104}},\ \bibinfo
  {pages} {022002} (\bibinfo {year} {2021}{\natexlab{a}})},\ \Eprint
  {https://arxiv.org/abs/2011.03545} {arXiv:2011.03545 [astro-ph.HE]}
  \BibitemShut {NoStop}%
\bibitem [{\citenamefont {Abbasi}\ \emph
  {et~al.}(2022{\natexlab{a}})\citenamefont {Abbasi} \emph
  {et~al.}}]{IceCube:2021uhz}%
  \BibitemOpen
  \bibfield  {author} {\bibinfo {author} {\bibfnamefont {R.~U.}\ \bibnamefont
  {Abbasi}} \emph {et~al.} (\bibinfo {collaboration} {IceCube}),\ }\bibfield
  {title} {\bibinfo {title} {{Improved Characterization of the Astrophysical
  Muon\textendash{}neutrino Flux with 9.5 Years of IceCube Data}},\ }\href
  {https://doi.org/10.3847/1538-4357/ac4d29} {\bibfield  {journal} {\bibinfo
  {journal} {Astrophys. J.}\ }\textbf {\bibinfo {volume} {928}},\ \bibinfo
  {pages} {50} (\bibinfo {year} {2022}{\natexlab{a}})},\ \Eprint
  {https://arxiv.org/abs/2111.10299} {arXiv:2111.10299 [astro-ph.HE]}
  \BibitemShut {NoStop}%
\bibitem [{\citenamefont {Mena}\ \emph {et~al.}(2014)\citenamefont {Mena},
  \citenamefont {Palomares-Ruiz},\ and\ \citenamefont
  {Vincent}}]{Mena:2014sja}%
  \BibitemOpen
  \bibfield  {author} {\bibinfo {author} {\bibfnamefont {O.}~\bibnamefont
  {Mena}}, \bibinfo {author} {\bibfnamefont {S.}~\bibnamefont
  {Palomares-Ruiz}},\ and\ \bibinfo {author} {\bibfnamefont {A.~C.}\
  \bibnamefont {Vincent}},\ }\bibfield  {title} {\bibinfo {title} {{Flavor
  Composition of the High-Energy Neutrino Events in IceCube}},\ }\href
  {https://doi.org/10.1103/PhysRevLett.113.091103} {\bibfield  {journal}
  {\bibinfo  {journal} {Phys. Rev. Lett.}\ }\textbf {\bibinfo {volume} {113}},\
  \bibinfo {pages} {091103} (\bibinfo {year} {2014})},\ \Eprint
  {https://arxiv.org/abs/1404.0017} {arXiv:1404.0017 [astro-ph.HE]}
  \BibitemShut {NoStop}%
\bibitem [{\citenamefont {Palomares-Ruiz}\ \emph {et~al.}(2015)\citenamefont
  {Palomares-Ruiz}, \citenamefont {Vincent},\ and\ \citenamefont
  {Mena}}]{Palomares-Ruiz:2015mka}%
  \BibitemOpen
  \bibfield  {author} {\bibinfo {author} {\bibfnamefont {S.}~\bibnamefont
  {Palomares-Ruiz}}, \bibinfo {author} {\bibfnamefont {A.~C.}\ \bibnamefont
  {Vincent}},\ and\ \bibinfo {author} {\bibfnamefont {O.}~\bibnamefont
  {Mena}},\ }\bibfield  {title} {\bibinfo {title} {{Spectral analysis of the
  high-energy IceCube neutrinos}},\ }\href
  {https://doi.org/10.1103/PhysRevD.91.103008} {\bibfield  {journal} {\bibinfo
  {journal} {Phys. Rev. D}\ }\textbf {\bibinfo {volume} {91}},\ \bibinfo
  {pages} {103008} (\bibinfo {year} {2015})},\ \Eprint
  {https://arxiv.org/abs/1502.02649} {arXiv:1502.02649 [astro-ph.HE]}
  \BibitemShut {NoStop}%
\bibitem [{\citenamefont {Aartsen}\ \emph
  {et~al.}(2015{\natexlab{c}})\citenamefont {Aartsen} \emph
  {et~al.}}]{IceCube:2015rro}%
  \BibitemOpen
  \bibfield  {author} {\bibinfo {author} {\bibfnamefont {M.~G.}\ \bibnamefont
  {Aartsen}} \emph {et~al.} (\bibinfo {collaboration} {IceCube}),\ }\bibfield
  {title} {\bibinfo {title} {{Flavor Ratio of Astrophysical Neutrinos above 35
  TeV in IceCube}},\ }\href {https://doi.org/10.1103/PhysRevLett.114.171102}
  {\bibfield  {journal} {\bibinfo  {journal} {Phys. Rev. Lett.}\ }\textbf
  {\bibinfo {volume} {114}},\ \bibinfo {pages} {171102} (\bibinfo {year}
  {2015}{\natexlab{c}})},\ \Eprint {https://arxiv.org/abs/1502.03376}
  {arXiv:1502.03376 [astro-ph.HE]} \BibitemShut {NoStop}%
\bibitem [{\citenamefont {Palladino}\ and\ \citenamefont
  {Vissani}(2015)}]{Palladino:2015vna}%
  \BibitemOpen
  \bibfield  {author} {\bibinfo {author} {\bibfnamefont {A.}~\bibnamefont
  {Palladino}}\ and\ \bibinfo {author} {\bibfnamefont {F.}~\bibnamefont
  {Vissani}},\ }\bibfield  {title} {\bibinfo {title} {{The natural
  parameterization of cosmic neutrino oscillations}},\ }\href
  {https://doi.org/10.1140/epjc/s10052-015-3664-6} {\bibfield  {journal}
  {\bibinfo  {journal} {Eur. Phys. J. C}\ }\textbf {\bibinfo {volume} {75}},\
  \bibinfo {pages} {433} (\bibinfo {year} {2015})},\ \Eprint
  {https://arxiv.org/abs/1504.05238} {arXiv:1504.05238 [hep-ph]} \BibitemShut
  {NoStop}%
\bibitem [{\citenamefont {Vincent}\ \emph {et~al.}(2016)\citenamefont
  {Vincent}, \citenamefont {Palomares-Ruiz},\ and\ \citenamefont
  {Mena}}]{Vincent:2016nut}%
  \BibitemOpen
  \bibfield  {author} {\bibinfo {author} {\bibfnamefont {A.~C.}\ \bibnamefont
  {Vincent}}, \bibinfo {author} {\bibfnamefont {S.}~\bibnamefont
  {Palomares-Ruiz}},\ and\ \bibinfo {author} {\bibfnamefont {O.}~\bibnamefont
  {Mena}},\ }\bibfield  {title} {\bibinfo {title} {{Analysis of the 4-year
  IceCube high-energy starting events}},\ }\href
  {https://doi.org/10.1103/PhysRevD.94.023009} {\bibfield  {journal} {\bibinfo
  {journal} {Phys. Rev. D}\ }\textbf {\bibinfo {volume} {94}},\ \bibinfo
  {pages} {023009} (\bibinfo {year} {2016})},\ \Eprint
  {https://arxiv.org/abs/1605.01556} {arXiv:1605.01556 [astro-ph.HE]}
  \BibitemShut {NoStop}%
\bibitem [{\citenamefont {Aartsen}\ \emph {et~al.}(2019)\citenamefont {Aartsen}
  \emph {et~al.}}]{IceCube:2018pgc}%
  \BibitemOpen
  \bibfield  {author} {\bibinfo {author} {\bibfnamefont {M.~G.}\ \bibnamefont
  {Aartsen}} \emph {et~al.} (\bibinfo {collaboration} {IceCube}),\ }\bibfield
  {title} {\bibinfo {title} {{Measurements using the inelasticity distribution
  of multi-TeV neutrino interactions in IceCube}},\ }\href
  {https://doi.org/10.1103/PhysRevD.99.032004} {\bibfield  {journal} {\bibinfo
  {journal} {Phys. Rev. D}\ }\textbf {\bibinfo {volume} {99}},\ \bibinfo
  {pages} {032004} (\bibinfo {year} {2019})},\ \Eprint
  {https://arxiv.org/abs/1808.07629} {arXiv:1808.07629 [hep-ex]} \BibitemShut
  {NoStop}%
\bibitem [{\citenamefont {Abbasi}\ \emph
  {et~al.}(2022{\natexlab{b}})\citenamefont {Abbasi} \emph
  {et~al.}}]{IceCube:2020fpi}%
  \BibitemOpen
  \bibfield  {author} {\bibinfo {author} {\bibfnamefont {R.}~\bibnamefont
  {Abbasi}} \emph {et~al.} (\bibinfo {collaboration} {IceCube}),\ }\bibfield
  {title} {\bibinfo {title} {{Detection of astrophysical tau neutrino
  candidates in IceCube}},\ }\href
  {https://doi.org/10.1140/epjc/s10052-022-10795-y} {\bibfield  {journal}
  {\bibinfo  {journal} {Eur. Phys. J. C}\ }\textbf {\bibinfo {volume} {82}},\
  \bibinfo {pages} {1031} (\bibinfo {year} {2022}{\natexlab{b}})},\ \Eprint
  {https://arxiv.org/abs/2011.03561} {arXiv:2011.03561 [hep-ex]} \BibitemShut
  {NoStop}%
\bibitem [{\citenamefont {Berezinsky}\ and\ \citenamefont
  {Zatsepin}(1969)}]{Berezinsky:1969erk}%
  \BibitemOpen
  \bibfield  {author} {\bibinfo {author} {\bibfnamefont {V.~S.}\ \bibnamefont
  {Berezinsky}}\ and\ \bibinfo {author} {\bibfnamefont {G.~T.}\ \bibnamefont
  {Zatsepin}},\ }\bibfield  {title} {\bibinfo {title} {{Cosmic rays at
  ultrahigh-energies (neutrino?)}},\ }\href
  {https://doi.org/10.1016/0370-2693(69)90341-4} {\bibfield  {journal}
  {\bibinfo  {journal} {Phys. Lett. B}\ }\textbf {\bibinfo {volume} {28}},\
  \bibinfo {pages} {423} (\bibinfo {year} {1969})}\BibitemShut {NoStop}%
\bibitem [{\citenamefont {Aartsen}\ \emph
  {et~al.}(2018{\natexlab{a}})\citenamefont {Aartsen} \emph
  {et~al.}}]{IceCube:2018fhm}%
  \BibitemOpen
  \bibfield  {author} {\bibinfo {author} {\bibfnamefont {M.~G.}\ \bibnamefont
  {Aartsen}} \emph {et~al.} (\bibinfo {collaboration} {IceCube}),\ }\bibfield
  {title} {\bibinfo {title} {{Differential limit on the extremely-high-energy
  cosmic neutrino flux in the presence of astrophysical background from nine
  years of IceCube data}},\ }\href {https://doi.org/10.1103/PhysRevD.98.062003}
  {\bibfield  {journal} {\bibinfo  {journal} {Phys. Rev. D}\ }\textbf {\bibinfo
  {volume} {98}},\ \bibinfo {pages} {062003} (\bibinfo {year}
  {2018}{\natexlab{a}})},\ \Eprint {https://arxiv.org/abs/1807.01820}
  {arXiv:1807.01820 [astro-ph.HE]} \BibitemShut {NoStop}%
\bibitem [{\citenamefont {Aab}\ \emph {et~al.}(2019)\citenamefont {Aab} \emph
  {et~al.}}]{PierreAuger:2019ens}%
  \BibitemOpen
  \bibfield  {author} {\bibinfo {author} {\bibfnamefont {A.}~\bibnamefont
  {Aab}} \emph {et~al.} (\bibinfo {collaboration} {Pierre Auger}),\ }\bibfield
  {title} {\bibinfo {title} {{Probing the origin of ultra-high-energy cosmic
  rays with neutrinos in the EeV energy range using the Pierre Auger
  Observatory}},\ }\href {https://doi.org/10.1088/1475-7516/2019/10/022}
  {\bibfield  {journal} {\bibinfo  {journal} {JCAP}\ }\textbf {\bibinfo
  {volume} {10}},\ \bibinfo {pages} {022}},\ \Eprint
  {https://arxiv.org/abs/1906.07422} {arXiv:1906.07422 [astro-ph.HE]}
  \BibitemShut {NoStop}%
\bibitem [{\citenamefont {Mammen~Abraham}\ \emph {et~al.}(2022)\citenamefont
  {Mammen~Abraham} \emph {et~al.}}]{MammenAbraham:2022xoc}%
  \BibitemOpen
  \bibfield  {author} {\bibinfo {author} {\bibfnamefont {R.}~\bibnamefont
  {Mammen~Abraham}} \emph {et~al.},\ }\bibfield  {title} {\bibinfo {title}
  {{Tau neutrinos in the next decade: from GeV to EeV}},\ }\href
  {https://doi.org/10.1088/1361-6471/ac89d2} {\bibfield  {journal} {\bibinfo
  {journal} {J. Phys. G}\ }\textbf {\bibinfo {volume} {49}},\ \bibinfo {pages}
  {110501} (\bibinfo {year} {2022})},\ \Eprint
  {https://arxiv.org/abs/2203.05591} {arXiv:2203.05591 [hep-ph]} \BibitemShut
  {NoStop}%
\bibitem [{\citenamefont {Valera}\ \emph {et~al.}(2022)\citenamefont {Valera},
  \citenamefont {Bustamante},\ and\ \citenamefont {Glaser}}]{Valera:2022ylt}%
  \BibitemOpen
  \bibfield  {author} {\bibinfo {author} {\bibfnamefont {V.~B.}\ \bibnamefont
  {Valera}}, \bibinfo {author} {\bibfnamefont {M.}~\bibnamefont {Bustamante}},\
  and\ \bibinfo {author} {\bibfnamefont {C.}~\bibnamefont {Glaser}},\
  }\bibfield  {title} {\bibinfo {title} {{The ultra-high-energy
  neutrino-nucleon cross section: measurement forecasts for an era of cosmic
  EeV-neutrino discovery}},\ }\href {https://doi.org/10.1007/JHEP06(2022)105}
  {\bibfield  {journal} {\bibinfo  {journal} {JHEP}\ }\textbf {\bibinfo
  {volume} {06}},\ \bibinfo {pages} {105}},\ \Eprint
  {https://arxiv.org/abs/2204.04237} {arXiv:2204.04237 [hep-ph]} \BibitemShut
  {NoStop}%
\bibitem [{\citenamefont {Fiorillo}\ \emph
  {et~al.}(2023{\natexlab{a}})\citenamefont {Fiorillo}, \citenamefont
  {Bustamante},\ and\ \citenamefont {Valera}}]{Fiorillo:2022ijt}%
  \BibitemOpen
  \bibfield  {author} {\bibinfo {author} {\bibfnamefont {D.~F.~G.}\
  \bibnamefont {Fiorillo}}, \bibinfo {author} {\bibfnamefont {M.}~\bibnamefont
  {Bustamante}},\ and\ \bibinfo {author} {\bibfnamefont {V.~B.}\ \bibnamefont
  {Valera}},\ }\bibfield  {title} {\bibinfo {title} {{Near-future discovery of
  point sources of ultra-high-energy neutrinos}},\ }\href
  {https://doi.org/10.1088/1475-7516/2023/03/026} {\bibfield  {journal}
  {\bibinfo  {journal} {JCAP}\ }\textbf {\bibinfo {volume} {03}},\ \bibinfo
  {pages} {026}},\ \Eprint {https://arxiv.org/abs/2205.15985} {arXiv:2205.15985
  [astro-ph.HE]} \BibitemShut {NoStop}%
\bibitem [{\citenamefont {Valera}\ \emph
  {et~al.}(2023{\natexlab{a}})\citenamefont {Valera}, \citenamefont
  {Bustamante},\ and\ \citenamefont {Glaser}}]{Valera:2022wmu}%
  \BibitemOpen
  \bibfield  {author} {\bibinfo {author} {\bibfnamefont {V.~B.}\ \bibnamefont
  {Valera}}, \bibinfo {author} {\bibfnamefont {M.}~\bibnamefont {Bustamante}},\
  and\ \bibinfo {author} {\bibfnamefont {C.}~\bibnamefont {Glaser}},\
  }\bibfield  {title} {\bibinfo {title} {{Near-future discovery of the diffuse
  flux of ultrahigh-energy cosmic neutrinos}},\ }\href
  {https://doi.org/10.1103/PhysRevD.107.043019} {\bibfield  {journal} {\bibinfo
   {journal} {Phys. Rev. D}\ }\textbf {\bibinfo {volume} {107}},\ \bibinfo
  {pages} {043019} (\bibinfo {year} {2023}{\natexlab{a}})},\ \Eprint
  {https://arxiv.org/abs/2210.03756} {arXiv:2210.03756 [astro-ph.HE]}
  \BibitemShut {NoStop}%
\bibitem [{\citenamefont {Fiorillo}\ \emph
  {et~al.}(2023{\natexlab{b}})\citenamefont {Fiorillo}, \citenamefont {Valera},
  \citenamefont {Bustamante},\ and\ \citenamefont {Winter}}]{Fiorillo:2023clw}%
  \BibitemOpen
  \bibfield  {author} {\bibinfo {author} {\bibfnamefont {D.~F.~G.}\
  \bibnamefont {Fiorillo}}, \bibinfo {author} {\bibfnamefont {V.~B.}\
  \bibnamefont {Valera}}, \bibinfo {author} {\bibfnamefont {M.}~\bibnamefont
  {Bustamante}},\ and\ \bibinfo {author} {\bibfnamefont {W.}~\bibnamefont
  {Winter}},\ }\bibfield  {title} {\bibinfo {title} {{Searches for dark matter
  decay with ultrahigh-energy neutrinos endure backgrounds}},\ }\href
  {https://doi.org/10.1103/PhysRevD.108.103012} {\bibfield  {journal} {\bibinfo
   {journal} {Phys. Rev. D}\ }\textbf {\bibinfo {volume} {108}},\ \bibinfo
  {pages} {103012} (\bibinfo {year} {2023}{\natexlab{b}})},\ \Eprint
  {https://arxiv.org/abs/2307.02538} {arXiv:2307.02538 [astro-ph.HE]}
  \BibitemShut {NoStop}%
\bibitem [{\citenamefont {Valera}\ \emph
  {et~al.}(2023{\natexlab{b}})\citenamefont {Valera}, \citenamefont
  {Bustamante},\ and\ \citenamefont {Mena}}]{Valera:2023ayh}%
  \BibitemOpen
  \bibfield  {author} {\bibinfo {author} {\bibfnamefont {V.~B.}\ \bibnamefont
  {Valera}}, \bibinfo {author} {\bibfnamefont {M.}~\bibnamefont {Bustamante}},\
  and\ \bibinfo {author} {\bibfnamefont {O.}~\bibnamefont {Mena}},\ }\bibfield
  {title} {\bibinfo {title} {{Joint measurement of the ultra-high-energy
  neutrino spectrum and cross section}},\ }\href@noop {} {\  (\bibinfo {year}
  {2023}{\natexlab{b}})},\ \Eprint {https://arxiv.org/abs/2308.07709}
  {arXiv:2308.07709 [astro-ph.HE]} \BibitemShut {NoStop}%
\bibitem [{\citenamefont {Aartsen}\ \emph
  {et~al.}(2017{\natexlab{a}})\citenamefont {Aartsen} \emph
  {et~al.}}]{IceCube:2016zyt}%
  \BibitemOpen
  \bibfield  {author} {\bibinfo {author} {\bibfnamefont {M.~G.}\ \bibnamefont
  {Aartsen}} \emph {et~al.} (\bibinfo {collaboration} {IceCube}),\ }\bibfield
  {title} {\bibinfo {title} {{The IceCube Neutrino Observatory: Instrumentation
  and Online Systems}},\ }\href
  {https://doi.org/10.1088/1748-0221/12/03/P03012} {\bibfield  {journal}
  {\bibinfo  {journal} {JINST}\ }\textbf {\bibinfo {volume} {12}}\bibfield
  {number} {\bibinfo  {number} { (03)},\ \bibinfo {pages} {P03012}},\ }\Eprint
  {https://arxiv.org/abs/1612.05093} {arXiv:1612.05093 [astro-ph.IM]}
  \BibitemShut {NoStop}%
\bibitem [{\citenamefont {Aartsen}\ \emph {et~al.}(2021)\citenamefont {Aartsen}
  \emph {et~al.}}]{IceCube-Gen2:2020qha}%
  \BibitemOpen
  \bibfield  {author} {\bibinfo {author} {\bibfnamefont {M.~G.}\ \bibnamefont
  {Aartsen}} \emph {et~al.} (\bibinfo {collaboration} {IceCube-Gen2}),\
  }\bibfield  {title} {\bibinfo {title} {{IceCube-Gen2: the window to the
  extreme Universe}},\ }\href {https://doi.org/10.1088/1361-6471/abbd48}
  {\bibfield  {journal} {\bibinfo  {journal} {J. Phys. G}\ }\textbf {\bibinfo
  {volume} {48}},\ \bibinfo {pages} {060501} (\bibinfo {year} {2021})},\
  \Eprint {https://arxiv.org/abs/2008.04323} {arXiv:2008.04323 [astro-ph.HE]}
  \BibitemShut {NoStop}%
\bibitem [{\citenamefont {Aguilar}\ \emph {et~al.}(2021)\citenamefont {Aguilar}
  \emph {et~al.}}]{RNO-G:2020rmc}%
  \BibitemOpen
  \bibfield  {author} {\bibinfo {author} {\bibfnamefont {J.~A.}\ \bibnamefont
  {Aguilar}} \emph {et~al.} (\bibinfo {collaboration} {RNO-G}),\ }\bibfield
  {title} {\bibinfo {title} {{Design and Sensitivity of the Radio Neutrino
  Observatory in Greenland (RNO-G)}},\ }\href
  {https://doi.org/10.1088/1748-0221/16/03/P03025} {\bibfield  {journal}
  {\bibinfo  {journal} {JINST}\ }\textbf {\bibinfo {volume} {16}}\bibfield
  {number} {\bibinfo  {number} { (03)},\ \bibinfo {pages} {P03025}},\ }\Eprint
  {https://arxiv.org/abs/2010.12279} {arXiv:2010.12279 [astro-ph.IM]}
  \BibitemShut {NoStop}%
\bibitem [{\citenamefont {Adams}\ \emph {et~al.}(2017)\citenamefont {Adams}
  \emph {et~al.}}]{Adams:2017fjh}%
  \BibitemOpen
  \bibfield  {author} {\bibinfo {author} {\bibfnamefont {J.~H.}\ \bibnamefont
  {Adams}} \emph {et~al.},\ }\bibfield  {title} {\bibinfo {title} {{White paper
  on EUSO-SPB2}},\ }\href@noop {} {\  (\bibinfo {year} {2017})},\ \Eprint
  {https://arxiv.org/abs/1703.04513} {arXiv:1703.04513 [astro-ph.HE]}
  \BibitemShut {NoStop}%
\bibitem [{\citenamefont {\'Alvarez-Mu\~niz}\ \emph {et~al.}(2020)\citenamefont
  {\'Alvarez-Mu\~niz} \emph {et~al.}}]{GRAND:2018iaj}%
  \BibitemOpen
  \bibfield  {author} {\bibinfo {author} {\bibfnamefont {J.}~\bibnamefont
  {\'Alvarez-Mu\~niz}} \emph {et~al.} (\bibinfo {collaboration} {GRAND}),\
  }\bibfield  {title} {\bibinfo {title} {{The Giant Radio Array for Neutrino
  Detection (GRAND): Science and Design}},\ }\href
  {https://doi.org/10.1007/s11433-018-9385-7} {\bibfield  {journal} {\bibinfo
  {journal} {Sci. China Phys. Mech. Astron.}\ }\textbf {\bibinfo {volume}
  {63}},\ \bibinfo {pages} {219501} (\bibinfo {year} {2020})},\ \Eprint
  {https://arxiv.org/abs/1810.09994} {arXiv:1810.09994 [astro-ph.HE]}
  \BibitemShut {NoStop}%
\bibitem [{\citenamefont {Deaconu}(2020)}]{Deaconu:2019rdx}%
  \BibitemOpen
  \bibfield  {author} {\bibinfo {author} {\bibfnamefont {C.}~\bibnamefont
  {Deaconu}} (\bibinfo {collaboration} {ANITA}),\ }\bibfield  {title} {\bibinfo
  {title} {{Searches for Ultra-High Energy Neutrinos with ANITA}},\ }\href
  {https://doi.org/10.22323/1.358.0867} {\bibfield  {journal} {\bibinfo
  {journal} {PoS}\ }\textbf {\bibinfo {volume} {ICRC2019}},\ \bibinfo {pages}
  {867} (\bibinfo {year} {2020})},\ \Eprint {https://arxiv.org/abs/1908.00923}
  {arXiv:1908.00923 [astro-ph.HE]} \BibitemShut {NoStop}%
\bibitem [{\citenamefont {Prohira}\ \emph {et~al.}(2020)\citenamefont {Prohira}
  \emph {et~al.}}]{Prohira:2019glh}%
  \BibitemOpen
  \bibfield  {author} {\bibinfo {author} {\bibfnamefont {S.}~\bibnamefont
  {Prohira}} \emph {et~al.},\ }\bibfield  {title} {\bibinfo {title}
  {{Observation of Radar Echoes From High-Energy Particle Cascades}},\ }\href
  {https://doi.org/10.1103/PhysRevLett.124.091101} {\bibfield  {journal}
  {\bibinfo  {journal} {Phys. Rev. Lett.}\ }\textbf {\bibinfo {volume} {124}},\
  \bibinfo {pages} {091101} (\bibinfo {year} {2020})},\ \Eprint
  {https://arxiv.org/abs/1910.12830} {arXiv:1910.12830 [astro-ph.HE]}
  \BibitemShut {NoStop}%
\bibitem [{\citenamefont {Nam}\ \emph {et~al.}(2020)\citenamefont {Nam} \emph
  {et~al.}}]{Nam:2020hng}%
  \BibitemOpen
  \bibfield  {author} {\bibinfo {author} {\bibfnamefont {J.}~\bibnamefont
  {Nam}} \emph {et~al.},\ }\bibfield  {title} {\bibinfo {title}
  {{High-elevation synoptic radio array for detection of upward moving
  air-showers, deployed in the Antarctic mountains}},\ }\href
  {https://doi.org/10.22323/1.358.0967} {\bibfield  {journal} {\bibinfo
  {journal} {PoS}\ }\textbf {\bibinfo {volume} {ICRC2019}},\ \bibinfo {pages}
  {967} (\bibinfo {year} {2020})}\BibitemShut {NoStop}%
\bibitem [{\citenamefont {Wissel}\ \emph {et~al.}(2020)\citenamefont {Wissel}
  \emph {et~al.}}]{Wissel:2020sec}%
  \BibitemOpen
  \bibfield  {author} {\bibinfo {author} {\bibfnamefont {S.}~\bibnamefont
  {Wissel}} \emph {et~al.},\ }\bibfield  {title} {\bibinfo {title} {{Prospects
  for high-elevation radio detection of \ensuremath{>}100 PeV tau neutrinos}},\
  }\href {https://doi.org/10.1088/1475-7516/2020/11/065} {\bibfield  {journal}
  {\bibinfo  {journal} {JCAP}\ }\textbf {\bibinfo {volume} {11}},\ \bibinfo
  {pages} {065}},\ \Eprint {https://arxiv.org/abs/2004.12718} {arXiv:2004.12718
  [astro-ph.IM]} \BibitemShut {NoStop}%
\bibitem [{\citenamefont {Otte}(2019)}]{Otte:2018uxj}%
  \BibitemOpen
  \bibfield  {author} {\bibinfo {author} {\bibfnamefont {A.~N.}\ \bibnamefont
  {Otte}},\ }\bibfield  {title} {\bibinfo {title} {{Studies of an air-shower
  imaging system for the detection of ultrahigh-energy neutrinos}},\ }\href
  {https://doi.org/10.1103/PhysRevD.99.083012} {\bibfield  {journal} {\bibinfo
  {journal} {Phys. Rev. D}\ }\textbf {\bibinfo {volume} {99}},\ \bibinfo
  {pages} {083012} (\bibinfo {year} {2019})},\ \Eprint
  {https://arxiv.org/abs/1811.09287} {arXiv:1811.09287 [astro-ph.IM]}
  \BibitemShut {NoStop}%
\bibitem [{\citenamefont {Olinto}\ \emph {et~al.}(2021)\citenamefont {Olinto}
  \emph {et~al.}}]{POEMMA:2020ykm}%
  \BibitemOpen
  \bibfield  {author} {\bibinfo {author} {\bibfnamefont {A.~V.}\ \bibnamefont
  {Olinto}} \emph {et~al.} (\bibinfo {collaboration} {POEMMA}),\ }\bibfield
  {title} {\bibinfo {title} {{The POEMMA (Probe of Extreme Multi-Messenger
  Astrophysics) observatory}},\ }\href
  {https://doi.org/10.1088/1475-7516/2021/06/007} {\bibfield  {journal}
  {\bibinfo  {journal} {JCAP}\ }\textbf {\bibinfo {volume} {06}},\ \bibinfo
  {pages} {007}},\ \Eprint {https://arxiv.org/abs/2012.07945} {arXiv:2012.07945
  [astro-ph.IM]} \BibitemShut {NoStop}%
\bibitem [{\citenamefont {Wang}\ \emph {et~al.}(2013)\citenamefont {Wang},
  \citenamefont {Chen}, \citenamefont {Huang},\ and\ \citenamefont
  {Nam}}]{Wang:2013njo}%
  \BibitemOpen
  \bibfield  {author} {\bibinfo {author} {\bibfnamefont {S.-H.}\ \bibnamefont
  {Wang}}, \bibinfo {author} {\bibfnamefont {P.}~\bibnamefont {Chen}}, \bibinfo
  {author} {\bibfnamefont {M.}~\bibnamefont {Huang}},\ and\ \bibinfo {author}
  {\bibfnamefont {J.}~\bibnamefont {Nam}},\ }\bibfield  {title} {\bibinfo
  {title} {{Feasibility of Determining Diffuse Ultra-High Energy Cosmic
  Neutrino Flavor Ratio through ARA Neutrino Observatory}},\ }\href
  {https://doi.org/10.1088/1475-7516/2013/11/062} {\bibfield  {journal}
  {\bibinfo  {journal} {JCAP}\ }\textbf {\bibinfo {volume} {11}},\ \bibinfo
  {pages} {062}},\ \Eprint {https://arxiv.org/abs/1302.1586} {arXiv:1302.1586
  [astro-ph.HE]} \BibitemShut {NoStop}%
\bibitem [{\citenamefont {Stj\"arnholm}\ \emph {et~al.}(2021)\citenamefont
  {Stj\"arnholm}, \citenamefont {Ericsson},\ and\ \citenamefont
  {Glaser}}]{Stjarnholm:2021xpj}%
  \BibitemOpen
  \bibfield  {author} {\bibinfo {author} {\bibfnamefont {S.}~\bibnamefont
  {Stj\"arnholm}}, \bibinfo {author} {\bibfnamefont {O.}~\bibnamefont
  {Ericsson}},\ and\ \bibinfo {author} {\bibfnamefont {C.}~\bibnamefont
  {Glaser}},\ }\bibfield  {title} {\bibinfo {title} {{Neutrino direction and
  flavor reconstruction from radio detector data using deep convolutional
  neural networks}},\ }\href {https://doi.org/10.22323/1.395.1055} {\bibfield
  {journal} {\bibinfo  {journal} {PoS}\ }\textbf {\bibinfo {volume}
  {ICRC2021}},\ \bibinfo {pages} {1055} (\bibinfo {year} {2021})}\BibitemShut
  {NoStop}%
\bibitem [{\citenamefont {Glaser}\ \emph {et~al.}(2021)\citenamefont {Glaser},
  \citenamefont {Garc\'\i{}a-Fern\'andez},\ and\ \citenamefont
  {Nelles}}]{Glaser:2021hfi}%
  \BibitemOpen
  \bibfield  {author} {\bibinfo {author} {\bibfnamefont {C.}~\bibnamefont
  {Glaser}}, \bibinfo {author} {\bibfnamefont {D.}~\bibnamefont
  {Garc\'\i{}a-Fern\'andez}},\ and\ \bibinfo {author} {\bibfnamefont
  {A.}~\bibnamefont {Nelles}},\ }\bibfield  {title} {\bibinfo {title}
  {{Prospects for neutrino-flavor physics with in-ice radio detectors}},\
  }\href {https://doi.org/10.22323/1.395.1231} {\bibfield  {journal} {\bibinfo
  {journal} {PoS}\ }\textbf {\bibinfo {volume} {ICRC2021}},\ \bibinfo {pages}
  {1231} (\bibinfo {year} {2021})}\BibitemShut {NoStop}%
\bibitem [{\citenamefont {Coleman}\ \emph {et~al.}(2024)\citenamefont
  {Coleman}, \citenamefont {Ericsson}, \citenamefont {Glaser},\ and\
  \citenamefont {Bustamante}}]{Coleman:2024scd}%
  \BibitemOpen
  \bibfield  {author} {\bibinfo {author} {\bibfnamefont {A.}~\bibnamefont
  {Coleman}}, \bibinfo {author} {\bibfnamefont {O.}~\bibnamefont {Ericsson}},
  \bibinfo {author} {\bibfnamefont {C.}~\bibnamefont {Glaser}},\ and\ \bibinfo
  {author} {\bibfnamefont {M.}~\bibnamefont {Bustamante}},\ }\bibfield  {title}
  {\bibinfo {title} {{Flavor composition of ultrahigh-energy cosmic neutrinos:
  Measurement forecasts for in-ice radio-based EeV neutrino telescopes}},\
  }\href {https://doi.org/10.1103/PhysRevD.110.023044} {\bibfield  {journal}
  {\bibinfo  {journal} {Phys. Rev. D}\ }\textbf {\bibinfo {volume} {110}},\
  \bibinfo {pages} {023044} (\bibinfo {year} {2024})},\ \Eprint
  {https://arxiv.org/abs/2402.02432} {arXiv:2402.02432 [astro-ph.HE]}
  \BibitemShut {NoStop}%
\bibitem [{\citenamefont {Abbasi}\ \emph
  {et~al.}(2021{\natexlab{b}})\citenamefont {Abbasi} \emph
  {et~al.}}]{IceCube-Gen2:2021rkf}%
  \BibitemOpen
  \bibfield  {author} {\bibinfo {author} {\bibfnamefont {R.~U.}\ \bibnamefont
  {Abbasi}} \emph {et~al.} (\bibinfo {collaboration} {IceCube-Gen2}),\
  }\bibfield  {title} {\bibinfo {title} {{Sensitivity studies for the
  IceCube-Gen2 radio array}},\ }\href {https://doi.org/10.22323/1.395.1183}
  {\bibfield  {journal} {\bibinfo  {journal} {PoS}\ }\textbf {\bibinfo {volume}
  {ICRC2021}},\ \bibinfo {pages} {1183} (\bibinfo {year}
  {2021}{\natexlab{b}})},\ \Eprint {https://arxiv.org/abs/2107.08910}
  {arXiv:2107.08910 [astro-ph.HE]} \BibitemShut {NoStop}%
\bibitem [{\citenamefont {Greisen}(1966)}]{Greisen:1966jv}%
  \BibitemOpen
  \bibfield  {author} {\bibinfo {author} {\bibfnamefont {K.}~\bibnamefont
  {Greisen}},\ }\bibfield  {title} {\bibinfo {title} {{End to the cosmic ray
  spectrum?}},\ }\href {https://doi.org/10.1103/PhysRevLett.16.748} {\bibfield
  {journal} {\bibinfo  {journal} {Phys. Rev. Lett.}\ }\textbf {\bibinfo
  {volume} {16}},\ \bibinfo {pages} {748} (\bibinfo {year} {1966})}\BibitemShut
  {NoStop}%
\bibitem [{\citenamefont {Zatsepin}\ and\ \citenamefont
  {Kuzmin}(1966)}]{Zatsepin:1966jv}%
  \BibitemOpen
  \bibfield  {author} {\bibinfo {author} {\bibfnamefont {G.~T.}\ \bibnamefont
  {Zatsepin}}\ and\ \bibinfo {author} {\bibfnamefont {V.~A.}\ \bibnamefont
  {Kuzmin}},\ }\bibfield  {title} {\bibinfo {title} {{Upper limit of the
  spectrum of cosmic rays}},\ }\href@noop {} {\bibfield  {journal} {\bibinfo
  {journal} {JETP Lett.}\ }\textbf {\bibinfo {volume} {4}},\ \bibinfo {pages}
  {78} (\bibinfo {year} {1966})}\BibitemShut {NoStop}%
\bibitem [{\citenamefont {Mannheim}(1995)}]{Mannheim:1995mm}%
  \BibitemOpen
  \bibfield  {author} {\bibinfo {author} {\bibfnamefont {K.}~\bibnamefont
  {Mannheim}},\ }\bibfield  {title} {\bibinfo {title} {{High-energy neutrinos
  from extragalactic jets}},\ }\href
  {https://doi.org/10.1016/0927-6505(94)00044-4} {\bibfield  {journal}
  {\bibinfo  {journal} {Astropart. Phys.}\ }\textbf {\bibinfo {volume} {3}},\
  \bibinfo {pages} {295} (\bibinfo {year} {1995})}\BibitemShut {NoStop}%
\bibitem [{\citenamefont {Atoyan}\ and\ \citenamefont
  {Dermer}(2001)}]{Atoyan:2001ey}%
  \BibitemOpen
  \bibfield  {author} {\bibinfo {author} {\bibfnamefont {A.~M.}\ \bibnamefont
  {Atoyan}}\ and\ \bibinfo {author} {\bibfnamefont {C.~D.}\ \bibnamefont
  {Dermer}},\ }\bibfield  {title} {\bibinfo {title} {{High-energy neutrinos
  from photomeson processes in blazars}},\ }\href
  {https://doi.org/10.1103/PhysRevLett.87.221102} {\bibfield  {journal}
  {\bibinfo  {journal} {Phys. Rev. Lett.}\ }\textbf {\bibinfo {volume} {87}},\
  \bibinfo {pages} {221102} (\bibinfo {year} {2001})},\ \Eprint
  {https://arxiv.org/abs/astro-ph/0108053} {arXiv:astro-ph/0108053}
  \BibitemShut {NoStop}%
\bibitem [{\citenamefont {Atoyan}\ and\ \citenamefont
  {Dermer}(2003)}]{Atoyan:2002gu}%
  \BibitemOpen
  \bibfield  {author} {\bibinfo {author} {\bibfnamefont {A.~M.}\ \bibnamefont
  {Atoyan}}\ and\ \bibinfo {author} {\bibfnamefont {C.~D.}\ \bibnamefont
  {Dermer}},\ }\bibfield  {title} {\bibinfo {title} {{Neutral beams from blazar
  jets}},\ }\href {https://doi.org/10.1086/346261} {\bibfield  {journal}
  {\bibinfo  {journal} {Astrophys. J.}\ }\textbf {\bibinfo {volume} {586}},\
  \bibinfo {pages} {79} (\bibinfo {year} {2003})},\ \Eprint
  {https://arxiv.org/abs/astro-ph/0209231} {arXiv:astro-ph/0209231}
  \BibitemShut {NoStop}%
\bibitem [{\citenamefont {{\'A}lvarez-Mu{\~n}iz}\ and\ \citenamefont
  {M\'esz\'aross}(2004)}]{Alvarez-Muniz:2004xlu}%
  \BibitemOpen
  \bibfield  {author} {\bibinfo {author} {\bibfnamefont {J.}~\bibnamefont
  {{\'A}lvarez-Mu{\~n}iz}}\ and\ \bibinfo {author} {\bibfnamefont
  {P.}~\bibnamefont {M\'esz\'aross}},\ }\bibfield  {title} {\bibinfo {title}
  {{High energy neutrinos from radio-quiet AGNs}},\ }\href
  {https://doi.org/10.1103/PhysRevD.70.123001} {\bibfield  {journal} {\bibinfo
  {journal} {Phys. Rev. D}\ }\textbf {\bibinfo {volume} {70}},\ \bibinfo
  {pages} {123001} (\bibinfo {year} {2004})},\ \Eprint
  {https://arxiv.org/abs/astro-ph/0409034} {arXiv:astro-ph/0409034}
  \BibitemShut {NoStop}%
\bibitem [{\citenamefont {Murase}\ \emph {et~al.}(2014)\citenamefont {Murase},
  \citenamefont {Inoue},\ and\ \citenamefont {Dermer}}]{Murase:2014foa}%
  \BibitemOpen
  \bibfield  {author} {\bibinfo {author} {\bibfnamefont {K.}~\bibnamefont
  {Murase}}, \bibinfo {author} {\bibfnamefont {Y.}~\bibnamefont {Inoue}},\ and\
  \bibinfo {author} {\bibfnamefont {C.~D.}\ \bibnamefont {Dermer}},\ }\bibfield
   {title} {\bibinfo {title} {{Diffuse Neutrino Intensity from the Inner Jets
  of Active Galactic Nuclei: Impacts of External Photon Fields and the Blazar
  Sequence}},\ }\href {https://doi.org/10.1103/PhysRevD.90.023007} {\bibfield
  {journal} {\bibinfo  {journal} {Phys. Rev. D}\ }\textbf {\bibinfo {volume}
  {90}},\ \bibinfo {pages} {023007} (\bibinfo {year} {2014})},\ \Eprint
  {https://arxiv.org/abs/1403.4089} {arXiv:1403.4089 [astro-ph.HE]}
  \BibitemShut {NoStop}%
\bibitem [{\citenamefont {Kimura}\ \emph {et~al.}(2015)\citenamefont {Kimura},
  \citenamefont {Murase},\ and\ \citenamefont {Toma}}]{Kimura:2014jba}%
  \BibitemOpen
  \bibfield  {author} {\bibinfo {author} {\bibfnamefont {S.~S.}\ \bibnamefont
  {Kimura}}, \bibinfo {author} {\bibfnamefont {K.}~\bibnamefont {Murase}},\
  and\ \bibinfo {author} {\bibfnamefont {K.}~\bibnamefont {Toma}},\ }\bibfield
  {title} {\bibinfo {title} {{Neutrino and Cosmic-Ray Emission and Cumulative
  Background from Radiatively Inefficient Accretion Flows in Low-Luminosity
  Active Galactic Nuclei}},\ }\href
  {https://doi.org/10.1088/0004-637X/806/2/159} {\bibfield  {journal} {\bibinfo
   {journal} {Astrophys. J.}\ }\textbf {\bibinfo {volume} {806}},\ \bibinfo
  {pages} {159} (\bibinfo {year} {2015})},\ \Eprint
  {https://arxiv.org/abs/1411.3588} {arXiv:1411.3588 [astro-ph.HE]}
  \BibitemShut {NoStop}%
\bibitem [{\citenamefont {Padovani}\ \emph {et~al.}(2015)\citenamefont
  {Padovani}, \citenamefont {Petropoulou}, \citenamefont {Giommi},\ and\
  \citenamefont {Resconi}}]{Padovani:2015mba}%
  \BibitemOpen
  \bibfield  {author} {\bibinfo {author} {\bibfnamefont {P.}~\bibnamefont
  {Padovani}}, \bibinfo {author} {\bibfnamefont {M.}~\bibnamefont
  {Petropoulou}}, \bibinfo {author} {\bibfnamefont {P.}~\bibnamefont
  {Giommi}},\ and\ \bibinfo {author} {\bibfnamefont {E.}~\bibnamefont
  {Resconi}},\ }\bibfield  {title} {\bibinfo {title} {{A simplified view of
  blazars: the neutrino background}},\ }\href
  {https://doi.org/10.1093/mnras/stv1467} {\bibfield  {journal} {\bibinfo
  {journal} {Mon. Not. Roy. Astron. Soc.}\ }\textbf {\bibinfo {volume} {452}},\
  \bibinfo {pages} {1877} (\bibinfo {year} {2015})},\ \Eprint
  {https://arxiv.org/abs/1506.09135} {arXiv:1506.09135 [astro-ph.HE]}
  \BibitemShut {NoStop}%
\bibitem [{\citenamefont {Palladino}\ \emph {et~al.}(2019)\citenamefont
  {Palladino}, \citenamefont {Rodrigues}, \citenamefont {Gao},\ and\
  \citenamefont {Winter}}]{Palladino:2018lov}%
  \BibitemOpen
  \bibfield  {author} {\bibinfo {author} {\bibfnamefont {A.}~\bibnamefont
  {Palladino}}, \bibinfo {author} {\bibfnamefont {X.}~\bibnamefont
  {Rodrigues}}, \bibinfo {author} {\bibfnamefont {S.}~\bibnamefont {Gao}},\
  and\ \bibinfo {author} {\bibfnamefont {W.}~\bibnamefont {Winter}},\
  }\bibfield  {title} {\bibinfo {title} {{Interpretation of the diffuse
  astrophysical neutrino flux in terms of the blazar sequence}},\ }\href
  {https://doi.org/10.3847/1538-4357/aaf507} {\bibfield  {journal} {\bibinfo
  {journal} {Astrophys. J.}\ }\textbf {\bibinfo {volume} {871}},\ \bibinfo
  {pages} {41} (\bibinfo {year} {2019})},\ \Eprint
  {https://arxiv.org/abs/1806.04769} {arXiv:1806.04769 [astro-ph.HE]}
  \BibitemShut {NoStop}%
\bibitem [{\citenamefont {Rodrigues}\ \emph {et~al.}(2021)\citenamefont
  {Rodrigues}, \citenamefont {Heinze}, \citenamefont {Palladino}, \citenamefont
  {van Vliet},\ and\ \citenamefont {Winter}}]{Rodrigues:2020pli}%
  \BibitemOpen
  \bibfield  {author} {\bibinfo {author} {\bibfnamefont {X.}~\bibnamefont
  {Rodrigues}}, \bibinfo {author} {\bibfnamefont {J.}~\bibnamefont {Heinze}},
  \bibinfo {author} {\bibfnamefont {A.}~\bibnamefont {Palladino}}, \bibinfo
  {author} {\bibfnamefont {A.}~\bibnamefont {van Vliet}},\ and\ \bibinfo
  {author} {\bibfnamefont {W.}~\bibnamefont {Winter}},\ }\bibfield  {title}
  {\bibinfo {title} {{Active Galactic Nuclei Jets as the Origin of
  Ultrahigh-Energy Cosmic Rays and Perspectives for the Detection of
  Astrophysical Source Neutrinos at EeV Energies}},\ }\href
  {https://doi.org/10.1103/PhysRevLett.126.191101} {\bibfield  {journal}
  {\bibinfo  {journal} {Phys. Rev. Lett.}\ }\textbf {\bibinfo {volume} {126}},\
  \bibinfo {pages} {191101} (\bibinfo {year} {2021})},\ \Eprint
  {https://arxiv.org/abs/2003.08392} {arXiv:2003.08392 [astro-ph.HE]}
  \BibitemShut {NoStop}%
\bibitem [{\citenamefont {Righi}\ \emph {et~al.}(2020)\citenamefont {Righi},
  \citenamefont {Palladino}, \citenamefont {Tavecchio},\ and\ \citenamefont
  {Vissani}}]{Righi:2020ufi}%
  \BibitemOpen
  \bibfield  {author} {\bibinfo {author} {\bibfnamefont {C.}~\bibnamefont
  {Righi}}, \bibinfo {author} {\bibfnamefont {A.}~\bibnamefont {Palladino}},
  \bibinfo {author} {\bibfnamefont {F.}~\bibnamefont {Tavecchio}},\ and\
  \bibinfo {author} {\bibfnamefont {F.}~\bibnamefont {Vissani}},\ }\bibfield
  {title} {\bibinfo {title} {{EeV astrophysical neutrinos from flat spectrum
  radio quasars}},\ }\href {https://doi.org/10.1051/0004-6361/202038301}
  {\bibfield  {journal} {\bibinfo  {journal} {Astron. Astrophys.}\ }\textbf
  {\bibinfo {volume} {642}},\ \bibinfo {pages} {A92} (\bibinfo {year}
  {2020})},\ \Eprint {https://arxiv.org/abs/2003.08701} {arXiv:2003.08701
  [astro-ph.HE]} \BibitemShut {NoStop}%
\bibitem [{\citenamefont {Kimura}\ \emph {et~al.}(2021)\citenamefont {Kimura},
  \citenamefont {Murase},\ and\ \citenamefont {M\'esz\'aros}}]{Kimura:2020thg}%
  \BibitemOpen
  \bibfield  {author} {\bibinfo {author} {\bibfnamefont {S.~S.}\ \bibnamefont
  {Kimura}}, \bibinfo {author} {\bibfnamefont {K.}~\bibnamefont {Murase}},\
  and\ \bibinfo {author} {\bibfnamefont {P.}~\bibnamefont {M\'esz\'aros}},\
  }\bibfield  {title} {\bibinfo {title} {{Soft gamma rays from low accreting
  supermassive black holes and connection to energetic neutrinos}},\ }\href
  {https://doi.org/10.1038/s41467-021-25111-7} {\bibfield  {journal} {\bibinfo
  {journal} {Nature Commun.}\ }\textbf {\bibinfo {volume} {12}},\ \bibinfo
  {pages} {5615} (\bibinfo {year} {2021})},\ \Eprint
  {https://arxiv.org/abs/2005.01934} {arXiv:2005.01934 [astro-ph.HE]}
  \BibitemShut {NoStop}%
\bibitem [{\citenamefont {Neronov}\ and\ \citenamefont
  {Semikoz}(2021)}]{Neronov:2020fww}%
  \BibitemOpen
  \bibfield  {author} {\bibinfo {author} {\bibfnamefont {A.}~\bibnamefont
  {Neronov}}\ and\ \bibinfo {author} {\bibfnamefont {D.}~\bibnamefont
  {Semikoz}},\ }\bibfield  {title} {\bibinfo {title} {{Radio-to-Gamma-Ray
  Synchrotron and Neutrino Emission from Proton\textendash{}Proton Interactions
  in Active Galactic Nuclei}},\ }\href
  {https://doi.org/10.1134/S0021364021020028} {\bibfield  {journal} {\bibinfo
  {journal} {JETP Lett.}\ }\textbf {\bibinfo {volume} {113}},\ \bibinfo {pages}
  {69} (\bibinfo {year} {2021})},\ \Eprint {https://arxiv.org/abs/2012.04425}
  {arXiv:2012.04425 [astro-ph.HE]} \BibitemShut {NoStop}%
\bibitem [{\citenamefont {Paczynski}\ and\ \citenamefont
  {Xu}(1994)}]{Paczynski:1994uv}%
  \BibitemOpen
  \bibfield  {author} {\bibinfo {author} {\bibfnamefont {B.}~\bibnamefont
  {Paczynski}}\ and\ \bibinfo {author} {\bibfnamefont {G.-H.}\ \bibnamefont
  {Xu}},\ }\bibfield  {title} {\bibinfo {title} {{Neutrino bursts from
  gamma-ray bursts}},\ }\href {https://doi.org/10.1086/174178} {\bibfield
  {journal} {\bibinfo  {journal} {Astrophys. J.}\ }\textbf {\bibinfo {volume}
  {427}},\ \bibinfo {pages} {708} (\bibinfo {year} {1994})}\BibitemShut
  {NoStop}%
\bibitem [{\citenamefont {Waxman}\ and\ \citenamefont
  {Bahcall}(1997)}]{Waxman:1997ti}%
  \BibitemOpen
  \bibfield  {author} {\bibinfo {author} {\bibfnamefont {E.}~\bibnamefont
  {Waxman}}\ and\ \bibinfo {author} {\bibfnamefont {J.~N.}\ \bibnamefont
  {Bahcall}},\ }\bibfield  {title} {\bibinfo {title} {{High-energy neutrinos
  from cosmological gamma-ray burst fireballs}},\ }\href
  {https://doi.org/10.1103/PhysRevLett.78.2292} {\bibfield  {journal} {\bibinfo
   {journal} {Phys. Rev. Lett.}\ }\textbf {\bibinfo {volume} {78}},\ \bibinfo
  {pages} {2292} (\bibinfo {year} {1997})},\ \Eprint
  {https://arxiv.org/abs/astro-ph/9701231} {arXiv:astro-ph/9701231}
  \BibitemShut {NoStop}%
\bibitem [{\citenamefont {Murase}\ \emph {et~al.}(2006)\citenamefont {Murase},
  \citenamefont {Ioka}, \citenamefont {Nagataki},\ and\ \citenamefont
  {Nakamura}}]{Murase:2006mm}%
  \BibitemOpen
  \bibfield  {author} {\bibinfo {author} {\bibfnamefont {K.}~\bibnamefont
  {Murase}}, \bibinfo {author} {\bibfnamefont {K.}~\bibnamefont {Ioka}},
  \bibinfo {author} {\bibfnamefont {S.}~\bibnamefont {Nagataki}},\ and\
  \bibinfo {author} {\bibfnamefont {T.}~\bibnamefont {Nakamura}},\ }\bibfield
  {title} {\bibinfo {title} {{High Energy Neutrinos and Cosmic-Rays from
  Low-Luminosity Gamma-Ray Bursts?}},\ }\href {https://doi.org/10.1086/509323}
  {\bibfield  {journal} {\bibinfo  {journal} {Astrophys. J. Lett.}\ }\textbf
  {\bibinfo {volume} {651}},\ \bibinfo {pages} {L5} (\bibinfo {year} {2006})},\
  \Eprint {https://arxiv.org/abs/astro-ph/0607104} {arXiv:astro-ph/0607104}
  \BibitemShut {NoStop}%
\bibitem [{\citenamefont {Bustamante}\ \emph
  {et~al.}(2015{\natexlab{b}})\citenamefont {Bustamante}, \citenamefont
  {Baerwald}, \citenamefont {Murase},\ and\ \citenamefont
  {Winter}}]{Bustamante:2014oka}%
  \BibitemOpen
  \bibfield  {author} {\bibinfo {author} {\bibfnamefont {M.}~\bibnamefont
  {Bustamante}}, \bibinfo {author} {\bibfnamefont {P.}~\bibnamefont
  {Baerwald}}, \bibinfo {author} {\bibfnamefont {K.}~\bibnamefont {Murase}},\
  and\ \bibinfo {author} {\bibfnamefont {W.}~\bibnamefont {Winter}},\
  }\bibfield  {title} {\bibinfo {title} {{Neutrino and cosmic-ray emission from
  multiple internal shocks in gamma-ray bursts}},\ }\href
  {https://doi.org/10.1038/ncomms7783} {\bibfield  {journal} {\bibinfo
  {journal} {Nature Commun.}\ }\textbf {\bibinfo {volume} {6}},\ \bibinfo
  {pages} {6783} (\bibinfo {year} {2015}{\natexlab{b}})},\ \Eprint
  {https://arxiv.org/abs/1409.2874} {arXiv:1409.2874 [astro-ph.HE]}
  \BibitemShut {NoStop}%
\bibitem [{\citenamefont {Senno}\ \emph {et~al.}(2016)\citenamefont {Senno},
  \citenamefont {Murase},\ and\ \citenamefont {M\'esz\'aros}}]{Senno:2015tsn}%
  \BibitemOpen
  \bibfield  {author} {\bibinfo {author} {\bibfnamefont {N.}~\bibnamefont
  {Senno}}, \bibinfo {author} {\bibfnamefont {K.}~\bibnamefont {Murase}},\ and\
  \bibinfo {author} {\bibfnamefont {P.}~\bibnamefont {M\'esz\'aros}},\
  }\bibfield  {title} {\bibinfo {title} {{Choked Jets and Low-Luminosity
  Gamma-Ray Bursts as Hidden Neutrino Sources}},\ }\href
  {https://doi.org/10.1103/PhysRevD.93.083003} {\bibfield  {journal} {\bibinfo
  {journal} {Phys. Rev. D}\ }\textbf {\bibinfo {volume} {93}},\ \bibinfo
  {pages} {083003} (\bibinfo {year} {2016})},\ \Eprint
  {https://arxiv.org/abs/1512.08513} {arXiv:1512.08513 [astro-ph.HE]}
  \BibitemShut {NoStop}%
\bibitem [{\citenamefont {Pitik}\ \emph {et~al.}(2021)\citenamefont {Pitik},
  \citenamefont {Tamborra},\ and\ \citenamefont {Petropoulou}}]{Pitik:2021xhb}%
  \BibitemOpen
  \bibfield  {author} {\bibinfo {author} {\bibfnamefont {T.}~\bibnamefont
  {Pitik}}, \bibinfo {author} {\bibfnamefont {I.}~\bibnamefont {Tamborra}},\
  and\ \bibinfo {author} {\bibfnamefont {M.}~\bibnamefont {Petropoulou}},\
  }\bibfield  {title} {\bibinfo {title} {{Neutrino signal dependence on
  gamma-ray burst emission mechanism}},\ }\href
  {https://doi.org/10.1088/1475-7516/2021/05/034} {\bibfield  {journal}
  {\bibinfo  {journal} {JCAP}\ }\textbf {\bibinfo {volume} {05}},\ \bibinfo
  {pages} {034}},\ \Eprint {https://arxiv.org/abs/2102.02223} {arXiv:2102.02223
  [astro-ph.HE]} \BibitemShut {NoStop}%
\bibitem [{\citenamefont {Rudolph}\ \emph
  {et~al.}(2023{\natexlab{a}})\citenamefont {Rudolph}, \citenamefont
  {Petropoulou}, \citenamefont {Winter},\ and\ \citenamefont
  {Bo\v{s}njak}}]{Rudolph:2022dky}%
  \BibitemOpen
  \bibfield  {author} {\bibinfo {author} {\bibfnamefont {A.}~\bibnamefont
  {Rudolph}}, \bibinfo {author} {\bibfnamefont {M.}~\bibnamefont
  {Petropoulou}}, \bibinfo {author} {\bibfnamefont {W.}~\bibnamefont
  {Winter}},\ and\ \bibinfo {author} {\bibfnamefont {Z.}~\bibnamefont
  {Bo\v{s}njak}},\ }\bibfield  {title} {\bibinfo {title} {{Multi-messenger
  Model for the Prompt Emission from GRB 221009A}},\ }\href
  {https://doi.org/10.3847/2041-8213/acb6d7} {\bibfield  {journal} {\bibinfo
  {journal} {Astrophys. J. Lett.}\ }\textbf {\bibinfo {volume} {944}},\
  \bibinfo {pages} {L34} (\bibinfo {year} {2023}{\natexlab{a}})},\ \Eprint
  {https://arxiv.org/abs/2212.00766} {arXiv:2212.00766 [astro-ph.HE]}
  \BibitemShut {NoStop}%
\bibitem [{\citenamefont {Rudolph}\ \emph
  {et~al.}(2023{\natexlab{b}})\citenamefont {Rudolph}, \citenamefont
  {Petropoulou}, \citenamefont {Bo\v{s}njak},\ and\ \citenamefont
  {Winter}}]{Rudolph:2022ppp}%
  \BibitemOpen
  \bibfield  {author} {\bibinfo {author} {\bibfnamefont {A.}~\bibnamefont
  {Rudolph}}, \bibinfo {author} {\bibfnamefont {M.}~\bibnamefont
  {Petropoulou}}, \bibinfo {author} {\bibfnamefont {Z.}~\bibnamefont
  {Bo\v{s}njak}},\ and\ \bibinfo {author} {\bibfnamefont {W.}~\bibnamefont
  {Winter}},\ }\bibfield  {title} {\bibinfo {title} {{Multicollision Internal
  Shock Lepto-hadronic Models for Energetic Gamma-Ray Bursts (GRBs)}},\ }\href
  {https://doi.org/10.3847/1538-4357/acc861} {\bibfield  {journal} {\bibinfo
  {journal} {Astrophys. J.}\ }\textbf {\bibinfo {volume} {950}},\ \bibinfo
  {pages} {28} (\bibinfo {year} {2023}{\natexlab{b}})},\ \Eprint
  {https://arxiv.org/abs/2212.00765} {arXiv:2212.00765 [astro-ph.HE]}
  \BibitemShut {NoStop}%
\bibitem [{\citenamefont {Fang}\ \emph {et~al.}(2014)\citenamefont {Fang},
  \citenamefont {Kotera}, \citenamefont {Murase},\ and\ \citenamefont
  {Olinto}}]{Fang:2013vla}%
  \BibitemOpen
  \bibfield  {author} {\bibinfo {author} {\bibfnamefont {K.}~\bibnamefont
  {Fang}}, \bibinfo {author} {\bibfnamefont {K.}~\bibnamefont {Kotera}},
  \bibinfo {author} {\bibfnamefont {K.}~\bibnamefont {Murase}},\ and\ \bibinfo
  {author} {\bibfnamefont {A.~V.}\ \bibnamefont {Olinto}},\ }\bibfield  {title}
  {\bibinfo {title} {{Testing the Newborn Pulsar Origin of Ultrahigh Energy
  Cosmic Rays with EeV Neutrinos}},\ }\href
  {https://doi.org/10.1103/PhysRevD.90.103005} {\bibfield  {journal} {\bibinfo
  {journal} {Phys. Rev. D}\ }\textbf {\bibinfo {volume} {90}},\ \bibinfo
  {pages} {103005} (\bibinfo {year} {2014})},\ \bibinfo {note} {[Erratum:
  Phys.~Rev.~D 92, 129901 (2015)]},\ \Eprint {https://arxiv.org/abs/1311.2044}
  {arXiv:1311.2044 [astro-ph.HE]} \BibitemShut {NoStop}%
\bibitem [{\citenamefont {Fang}(2015)}]{Fang:2014qva}%
  \BibitemOpen
  \bibfield  {author} {\bibinfo {author} {\bibfnamefont {K.}~\bibnamefont
  {Fang}},\ }\bibfield  {title} {\bibinfo {title} {{High-Energy Neutrino
  Signatures of Newborn Pulsars In the Local Universe}},\ }\href
  {https://doi.org/10.1088/1475-7516/2015/06/004} {\bibfield  {journal}
  {\bibinfo  {journal} {JCAP}\ }\textbf {\bibinfo {volume} {06}},\ \bibinfo
  {pages} {004}},\ \Eprint {https://arxiv.org/abs/1411.2174} {arXiv:1411.2174
  [astro-ph.HE]} \BibitemShut {NoStop}%
\bibitem [{\citenamefont {Farrar}\ and\ \citenamefont
  {Gruzinov}(2009)}]{Farrar:2008ex}%
  \BibitemOpen
  \bibfield  {author} {\bibinfo {author} {\bibfnamefont {G.~R.}\ \bibnamefont
  {Farrar}}\ and\ \bibinfo {author} {\bibfnamefont {A.}~\bibnamefont
  {Gruzinov}},\ }\bibfield  {title} {\bibinfo {title} {{Giant AGN Flares and
  Cosmic Ray Bursts}},\ }\href {https://doi.org/10.1088/0004-637X/693/1/329}
  {\bibfield  {journal} {\bibinfo  {journal} {Astrophys. J.}\ }\textbf
  {\bibinfo {volume} {693}},\ \bibinfo {pages} {329} (\bibinfo {year}
  {2009})},\ \Eprint {https://arxiv.org/abs/0802.1074} {arXiv:0802.1074
  [astro-ph]} \BibitemShut {NoStop}%
\bibitem [{\citenamefont {Wang}\ \emph {et~al.}(2011)\citenamefont {Wang},
  \citenamefont {Liu}, \citenamefont {Dai},\ and\ \citenamefont
  {Cheng}}]{Wang:2011ip}%
  \BibitemOpen
  \bibfield  {author} {\bibinfo {author} {\bibfnamefont {X.-Y.}\ \bibnamefont
  {Wang}}, \bibinfo {author} {\bibfnamefont {R.-Y.}\ \bibnamefont {Liu}},
  \bibinfo {author} {\bibfnamefont {Z.-G.}\ \bibnamefont {Dai}},\ and\ \bibinfo
  {author} {\bibfnamefont {K.~S.}\ \bibnamefont {Cheng}},\ }\bibfield  {title}
  {\bibinfo {title} {{Probing the tidal disruption flares of massive black
  holes with high-energy neutrinos}},\ }\href
  {https://doi.org/10.1103/PhysRevD.84.081301} {\bibfield  {journal} {\bibinfo
  {journal} {Phys. Rev. D}\ }\textbf {\bibinfo {volume} {84}},\ \bibinfo
  {pages} {081301} (\bibinfo {year} {2011})},\ \Eprint
  {https://arxiv.org/abs/1106.2426} {arXiv:1106.2426 [astro-ph.HE]}
  \BibitemShut {NoStop}%
\bibitem [{\citenamefont {Dai}\ and\ \citenamefont {Fang}(2017)}]{Dai:2016gtz}%
  \BibitemOpen
  \bibfield  {author} {\bibinfo {author} {\bibfnamefont {L.}~\bibnamefont
  {Dai}}\ and\ \bibinfo {author} {\bibfnamefont {K.}~\bibnamefont {Fang}},\
  }\bibfield  {title} {\bibinfo {title} {{Can tidal disruption events produce
  the IceCube neutrinos?}},\ }\href {https://doi.org/10.1093/mnras/stx863}
  {\bibfield  {journal} {\bibinfo  {journal} {Mon. Not. Roy. Astron. Soc.}\
  }\textbf {\bibinfo {volume} {469}},\ \bibinfo {pages} {1354} (\bibinfo {year}
  {2017})},\ \Eprint {https://arxiv.org/abs/1612.00011} {arXiv:1612.00011
  [astro-ph.HE]} \BibitemShut {NoStop}%
\bibitem [{\citenamefont {Senno}\ \emph {et~al.}(2017)\citenamefont {Senno},
  \citenamefont {Murase},\ and\ \citenamefont {M\'esz\'aros}}]{Senno:2016bso}%
  \BibitemOpen
  \bibfield  {author} {\bibinfo {author} {\bibfnamefont {N.}~\bibnamefont
  {Senno}}, \bibinfo {author} {\bibfnamefont {K.}~\bibnamefont {Murase}},\ and\
  \bibinfo {author} {\bibfnamefont {P.}~\bibnamefont {M\'esz\'aros}},\
  }\bibfield  {title} {\bibinfo {title} {{High-energy Neutrino Flares from
  X-Ray Bright and Dark Tidal Disruption Events}},\ }\href
  {https://doi.org/10.3847/1538-4357/aa6344} {\bibfield  {journal} {\bibinfo
  {journal} {Astrophys. J.}\ }\textbf {\bibinfo {volume} {838}},\ \bibinfo
  {pages} {3} (\bibinfo {year} {2017})},\ \Eprint
  {https://arxiv.org/abs/1612.00918} {arXiv:1612.00918 [astro-ph.HE]}
  \BibitemShut {NoStop}%
\bibitem [{\citenamefont {Lunardini}\ and\ \citenamefont
  {Winter}(2017)}]{Lunardini:2016xwi}%
  \BibitemOpen
  \bibfield  {author} {\bibinfo {author} {\bibfnamefont {C.}~\bibnamefont
  {Lunardini}}\ and\ \bibinfo {author} {\bibfnamefont {W.}~\bibnamefont
  {Winter}},\ }\bibfield  {title} {\bibinfo {title} {{High Energy Neutrinos
  from the Tidal Disruption of Stars}},\ }\href
  {https://doi.org/10.1103/PhysRevD.95.123001} {\bibfield  {journal} {\bibinfo
  {journal} {Phys. Rev. D}\ }\textbf {\bibinfo {volume} {95}},\ \bibinfo
  {pages} {123001} (\bibinfo {year} {2017})},\ \Eprint
  {https://arxiv.org/abs/1612.03160} {arXiv:1612.03160 [astro-ph.HE]}
  \BibitemShut {NoStop}%
\bibitem [{\citenamefont {Zhang}\ \emph {et~al.}(2017)\citenamefont {Zhang},
  \citenamefont {Murase}, \citenamefont {Oikonomou},\ and\ \citenamefont
  {Li}}]{Zhang:2017hom}%
  \BibitemOpen
  \bibfield  {author} {\bibinfo {author} {\bibfnamefont {B.~T.}\ \bibnamefont
  {Zhang}}, \bibinfo {author} {\bibfnamefont {K.}~\bibnamefont {Murase}},
  \bibinfo {author} {\bibfnamefont {F.}~\bibnamefont {Oikonomou}},\ and\
  \bibinfo {author} {\bibfnamefont {Z.}~\bibnamefont {Li}},\ }\bibfield
  {title} {\bibinfo {title} {{High-energy cosmic ray nuclei from tidal
  disruption events: Origin, survival, and implications}},\ }\href
  {https://doi.org/10.1103/PhysRevD.96.063007} {\bibfield  {journal} {\bibinfo
  {journal} {Phys. Rev. D}\ }\textbf {\bibinfo {volume} {96}},\ \bibinfo
  {pages} {063007} (\bibinfo {year} {2017})},\ \bibinfo {note} {[Addendum:
  Phys.~Rev.~D 96, 069902 (2017)]},\ \Eprint {https://arxiv.org/abs/1706.00391}
  {arXiv:1706.00391 [astro-ph.HE]} \BibitemShut {NoStop}%
\bibitem [{\citenamefont {Gu\'epin}\ \emph {et~al.}(2018)\citenamefont
  {Gu\'epin}, \citenamefont {Kotera}, \citenamefont {Barausse}, \citenamefont
  {Fang},\ and\ \citenamefont {Murase}}]{Guepin:2017abw}%
  \BibitemOpen
  \bibfield  {author} {\bibinfo {author} {\bibfnamefont {C.}~\bibnamefont
  {Gu\'epin}}, \bibinfo {author} {\bibfnamefont {K.}~\bibnamefont {Kotera}},
  \bibinfo {author} {\bibfnamefont {E.}~\bibnamefont {Barausse}}, \bibinfo
  {author} {\bibfnamefont {K.}~\bibnamefont {Fang}},\ and\ \bibinfo {author}
  {\bibfnamefont {K.}~\bibnamefont {Murase}},\ }\bibfield  {title} {\bibinfo
  {title} {{Ultra-High Energy Cosmic Rays and Neutrinos from Tidal Disruptions
  by Massive Black Holes}},\ }\href
  {https://doi.org/10.1051/0004-6361/201732392} {\bibfield  {journal} {\bibinfo
   {journal} {Astron. Astrophys.}\ }\textbf {\bibinfo {volume} {616}},\
  \bibinfo {pages} {A179} (\bibinfo {year} {2018})},\ \bibinfo {note}
  {[Erratum: Astron.~Astrophys.~636, C3 (2020)]},\ \Eprint
  {https://arxiv.org/abs/1711.11274} {arXiv:1711.11274 [astro-ph.HE]}
  \BibitemShut {NoStop}%
\bibitem [{\citenamefont {Winter}\ and\ \citenamefont
  {Lunardini}(2021)}]{Winter:2020ptf}%
  \BibitemOpen
  \bibfield  {author} {\bibinfo {author} {\bibfnamefont {W.}~\bibnamefont
  {Winter}}\ and\ \bibinfo {author} {\bibfnamefont {C.}~\bibnamefont
  {Lunardini}},\ }\bibfield  {title} {\bibinfo {title} {{A concordance scenario
  for the observed neutrino from a tidal disruption event}},\ }\href
  {https://doi.org/10.1038/s41550-021-01343-x} {\bibfield  {journal} {\bibinfo
  {journal} {Nature Astron.}\ }\textbf {\bibinfo {volume} {5}},\ \bibinfo
  {pages} {472} (\bibinfo {year} {2021})},\ \Eprint
  {https://arxiv.org/abs/2005.06097} {arXiv:2005.06097 [astro-ph.HE]}
  \BibitemShut {NoStop}%
\bibitem [{\citenamefont {Winter}\ and\ \citenamefont
  {Lunardini}(2023)}]{Winter:2022fpf}%
  \BibitemOpen
  \bibfield  {author} {\bibinfo {author} {\bibfnamefont {W.}~\bibnamefont
  {Winter}}\ and\ \bibinfo {author} {\bibfnamefont {C.}~\bibnamefont
  {Lunardini}},\ }\bibfield  {title} {\bibinfo {title} {{Interpretation of the
  Observed Neutrino Emission from Three Tidal Disruption Events}},\ }\href
  {https://doi.org/10.3847/1538-4357/acbe9e} {\bibfield  {journal} {\bibinfo
  {journal} {Astrophys. J.}\ }\textbf {\bibinfo {volume} {948}},\ \bibinfo
  {pages} {42} (\bibinfo {year} {2023})},\ \Eprint
  {https://arxiv.org/abs/2205.11538} {arXiv:2205.11538 [astro-ph.HE]}
  \BibitemShut {NoStop}%
\bibitem [{\citenamefont {Aloisio}\ \emph {et~al.}(2011)\citenamefont
  {Aloisio}, \citenamefont {Berezinsky},\ and\ \citenamefont
  {Gazizov}}]{Aloisio:2009sj}%
  \BibitemOpen
  \bibfield  {author} {\bibinfo {author} {\bibfnamefont {R.}~\bibnamefont
  {Aloisio}}, \bibinfo {author} {\bibfnamefont {V.}~\bibnamefont
  {Berezinsky}},\ and\ \bibinfo {author} {\bibfnamefont {A.}~\bibnamefont
  {Gazizov}},\ }\bibfield  {title} {\bibinfo {title} {{Ultra High Energy Cosmic
  Rays: The disappointing model}},\ }\href
  {https://doi.org/10.1016/j.astropartphys.2010.12.008} {\bibfield  {journal}
  {\bibinfo  {journal} {Astropart. Phys.}\ }\textbf {\bibinfo {volume} {34}},\
  \bibinfo {pages} {620} (\bibinfo {year} {2011})},\ \Eprint
  {https://arxiv.org/abs/0907.5194} {arXiv:0907.5194 [astro-ph.HE]}
  \BibitemShut {NoStop}%
\bibitem [{\citenamefont {Kotera}\ \emph {et~al.}(2010)\citenamefont {Kotera},
  \citenamefont {Allard},\ and\ \citenamefont {Olinto}}]{Kotera:2010yn}%
  \BibitemOpen
  \bibfield  {author} {\bibinfo {author} {\bibfnamefont {K.}~\bibnamefont
  {Kotera}}, \bibinfo {author} {\bibfnamefont {D.}~\bibnamefont {Allard}},\
  and\ \bibinfo {author} {\bibfnamefont {A.~V.}\ \bibnamefont {Olinto}},\
  }\bibfield  {title} {\bibinfo {title} {{Cosmogenic Neutrinos: parameter space
  and detectabilty from PeV to ZeV}},\ }\href
  {https://doi.org/10.1088/1475-7516/2010/10/013} {\bibfield  {journal}
  {\bibinfo  {journal} {JCAP}\ }\textbf {\bibinfo {volume} {10}},\ \bibinfo
  {pages} {013}},\ \Eprint {https://arxiv.org/abs/1009.1382} {arXiv:1009.1382
  [astro-ph.HE]} \BibitemShut {NoStop}%
\bibitem [{\citenamefont {Ahlers}\ and\ \citenamefont
  {Halzen}(2012)}]{Ahlers:2012rz}%
  \BibitemOpen
  \bibfield  {author} {\bibinfo {author} {\bibfnamefont {M.}~\bibnamefont
  {Ahlers}}\ and\ \bibinfo {author} {\bibfnamefont {F.}~\bibnamefont
  {Halzen}},\ }\bibfield  {title} {\bibinfo {title} {{Minimal Cosmogenic
  Neutrinos}},\ }\href {https://doi.org/10.1103/PhysRevD.86.083010} {\bibfield
  {journal} {\bibinfo  {journal} {Phys. Rev. D}\ }\textbf {\bibinfo {volume}
  {86}},\ \bibinfo {pages} {083010} (\bibinfo {year} {2012})},\ \Eprint
  {https://arxiv.org/abs/1208.4181} {arXiv:1208.4181 [astro-ph.HE]}
  \BibitemShut {NoStop}%
\bibitem [{\citenamefont {Heinze}\ \emph {et~al.}(2016)\citenamefont {Heinze},
  \citenamefont {Boncioli}, \citenamefont {Bustamante},\ and\ \citenamefont
  {Winter}}]{Heinze:2015hhp}%
  \BibitemOpen
  \bibfield  {author} {\bibinfo {author} {\bibfnamefont {J.}~\bibnamefont
  {Heinze}}, \bibinfo {author} {\bibfnamefont {D.}~\bibnamefont {Boncioli}},
  \bibinfo {author} {\bibfnamefont {M.}~\bibnamefont {Bustamante}},\ and\
  \bibinfo {author} {\bibfnamefont {W.}~\bibnamefont {Winter}},\ }\bibfield
  {title} {\bibinfo {title} {{Cosmogenic Neutrinos Challenge the Cosmic Ray
  Proton Dip Model}},\ }\href {https://doi.org/10.3847/0004-637X/825/2/122}
  {\bibfield  {journal} {\bibinfo  {journal} {Astrophys. J.}\ }\textbf
  {\bibinfo {volume} {825}},\ \bibinfo {pages} {122} (\bibinfo {year}
  {2016})},\ \Eprint {https://arxiv.org/abs/1512.05988} {arXiv:1512.05988
  [astro-ph.HE]} \BibitemShut {NoStop}%
\bibitem [{\citenamefont {Fang}\ and\ \citenamefont
  {Murase}(2018)}]{Fang:2017zjf}%
  \BibitemOpen
  \bibfield  {author} {\bibinfo {author} {\bibfnamefont {K.}~\bibnamefont
  {Fang}}\ and\ \bibinfo {author} {\bibfnamefont {K.}~\bibnamefont {Murase}},\
  }\bibfield  {title} {\bibinfo {title} {{Linking High-Energy Cosmic Particles
  by Black Hole Jets Embedded in Large-Scale Structures}},\ }\href
  {https://doi.org/10.1038/s41567-017-0025-4} {\bibfield  {journal} {\bibinfo
  {journal} {Nature Phys.}\ }\textbf {\bibinfo {volume} {14}},\ \bibinfo
  {pages} {396} (\bibinfo {year} {2018})},\ \Eprint
  {https://arxiv.org/abs/1704.00015} {arXiv:1704.00015 [astro-ph.HE]}
  \BibitemShut {NoStop}%
\bibitem [{\citenamefont {Romero-Wolf}\ and\ \citenamefont
  {Ave}(2018)}]{Romero-Wolf:2017xqe}%
  \BibitemOpen
  \bibfield  {author} {\bibinfo {author} {\bibfnamefont {A.}~\bibnamefont
  {Romero-Wolf}}\ and\ \bibinfo {author} {\bibfnamefont {M.}~\bibnamefont
  {Ave}},\ }\bibfield  {title} {\bibinfo {title} {{Bayesian Inference
  Constraints on Astrophysical Production of Ultra-high Energy Cosmic Rays and
  Cosmogenic Neutrino Flux Predictions}},\ }\href
  {https://doi.org/10.1088/1475-7516/2018/07/025} {\bibfield  {journal}
  {\bibinfo  {journal} {JCAP}\ }\textbf {\bibinfo {volume} {07}},\ \bibinfo
  {pages} {025}},\ \Eprint {https://arxiv.org/abs/1712.07290} {arXiv:1712.07290
  [astro-ph.HE]} \BibitemShut {NoStop}%
\bibitem [{\citenamefont {Alves~Batista}\ \emph {et~al.}(2019)\citenamefont
  {Alves~Batista}, \citenamefont {de~Almeida}, \citenamefont {Lago},\ and\
  \citenamefont {Kotera}}]{AlvesBatista:2018zui}%
  \BibitemOpen
  \bibfield  {author} {\bibinfo {author} {\bibfnamefont {R.}~\bibnamefont
  {Alves~Batista}}, \bibinfo {author} {\bibfnamefont {R.~M.}\ \bibnamefont
  {de~Almeida}}, \bibinfo {author} {\bibfnamefont {B.}~\bibnamefont {Lago}},\
  and\ \bibinfo {author} {\bibfnamefont {K.}~\bibnamefont {Kotera}},\
  }\bibfield  {title} {\bibinfo {title} {{Cosmogenic photon and neutrino fluxes
  in the Auger era}},\ }\href {https://doi.org/10.1088/1475-7516/2019/01/002}
  {\bibfield  {journal} {\bibinfo  {journal} {JCAP}\ }\textbf {\bibinfo
  {volume} {01}},\ \bibinfo {pages} {002}},\ \Eprint
  {https://arxiv.org/abs/1806.10879} {arXiv:1806.10879 [astro-ph.HE]}
  \BibitemShut {NoStop}%
\bibitem [{\citenamefont {Heinze}\ \emph {et~al.}(2019)\citenamefont {Heinze},
  \citenamefont {Fedynitch}, \citenamefont {Boncioli},\ and\ \citenamefont
  {Winter}}]{Heinze:2019jou}%
  \BibitemOpen
  \bibfield  {author} {\bibinfo {author} {\bibfnamefont {J.}~\bibnamefont
  {Heinze}}, \bibinfo {author} {\bibfnamefont {A.}~\bibnamefont {Fedynitch}},
  \bibinfo {author} {\bibfnamefont {D.}~\bibnamefont {Boncioli}},\ and\
  \bibinfo {author} {\bibfnamefont {W.}~\bibnamefont {Winter}},\ }\bibfield
  {title} {\bibinfo {title} {{A new view on Auger data and cosmogenic neutrinos
  in light of different nuclear disintegration and air-shower models}},\ }\href
  {https://doi.org/10.3847/1538-4357/ab05ce} {\bibfield  {journal} {\bibinfo
  {journal} {Astrophys. J.}\ }\textbf {\bibinfo {volume} {873}},\ \bibinfo
  {pages} {88} (\bibinfo {year} {2019})},\ \Eprint
  {https://arxiv.org/abs/1901.03338} {arXiv:1901.03338 [astro-ph.HE]}
  \BibitemShut {NoStop}%
\bibitem [{\citenamefont {Muzio}\ \emph {et~al.}(2019)\citenamefont {Muzio},
  \citenamefont {Unger},\ and\ \citenamefont {Farrar}}]{Muzio:2019leu}%
  \BibitemOpen
  \bibfield  {author} {\bibinfo {author} {\bibfnamefont {M.~S.}\ \bibnamefont
  {Muzio}}, \bibinfo {author} {\bibfnamefont {M.}~\bibnamefont {Unger}},\ and\
  \bibinfo {author} {\bibfnamefont {G.~R.}\ \bibnamefont {Farrar}},\ }\bibfield
   {title} {\bibinfo {title} {{Progress towards characterizing ultrahigh energy
  cosmic ray sources}},\ }\href {https://doi.org/10.1103/PhysRevD.100.103008}
  {\bibfield  {journal} {\bibinfo  {journal} {Phys. Rev. D}\ }\textbf {\bibinfo
  {volume} {100}},\ \bibinfo {pages} {103008} (\bibinfo {year} {2019})},\
  \Eprint {https://arxiv.org/abs/1906.06233} {arXiv:1906.06233 [astro-ph.HE]}
  \BibitemShut {NoStop}%
\bibitem [{\citenamefont {Anker}\ \emph {et~al.}(2020)\citenamefont {Anker}
  \emph {et~al.}}]{Anker:2020lre}%
  \BibitemOpen
  \bibfield  {author} {\bibinfo {author} {\bibfnamefont {A.}~\bibnamefont
  {Anker}} \emph {et~al.} (\bibinfo {collaboration} {{ARIANNA}}),\ }\bibfield
  {title} {\bibinfo {title} {{White Paper: ARIANNA-200 high energy neutrino
  telescope}},\ }\href@noop {} {\  (\bibinfo {year} {2020})},\ \Eprint
  {https://arxiv.org/abs/2004.09841} {arXiv:2004.09841 [astro-ph.IM]}
  \BibitemShut {NoStop}%
\bibitem [{\citenamefont {Muzio}\ \emph {et~al.}(2022)\citenamefont {Muzio},
  \citenamefont {Farrar},\ and\ \citenamefont {Unger}}]{Muzio:2021zud}%
  \BibitemOpen
  \bibfield  {author} {\bibinfo {author} {\bibfnamefont {M.~S.}\ \bibnamefont
  {Muzio}}, \bibinfo {author} {\bibfnamefont {G.~R.}\ \bibnamefont {Farrar}},\
  and\ \bibinfo {author} {\bibfnamefont {M.}~\bibnamefont {Unger}},\ }\bibfield
   {title} {\bibinfo {title} {{Probing the environments surrounding ultrahigh
  energy cosmic ray accelerators and their implications for astrophysical
  neutrinos}},\ }\href {https://doi.org/10.1103/PhysRevD.105.023022} {\bibfield
   {journal} {\bibinfo  {journal} {Phys. Rev. D}\ }\textbf {\bibinfo {volume}
  {105}},\ \bibinfo {pages} {023022} (\bibinfo {year} {2022})},\ \Eprint
  {https://arxiv.org/abs/2108.05512} {arXiv:2108.05512 [astro-ph.HE]}
  \BibitemShut {NoStop}%
\bibitem [{\citenamefont {Fiorillo}\ and\ \citenamefont
  {Bustamante}(2023)}]{Fiorillo:2022rft}%
  \BibitemOpen
  \bibfield  {author} {\bibinfo {author} {\bibfnamefont {D.~F.~G.}\
  \bibnamefont {Fiorillo}}\ and\ \bibinfo {author} {\bibfnamefont
  {M.}~\bibnamefont {Bustamante}},\ }\bibfield  {title} {\bibinfo {title}
  {{Bump hunting in the diffuse flux of high-energy cosmic neutrinos}},\ }\href
  {https://doi.org/10.1103/PhysRevD.107.083008} {\bibfield  {journal} {\bibinfo
   {journal} {Phys. Rev. D}\ }\textbf {\bibinfo {volume} {107}},\ \bibinfo
  {pages} {083008} (\bibinfo {year} {2023})},\ \Eprint
  {https://arxiv.org/abs/2301.00024} {arXiv:2301.00024 [astro-ph.HE]}
  \BibitemShut {NoStop}%
\bibitem [{\citenamefont {Stecker}\ \emph {et~al.}(1991)\citenamefont
  {Stecker}, \citenamefont {Done}, \citenamefont {Salamon},\ and\ \citenamefont
  {Sommers}}]{Stecker:1991vm}%
  \BibitemOpen
  \bibfield  {author} {\bibinfo {author} {\bibfnamefont {F.~W.}\ \bibnamefont
  {Stecker}}, \bibinfo {author} {\bibfnamefont {C.}~\bibnamefont {Done}},
  \bibinfo {author} {\bibfnamefont {M.~H.}\ \bibnamefont {Salamon}},\ and\
  \bibinfo {author} {\bibfnamefont {P.}~\bibnamefont {Sommers}},\ }\bibfield
  {title} {\bibinfo {title} {{High-energy neutrinos from active galactic
  nuclei}},\ }\href {https://doi.org/10.1103/PhysRevLett.66.2697} {\bibfield
  {journal} {\bibinfo  {journal} {Phys. Rev. Lett.}\ }\textbf {\bibinfo
  {volume} {66}},\ \bibinfo {pages} {2697} (\bibinfo {year} {1991})},\ \bibinfo
  {note} {[Erratum: Phys.Rev.Lett. 69, 2738 (1992)]}\BibitemShut {NoStop}%
\bibitem [{\citenamefont {Learned}\ and\ \citenamefont
  {Mannheim}(2000)}]{Learned:2000sw}%
  \BibitemOpen
  \bibfield  {author} {\bibinfo {author} {\bibfnamefont {J.~G.}\ \bibnamefont
  {Learned}}\ and\ \bibinfo {author} {\bibfnamefont {K.}~\bibnamefont
  {Mannheim}},\ }\bibfield  {title} {\bibinfo {title} {{High-energy neutrino
  astrophysics}},\ }\href {https://doi.org/10.1146/annurev.nucl.50.1.679}
  {\bibfield  {journal} {\bibinfo  {journal} {Ann. Rev. Nucl. Part. Sci.}\
  }\textbf {\bibinfo {volume} {50}},\ \bibinfo {pages} {679} (\bibinfo {year}
  {2000})}\BibitemShut {NoStop}%
\bibitem [{\citenamefont {Winter}(2012)}]{Winter:2012xq}%
  \BibitemOpen
  \bibfield  {author} {\bibinfo {author} {\bibfnamefont {W.}~\bibnamefont
  {Winter}},\ }\bibfield  {title} {\bibinfo {title} {{Neutrinos from Cosmic
  Accelerators Including Magnetic Field and Flavor Effects}},\ }\href
  {https://doi.org/10.1155/2012/586413} {\bibfield  {journal} {\bibinfo
  {journal} {Adv. High Energy Phys.}\ }\textbf {\bibinfo {volume} {2012}},\
  \bibinfo {pages} {586413} (\bibinfo {year} {2012})},\ \Eprint
  {https://arxiv.org/abs/1201.5462} {arXiv:1201.5462 [astro-ph.HE]}
  \BibitemShut {NoStop}%
\bibitem [{\citenamefont {Fiorillo}\ \emph
  {et~al.}(2021{\natexlab{b}})\citenamefont {Fiorillo}, \citenamefont
  {Van~Vliet}, \citenamefont {Morisi},\ and\ \citenamefont
  {Winter}}]{Fiorillo:2021hty}%
  \BibitemOpen
  \bibfield  {author} {\bibinfo {author} {\bibfnamefont {D.~F.~G.}\
  \bibnamefont {Fiorillo}}, \bibinfo {author} {\bibfnamefont {A.}~\bibnamefont
  {Van~Vliet}}, \bibinfo {author} {\bibfnamefont {S.}~\bibnamefont {Morisi}},\
  and\ \bibinfo {author} {\bibfnamefont {W.}~\bibnamefont {Winter}},\
  }\bibfield  {title} {\bibinfo {title} {{Unified thermal model for
  photohadronic neutrino production in astrophysical sources}},\ }\href
  {https://doi.org/10.1088/1475-7516/2021/07/028} {\bibfield  {journal}
  {\bibinfo  {journal} {JCAP}\ }\textbf {\bibinfo {volume} {07}},\ \bibinfo
  {pages} {028}},\ \Eprint {https://arxiv.org/abs/2103.16577} {arXiv:2103.16577
  [astro-ph.HE]} \BibitemShut {NoStop}%
\bibitem [{\citenamefont {Pontecorvo}(1957)}]{Pontecorvo:1957qd}%
  \BibitemOpen
  \bibfield  {author} {\bibinfo {author} {\bibfnamefont {B.}~\bibnamefont
  {Pontecorvo}},\ }\bibfield  {title} {\bibinfo {title} {{Inverse beta
  processes and nonconservation of lepton charge}},\ }\href@noop {} {\bibfield
  {journal} {\bibinfo  {journal} {Zh. Eksp. Teor. Fiz.}\ }\textbf {\bibinfo
  {volume} {34}},\ \bibinfo {pages} {247} (\bibinfo {year} {1957})}\BibitemShut
  {NoStop}%
\bibitem [{\citenamefont {Maki}\ \emph {et~al.}(1962)\citenamefont {Maki},
  \citenamefont {Nakagawa},\ and\ \citenamefont {Sakata}}]{Maki:1962mu}%
  \BibitemOpen
  \bibfield  {author} {\bibinfo {author} {\bibfnamefont {Z.}~\bibnamefont
  {Maki}}, \bibinfo {author} {\bibfnamefont {M.}~\bibnamefont {Nakagawa}},\
  and\ \bibinfo {author} {\bibfnamefont {S.}~\bibnamefont {Sakata}},\
  }\bibfield  {title} {\bibinfo {title} {{Remarks on the unified model of
  elementary particles}},\ }\href {https://doi.org/10.1143/PTP.28.870}
  {\bibfield  {journal} {\bibinfo  {journal} {Prog. Theor. Phys.}\ }\textbf
  {\bibinfo {volume} {28}},\ \bibinfo {pages} {870} (\bibinfo {year}
  {1962})}\BibitemShut {NoStop}%
\bibitem [{\citenamefont {Esteban}\ \emph {et~al.}(2020)\citenamefont
  {Esteban}, \citenamefont {Gonz{\'a}lez-Garc{\'i}a}, \citenamefont {Maltoni},
  \citenamefont {Schwetz},\ and\ \citenamefont {Zhou}}]{Esteban:2020cvm}%
  \BibitemOpen
  \bibfield  {author} {\bibinfo {author} {\bibfnamefont {I.}~\bibnamefont
  {Esteban}}, \bibinfo {author} {\bibfnamefont {M.~C.}\ \bibnamefont
  {Gonz{\'a}lez-Garc{\'i}a}}, \bibinfo {author} {\bibfnamefont
  {M.}~\bibnamefont {Maltoni}}, \bibinfo {author} {\bibfnamefont
  {T.}~\bibnamefont {Schwetz}},\ and\ \bibinfo {author} {\bibfnamefont
  {A.}~\bibnamefont {Zhou}},\ }\bibfield  {title} {\bibinfo {title} {{The fate
  of hints: updated global analysis of three-flavor neutrino oscillations}},\
  }\href {https://doi.org/10.1007/JHEP09(2020)178} {\bibfield  {journal}
  {\bibinfo  {journal} {JHEP}\ }\textbf {\bibinfo {volume} {09}},\ \bibinfo
  {pages} {178}},\ \Eprint {https://arxiv.org/abs/2007.14792} {arXiv:2007.14792
  [hep-ph]} \BibitemShut {NoStop}%
\bibitem [{\citenamefont {Esteban}\ \emph {et~al.}(2022)\citenamefont
  {Esteban}, \citenamefont {Gonz{\'a}lez-Garc{\'i}a}, \citenamefont {Maltoni},
  \citenamefont {Schwetz},\ and\ \citenamefont {Zhou}}]{NuFit5.2}%
  \BibitemOpen
  \bibfield  {author} {\bibinfo {author} {\bibfnamefont {I.}~\bibnamefont
  {Esteban}}, \bibinfo {author} {\bibfnamefont {M.~C.}\ \bibnamefont
  {Gonz{\'a}lez-Garc{\'i}a}}, \bibinfo {author} {\bibfnamefont
  {M.}~\bibnamefont {Maltoni}}, \bibinfo {author} {\bibfnamefont
  {T.}~\bibnamefont {Schwetz}},\ and\ \bibinfo {author} {\bibfnamefont
  {A.}~\bibnamefont {Zhou}},\ }\href@noop {} {}\bibinfo {howpublished}
  {\url{http://www.nu-fit.org/}} (\bibinfo {year} {2022}),\ \bibinfo {note}
  {{{NuFit~5.2}}}\BibitemShut {NoStop}%
\bibitem [{\citenamefont {Aghanim}\ \emph {et~al.}(2020)\citenamefont {Aghanim}
  \emph {et~al.}}]{Planck:2018vyg}%
  \BibitemOpen
  \bibfield  {author} {\bibinfo {author} {\bibfnamefont {N.}~\bibnamefont
  {Aghanim}} \emph {et~al.} (\bibinfo {collaboration} {Planck}),\ }\bibfield
  {title} {\bibinfo {title} {{Planck 2018 results. VI. Cosmological
  parameters}},\ }\href {https://doi.org/10.1051/0004-6361/201833910}
  {\bibfield  {journal} {\bibinfo  {journal} {Astron. Astrophys.}\ }\textbf
  {\bibinfo {volume} {641}},\ \bibinfo {pages} {A6} (\bibinfo {year} {2020})},\
  \bibinfo {note} {[Erratum: Astron.~Astrophys. 652, C4 (2021)]},\ \Eprint
  {https://arxiv.org/abs/1807.06209} {arXiv:1807.06209 [astro-ph.CO]}
  \BibitemShut {NoStop}%
\bibitem [{\citenamefont {Cole}\ \emph {et~al.}(2001)\citenamefont {Cole} \emph
  {et~al.}}]{2dFGRS:2000opy}%
  \BibitemOpen
  \bibfield  {author} {\bibinfo {author} {\bibfnamefont {S.}~\bibnamefont
  {Cole}} \emph {et~al.} (\bibinfo {collaboration} {2dFGRS}),\ }\bibfield
  {title} {\bibinfo {title} {{The 2dF Galaxy Redshift Survey: Near infrared
  galaxy luminosity functions}},\ }\href
  {https://doi.org/10.1046/j.1365-8711.2001.04591.x} {\bibfield  {journal}
  {\bibinfo  {journal} {Mon. Not. Roy. Astron. Soc.}\ }\textbf {\bibinfo
  {volume} {326}},\ \bibinfo {pages} {255} (\bibinfo {year} {2001})},\ \Eprint
  {https://arxiv.org/abs/astro-ph/0012429} {arXiv:astro-ph/0012429}
  \BibitemShut {NoStop}%
\bibitem [{\citenamefont {Hopkins}\ and\ \citenamefont
  {Beacom}(2006)}]{Hopkins:2006bw}%
  \BibitemOpen
  \bibfield  {author} {\bibinfo {author} {\bibfnamefont {A.~M.}\ \bibnamefont
  {Hopkins}}\ and\ \bibinfo {author} {\bibfnamefont {J.~F.}\ \bibnamefont
  {Beacom}},\ }\bibfield  {title} {\bibinfo {title} {{On the normalisation of
  the cosmic star formation history}},\ }\href {https://doi.org/10.1086/506610}
  {\bibfield  {journal} {\bibinfo  {journal} {Astrophys. J.}\ }\textbf
  {\bibinfo {volume} {651}},\ \bibinfo {pages} {142} (\bibinfo {year}
  {2006})},\ \Eprint {https://arxiv.org/abs/astro-ph/0601463}
  {arXiv:astro-ph/0601463} \BibitemShut {NoStop}%
\bibitem [{\citenamefont {Aartsen}\ \emph
  {et~al.}(2018{\natexlab{b}})\citenamefont {Aartsen} \emph
  {et~al.}}]{IceCube:2017qyp}%
  \BibitemOpen
  \bibfield  {author} {\bibinfo {author} {\bibfnamefont {M.~G.}\ \bibnamefont
  {Aartsen}} \emph {et~al.} (\bibinfo {collaboration} {IceCube}),\ }\bibfield
  {title} {\bibinfo {title} {{Neutrino Interferometry for High-Precision Tests
  of Lorentz Symmetry with IceCube}},\ }\href
  {https://doi.org/10.1038/s41567-018-0172-2} {\bibfield  {journal} {\bibinfo
  {journal} {Nature Phys.}\ }\textbf {\bibinfo {volume} {14}},\ \bibinfo
  {pages} {961} (\bibinfo {year} {2018}{\natexlab{b}})},\ \Eprint
  {https://arxiv.org/abs/1709.03434} {arXiv:1709.03434 [hep-ex]} \BibitemShut
  {NoStop}%
\bibitem [{\citenamefont {Abbasi}\ \emph
  {et~al.}(2022{\natexlab{c}})\citenamefont {Abbasi} \emph
  {et~al.}}]{IceCube:2021tdn}%
  \BibitemOpen
  \bibfield  {author} {\bibinfo {author} {\bibfnamefont {R.}~\bibnamefont
  {Abbasi}} \emph {et~al.} (\bibinfo {collaboration} {IceCube}),\ }\bibfield
  {title} {\bibinfo {title} {{Search for quantum gravity using astrophysical
  neutrino flavour with IceCube}},\ }\href
  {https://doi.org/10.1038/s41567-022-01762-1} {\bibfield  {journal} {\bibinfo
  {journal} {Nature Phys.}\ }\textbf {\bibinfo {volume} {18}},\ \bibinfo
  {pages} {1287} (\bibinfo {year} {2022}{\natexlab{c}})},\ \Eprint
  {https://arxiv.org/abs/2111.04654} {arXiv:2111.04654 [hep-ex]} \BibitemShut
  {NoStop}%
\bibitem [{\citenamefont {Schr\"oder}(2017)}]{Schroder:2016hrv}%
  \BibitemOpen
  \bibfield  {author} {\bibinfo {author} {\bibfnamefont {F.~G.}\ \bibnamefont
  {Schr\"oder}},\ }\bibfield  {title} {\bibinfo {title} {{Radio detection of
  Cosmic-Ray Air Showers and High-Energy Neutrinos}},\ }\href
  {https://doi.org/10.1016/j.ppnp.2016.12.002} {\bibfield  {journal} {\bibinfo
  {journal} {Prog. Part. Nucl. Phys.}\ }\textbf {\bibinfo {volume} {93}},\
  \bibinfo {pages} {1} (\bibinfo {year} {2017})},\ \Eprint
  {https://arxiv.org/abs/1607.08781} {arXiv:1607.08781 [astro-ph.IM]}
  \BibitemShut {NoStop}%
\bibitem [{\citenamefont {Aartsen}\ \emph
  {et~al.}(2017{\natexlab{b}})\citenamefont {Aartsen} \emph
  {et~al.}}]{IceCube:2017roe}%
  \BibitemOpen
  \bibfield  {author} {\bibinfo {author} {\bibfnamefont {M.~G.}\ \bibnamefont
  {Aartsen}} \emph {et~al.} (\bibinfo {collaboration} {IceCube}),\ }\bibfield
  {title} {\bibinfo {title} {{Measurement of the multi-TeV neutrino cross
  section with IceCube using Earth absorption}},\ }\href
  {https://doi.org/10.1038/nature24459} {\bibfield  {journal} {\bibinfo
  {journal} {Nature}\ }\textbf {\bibinfo {volume} {551}},\ \bibinfo {pages}
  {596} (\bibinfo {year} {2017}{\natexlab{b}})},\ \Eprint
  {https://arxiv.org/abs/1711.08119} {arXiv:1711.08119 [hep-ex]} \BibitemShut
  {NoStop}%
\bibitem [{\citenamefont {Bustamante}\ and\ \citenamefont
  {Connolly}(2019)}]{Bustamante:2017xuy}%
  \BibitemOpen
  \bibfield  {author} {\bibinfo {author} {\bibfnamefont {M.}~\bibnamefont
  {Bustamante}}\ and\ \bibinfo {author} {\bibfnamefont {A.}~\bibnamefont
  {Connolly}},\ }\bibfield  {title} {\bibinfo {title} {{Extracting the
  Energy-Dependent Neutrino-Nucleon Cross Section above 10 TeV Using IceCube
  Showers}},\ }\href {https://doi.org/10.1103/PhysRevLett.122.041101}
  {\bibfield  {journal} {\bibinfo  {journal} {Phys. Rev. Lett.}\ }\textbf
  {\bibinfo {volume} {122}},\ \bibinfo {pages} {041101} (\bibinfo {year}
  {2019})},\ \Eprint {https://arxiv.org/abs/1711.11043} {arXiv:1711.11043
  [astro-ph.HE]} \BibitemShut {NoStop}%
\bibitem [{\citenamefont {Abbasi}\ \emph {et~al.}(2020)\citenamefont {Abbasi}
  \emph {et~al.}}]{IceCube:2020rnc}%
  \BibitemOpen
  \bibfield  {author} {\bibinfo {author} {\bibfnamefont {R.}~\bibnamefont
  {Abbasi}} \emph {et~al.} (\bibinfo {collaboration} {IceCube}),\ }\bibfield
  {title} {\bibinfo {title} {{Measurement of the high-energy all-flavor
  neutrino-nucleon cross section with IceCube}},\ }\href
  {https://doi.org/10.1103/PhysRevD.104.022001} {\bibfield  {journal} {\bibinfo
   {journal} {Phys. Rev. D}\ }\textbf {\bibinfo {volume} {104}},\ \bibinfo
  {pages} {022001} (\bibinfo {year} {2020})},\ \Eprint
  {https://arxiv.org/abs/2011.03560} {arXiv:2011.03560 [hep-ex]} \BibitemShut
  {NoStop}%
\bibitem [{\citenamefont {Askar'yan}(1961)}]{Askaryan:1961pfb}%
  \BibitemOpen
  \bibfield  {author} {\bibinfo {author} {\bibfnamefont {G.~A.}\ \bibnamefont
  {Askar'yan}},\ }\bibfield  {title} {\bibinfo {title} {{Excess negative charge
  of an electron-photon shower and its coherent radio emission}},\ }\href@noop
  {} {\bibfield  {journal} {\bibinfo  {journal} {Zh. Eksp. Teor. Fiz.}\
  }\textbf {\bibinfo {volume} {41}},\ \bibinfo {pages} {616} (\bibinfo {year}
  {1961})}\BibitemShut {NoStop}%
\bibitem [{\citenamefont {Halzen}\ \emph {et~al.}(1991)\citenamefont {Halzen},
  \citenamefont {Zas},\ and\ \citenamefont {Stanev}}]{Halzen:1990vt}%
  \BibitemOpen
  \bibfield  {author} {\bibinfo {author} {\bibfnamefont {F.}~\bibnamefont
  {Halzen}}, \bibinfo {author} {\bibfnamefont {E.}~\bibnamefont {Zas}},\ and\
  \bibinfo {author} {\bibfnamefont {T.}~\bibnamefont {Stanev}},\ }\bibfield
  {title} {\bibinfo {title} {{Radiodetection of cosmic neutrinos: A Numerical,
  real time analysis}},\ }\href {https://doi.org/10.1016/0370-2693(91)91920-Q}
  {\bibfield  {journal} {\bibinfo  {journal} {Phys. Lett. B}\ }\textbf
  {\bibinfo {volume} {257}},\ \bibinfo {pages} {432} (\bibinfo {year}
  {1991})}\BibitemShut {NoStop}%
\bibitem [{\citenamefont {Zas}\ \emph {et~al.}(1992)\citenamefont {Zas},
  \citenamefont {Halzen},\ and\ \citenamefont {Stanev}}]{Zas:1991jv}%
  \BibitemOpen
  \bibfield  {author} {\bibinfo {author} {\bibfnamefont {E.}~\bibnamefont
  {Zas}}, \bibinfo {author} {\bibfnamefont {F.}~\bibnamefont {Halzen}},\ and\
  \bibinfo {author} {\bibfnamefont {T.}~\bibnamefont {Stanev}},\ }\bibfield
  {title} {\bibinfo {title} {{Electromagnetic pulses from high-energy showers:
  Implications for neutrino detection}},\ }\href
  {https://doi.org/10.1103/PhysRevD.45.362} {\bibfield  {journal} {\bibinfo
  {journal} {Phys. Rev. D}\ }\textbf {\bibinfo {volume} {45}},\ \bibinfo
  {pages} {362} (\bibinfo {year} {1992})}\BibitemShut {NoStop}%
\bibitem [{\citenamefont {Fargion}\ \emph {et~al.}(1999)\citenamefont
  {Fargion}, \citenamefont {Aiello},\ and\ \citenamefont
  {Conversano}}]{Fargion:1999se}%
  \BibitemOpen
  \bibfield  {author} {\bibinfo {author} {\bibfnamefont {D.}~\bibnamefont
  {Fargion}}, \bibinfo {author} {\bibfnamefont {A.}~\bibnamefont {Aiello}},\
  and\ \bibinfo {author} {\bibfnamefont {R.}~\bibnamefont {Conversano}},\
  }\bibfield  {title} {\bibinfo {title} {{Horizontal tau air showers from
  mountains in deep valley: Traces of UHECR neutrino tau}},\ }in\ \href@noop {}
  {\emph {\bibinfo {booktitle} {{26th International Cosmic Ray Conference}}}}\
  (\bibinfo {year} {1999})\ \Eprint {https://arxiv.org/abs/astro-ph/9906450}
  {arXiv:astro-ph/9906450} \BibitemShut {NoStop}%
\bibitem [{\citenamefont {Kahn}\ \emph {et~al.}(1966)\citenamefont {Kahn},
  \citenamefont {Lerche},\ and\ \citenamefont {Lovell}}]{geomagnetic}%
  \BibitemOpen
  \bibfield  {author} {\bibinfo {author} {\bibfnamefont {F.~D.}\ \bibnamefont
  {Kahn}}, \bibinfo {author} {\bibfnamefont {I.}~\bibnamefont {Lerche}},\ and\
  \bibinfo {author} {\bibfnamefont {A.~C.~B.}\ \bibnamefont {Lovell}},\
  }\bibfield  {title} {\bibinfo {title} {Radiation from cosmic ray air
  showers}\ }\href {https://doi.org/https://doi.org/10.1098/rspa.1966.0007.}
  {https://doi.org/10.1098/rspa.1966.0007.} (\bibinfo {year}
  {1966})\BibitemShut {NoStop}%
\bibitem [{\citenamefont {van Santen}\ \emph {et~al.}(2022)\citenamefont {van
  Santen}, \citenamefont {Clark}, \citenamefont {Halliday}, \citenamefont
  {Hallmann},\ and\ \citenamefont {Nelles}}]{vanSanten:2022wss}%
  \BibitemOpen
  \bibfield  {author} {\bibinfo {author} {\bibfnamefont {J.}~\bibnamefont {van
  Santen}}, \bibinfo {author} {\bibfnamefont {B.~A.}\ \bibnamefont {Clark}},
  \bibinfo {author} {\bibfnamefont {R.}~\bibnamefont {Halliday}}, \bibinfo
  {author} {\bibfnamefont {S.}~\bibnamefont {Hallmann}},\ and\ \bibinfo
  {author} {\bibfnamefont {A.}~\bibnamefont {Nelles}},\ }\bibfield  {title}
  {\bibinfo {title} {{toise: a framework to describe the performance of
  high-energy neutrino detectors}},\ }\href
  {https://doi.org/10.1088/1748-0221/17/08/T08009} {\bibfield  {journal}
  {\bibinfo  {journal} {JINST}\ }\textbf {\bibinfo {volume} {17}}\bibfield
  {number} {\bibinfo  {number} { (08)},\ \bibinfo {pages} {T08009}},\ }\Eprint
  {https://arxiv.org/abs/2202.11120} {arXiv:2202.11120 [astro-ph.IM]}
  \BibitemShut {NoStop}%
\bibitem [{\citenamefont {Feldman}\ and\ \citenamefont
  {Cousins}(1998)}]{Feldman_1998}%
  \BibitemOpen
  \bibfield  {author} {\bibinfo {author} {\bibfnamefont {G.~J.}\ \bibnamefont
  {Feldman}}\ and\ \bibinfo {author} {\bibfnamefont {R.~D.}\ \bibnamefont
  {Cousins}},\ }\bibfield  {title} {\bibinfo {title} {Unified approach to the
  classical statistical analysis of small signals},\ }\href
  {https://doi.org/10.1103/physrevd.57.3873} {\bibfield  {journal} {\bibinfo
  {journal} {Physical Review D}\ }\textbf {\bibinfo {volume} {57}},\ \bibinfo
  {pages} {3873} (\bibinfo {year} {1998})}\BibitemShut {NoStop}%
\bibitem [{\citenamefont {Glaser}\ \emph {et~al.}(2020)\citenamefont {Glaser}
  \emph {et~al.}}]{Glaser:2019cws}%
  \BibitemOpen
  \bibfield  {author} {\bibinfo {author} {\bibfnamefont {C.}~\bibnamefont
  {Glaser}} \emph {et~al.},\ }\bibfield  {title} {\bibinfo {title} {{NuRadioMC:
  Simulating the radio emission of neutrinos from interaction to detector}},\
  }\href {https://doi.org/10.1140/epjc/s10052-020-7612-8} {\bibfield  {journal}
  {\bibinfo  {journal} {Eur. Phys. J. C}\ }\textbf {\bibinfo {volume} {80}},\
  \bibinfo {pages} {77} (\bibinfo {year} {2020})},\ \Eprint
  {https://arxiv.org/abs/1906.01670} {arXiv:1906.01670 [astro-ph.IM]}
  \BibitemShut {NoStop}%
\bibitem [{\citenamefont {Enberg}\ \emph {et~al.}(2009)\citenamefont {Enberg},
  \citenamefont {Reno},\ and\ \citenamefont {Sarcevic}}]{Enberg:2008jm}%
  \BibitemOpen
  \bibfield  {author} {\bibinfo {author} {\bibfnamefont {R.}~\bibnamefont
  {Enberg}}, \bibinfo {author} {\bibfnamefont {M.~H.}\ \bibnamefont {Reno}},\
  and\ \bibinfo {author} {\bibfnamefont {I.}~\bibnamefont {Sarcevic}},\
  }\bibfield  {title} {\bibinfo {title} {{High energy neutrinos from charm in
  astrophysical sources}},\ }\href {https://doi.org/10.1103/PhysRevD.79.053006}
  {\bibfield  {journal} {\bibinfo  {journal} {Phys. Rev. D}\ }\textbf {\bibinfo
  {volume} {79}},\ \bibinfo {pages} {053006} (\bibinfo {year} {2009})},\
  \Eprint {https://arxiv.org/abs/0808.2807} {arXiv:0808.2807 [astro-ph]}
  \BibitemShut {NoStop}%
\bibitem [{\citenamefont {Carpio}\ \emph {et~al.}(2020)\citenamefont {Carpio},
  \citenamefont {Murase}, \citenamefont {Reno}, \citenamefont {Sarcevic},\ and\
  \citenamefont {Stasto}}]{Carpio:2020wzg}%
  \BibitemOpen
  \bibfield  {author} {\bibinfo {author} {\bibfnamefont {J.~A.}\ \bibnamefont
  {Carpio}}, \bibinfo {author} {\bibfnamefont {K.}~\bibnamefont {Murase}},
  \bibinfo {author} {\bibfnamefont {M.~H.}\ \bibnamefont {Reno}}, \bibinfo
  {author} {\bibfnamefont {I.}~\bibnamefont {Sarcevic}},\ and\ \bibinfo
  {author} {\bibfnamefont {A.}~\bibnamefont {Stasto}},\ }\bibfield  {title}
  {\bibinfo {title} {{Charm contribution to ultrahigh-energy neutrinos from
  newborn magnetars}},\ }\href {https://doi.org/10.1103/PhysRevD.102.103001}
  {\bibfield  {journal} {\bibinfo  {journal} {Phys. Rev. D}\ }\textbf {\bibinfo
  {volume} {102}},\ \bibinfo {pages} {103001} (\bibinfo {year} {2020})},\
  \Eprint {https://arxiv.org/abs/2007.07945} {arXiv:2007.07945 [astro-ph.HE]}
  \BibitemShut {NoStop}%
\bibitem [{\citenamefont {Bhattacharya}\ \emph {et~al.}(2023)\citenamefont
  {Bhattacharya}, \citenamefont {Enberg}, \citenamefont {Reno},\ and\
  \citenamefont {Sarcevic}}]{Bhattacharya:2023mmp}%
  \BibitemOpen
  \bibfield  {author} {\bibinfo {author} {\bibfnamefont {A.}~\bibnamefont
  {Bhattacharya}}, \bibinfo {author} {\bibfnamefont {R.}~\bibnamefont
  {Enberg}}, \bibinfo {author} {\bibfnamefont {M.~H.}\ \bibnamefont {Reno}},\
  and\ \bibinfo {author} {\bibfnamefont {I.}~\bibnamefont {Sarcevic}},\
  }\bibfield  {title} {\bibinfo {title} {{Energy-dependent flavour ratios in
  neutrino telescopes from charm}},\ }\href@noop {} {\  (\bibinfo {year}
  {2023})},\ \Eprint {https://arxiv.org/abs/2309.09139} {arXiv:2309.09139
  [astro-ph.HE]} \BibitemShut {NoStop}%
\bibitem [{\citenamefont {H{\"u}mmer}\ \emph
  {et~al.}(2010{\natexlab{b}})\citenamefont {H{\"u}mmer}, \citenamefont
  {R{\"u}ger}, \citenamefont {Spanier},\ and\ \citenamefont
  {Winter}}]{Hummer:2010vx}%
  \BibitemOpen
  \bibfield  {author} {\bibinfo {author} {\bibfnamefont {S.}~\bibnamefont
  {H{\"u}mmer}}, \bibinfo {author} {\bibfnamefont {M.}~\bibnamefont
  {R{\"u}ger}}, \bibinfo {author} {\bibfnamefont {F.}~\bibnamefont {Spanier}},\
  and\ \bibinfo {author} {\bibfnamefont {W.}~\bibnamefont {Winter}},\
  }\bibfield  {title} {\bibinfo {title} {{Simplified models for photohadronic
  interactions in cosmic accelerators}},\ }\href
  {https://doi.org/10.1088/0004-637X/721/1/630} {\bibfield  {journal} {\bibinfo
   {journal} {Astrophys. J.}\ }\textbf {\bibinfo {volume} {721}},\ \bibinfo
  {pages} {630} (\bibinfo {year} {2010}{\natexlab{b}})},\ \Eprint
  {https://arxiv.org/abs/1002.1310} {arXiv:1002.1310 [astro-ph.HE]}
  \BibitemShut {NoStop}%
\bibitem [{\citenamefont {Arg\"uelles}\ \emph {et~al.}(2023)\citenamefont
  {Arg\"uelles} \emph {et~al.}}]{Arguelles:2022tki}%
  \BibitemOpen
  \bibfield  {author} {\bibinfo {author} {\bibfnamefont {C.~A.}\ \bibnamefont
  {Arg\"uelles}} \emph {et~al.},\ }\bibfield  {title} {\bibinfo {title}
  {{Snowmass white paper: beyond the standard model effects on neutrino flavor:
  Submitted to the proceedings of the US community study on the future of
  particle physics (Snowmass 2021)}},\ }\href
  {https://doi.org/10.1140/epjc/s10052-022-11049-7} {\bibfield  {journal}
  {\bibinfo  {journal} {Eur. Phys. J. C}\ }\textbf {\bibinfo {volume} {83}},\
  \bibinfo {pages} {15} (\bibinfo {year} {2023})},\ \Eprint
  {https://arxiv.org/abs/2203.10811} {arXiv:2203.10811 [hep-ph]} \BibitemShut
  {NoStop}%
\bibitem [{\citenamefont {Kostelecky}\ and\ \citenamefont
  {Russell}(2011)}]{Kostelecky:2008ts}%
  \BibitemOpen
  \bibfield  {author} {\bibinfo {author} {\bibfnamefont {V.~A.}\ \bibnamefont
  {Kostelecky}}\ and\ \bibinfo {author} {\bibfnamefont {N.}~\bibnamefont
  {Russell}},\ }\bibfield  {title} {\bibinfo {title} {{Data Tables for Lorentz
  and CPT Violation}},\ }\href {https://doi.org/10.1103/RevModPhys.83.11}
  {\bibfield  {journal} {\bibinfo  {journal} {Rev. Mod. Phys.}\ }\textbf
  {\bibinfo {volume} {83}},\ \bibinfo {pages} {11} (\bibinfo {year} {2011})},\
  \Eprint {https://arxiv.org/abs/0801.0287} {arXiv:0801.0287 [hep-ph]}
  \BibitemShut {NoStop}%
\bibitem [{\citenamefont {Colladay}\ and\ \citenamefont
  {Kostelecky}(1998)}]{Colladay:1998fq}%
  \BibitemOpen
  \bibfield  {author} {\bibinfo {author} {\bibfnamefont {D.}~\bibnamefont
  {Colladay}}\ and\ \bibinfo {author} {\bibfnamefont {V.~A.}\ \bibnamefont
  {Kostelecky}},\ }\bibfield  {title} {\bibinfo {title} {{Lorentz violating
  extension of the standard model}},\ }\href
  {https://doi.org/10.1103/PhysRevD.58.116002} {\bibfield  {journal} {\bibinfo
  {journal} {Phys. Rev. D}\ }\textbf {\bibinfo {volume} {58}},\ \bibinfo
  {pages} {116002} (\bibinfo {year} {1998})},\ \Eprint
  {https://arxiv.org/abs/hep-ph/9809521} {arXiv:hep-ph/9809521} \BibitemShut
  {NoStop}%
\bibitem [{\citenamefont {Kostelecky}\ and\ \citenamefont
  {Mewes}(2004)}]{Kostelecky:2003cr}%
  \BibitemOpen
  \bibfield  {author} {\bibinfo {author} {\bibfnamefont {V.~A.}\ \bibnamefont
  {Kostelecky}}\ and\ \bibinfo {author} {\bibfnamefont {M.}~\bibnamefont
  {Mewes}},\ }\bibfield  {title} {\bibinfo {title} {{Lorentz and CPT violation
  in neutrinos}},\ }\href {https://doi.org/10.1103/PhysRevD.69.016005}
  {\bibfield  {journal} {\bibinfo  {journal} {Phys. Rev. D}\ }\textbf {\bibinfo
  {volume} {69}},\ \bibinfo {pages} {016005} (\bibinfo {year} {2004})},\
  \Eprint {https://arxiv.org/abs/hep-ph/0309025} {arXiv:hep-ph/0309025}
  \BibitemShut {NoStop}%
\bibitem [{\citenamefont {D\'iaz}\ and\ \citenamefont
  {Kostelecky}(2012)}]{Diaz:2011ia}%
  \BibitemOpen
  \bibfield  {author} {\bibinfo {author} {\bibfnamefont {J.~S.}\ \bibnamefont
  {D\'iaz}}\ and\ \bibinfo {author} {\bibfnamefont {A.}~\bibnamefont
  {Kostelecky}},\ }\bibfield  {title} {\bibinfo {title} {{Lorentz- and
  CPT-violating models for neutrino oscillations}},\ }\href
  {https://doi.org/10.1103/PhysRevD.85.016013} {\bibfield  {journal} {\bibinfo
  {journal} {Phys. Rev. D}\ }\textbf {\bibinfo {volume} {85}},\ \bibinfo
  {pages} {016013} (\bibinfo {year} {2012})},\ \Eprint
  {https://arxiv.org/abs/1108.1799} {arXiv:1108.1799 [hep-ph]} \BibitemShut
  {NoStop}%
\bibitem [{\citenamefont {Cowan}\ \emph {et~al.}(2011)\citenamefont {Cowan},
  \citenamefont {Cranmer}, \citenamefont {Gross},\ and\ \citenamefont
  {Vitells}}]{Cowan:2010js}%
  \BibitemOpen
  \bibfield  {author} {\bibinfo {author} {\bibfnamefont {G.}~\bibnamefont
  {Cowan}}, \bibinfo {author} {\bibfnamefont {K.}~\bibnamefont {Cranmer}},
  \bibinfo {author} {\bibfnamefont {E.}~\bibnamefont {Gross}},\ and\ \bibinfo
  {author} {\bibfnamefont {O.}~\bibnamefont {Vitells}},\ }\bibfield  {title}
  {\bibinfo {title} {{Asymptotic formulae for likelihood-based tests of new
  physics}},\ }\href {https://doi.org/10.1140/epjc/s10052-011-1554-0}
  {\bibfield  {journal} {\bibinfo  {journal} {Eur. Phys. J. C}\ }\textbf
  {\bibinfo {volume} {71}},\ \bibinfo {pages} {1554} (\bibinfo {year}
  {2011})},\ \bibinfo {note} {[Erratum: Eur.~Phys.~J.~C 73, 2501 (2013)]},\
  \Eprint {https://arxiv.org/abs/1007.1727} {arXiv:1007.1727 [physics.data-an]}
  \BibitemShut {NoStop}%
\end{thebibliography}%


\onecolumngrid


\appendix






\onecolumngrid


\renewcommand{\tocname}{\vspace*{-0.8cm}}


\section{Additional plots of flavor composition}\label{app:additional_flavor_composition}
\twocolumngrid


\begin{figure}[t!]
 \centering
 \includegraphics[trim={0.34cm 0 0 0},clip,width=1.05\columnwidth]{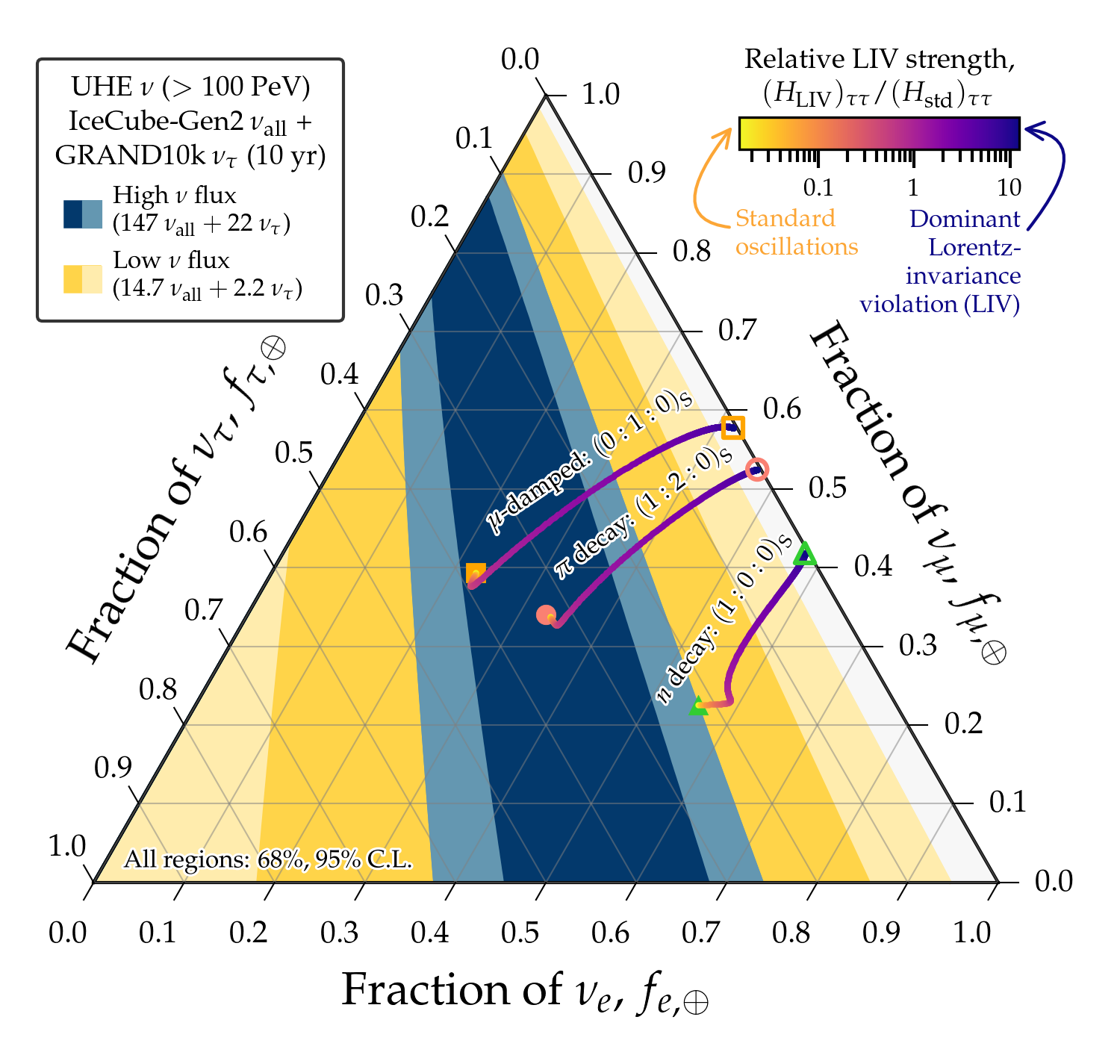}
 \caption{\textbf{\textit{Measurement of the UHE neutrino flavor composition at Earth, assuming $(\frac{1}{3},\frac{2}{3},0)_{\rm S}$ and using GRAND10k.}}  Same as \figu{ternary_grand50k} in the main text, but for GRAND10k.  Figure~\ref{fig:likelihood_sources_grand10k} shows the corresponding inferred flavor composition at the sources.}
 \label{fig:ternary_grand10k}
\end{figure}

\begin{figure}[t!]
 \centering
 \includegraphics[width=1.01\columnwidth]{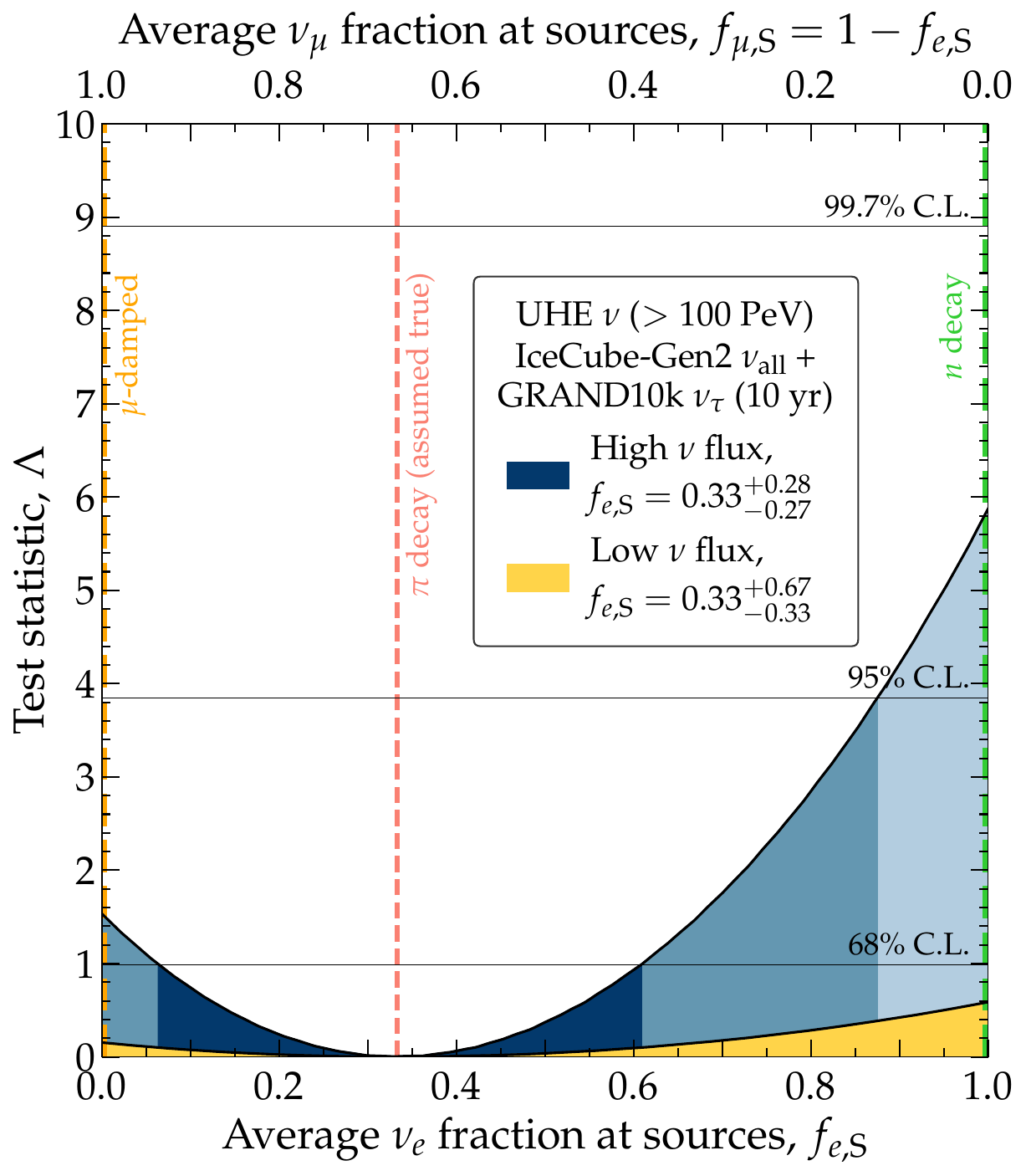}
 \caption{\textbf{\textit{Inferred flavor composition of UHE neutrinos at their sources, assuming $(\frac{1}{3},\frac{2}{3},0)_{\rm S}$, and using GRAND10k.}}  Same as \figu{likelihood_sources_grand10k} in the main text, but for GRAND10k.  The results are inferred from the flavor composition at Earth shown in \figu{ternary_grand10k}.}
 \label{fig:likelihood_sources_grand10k}
\end{figure}

\begin{figure*}[t!]
 \centering
 \includegraphics[trim={0.34cm 0 0 0},clip,width=1.03\columnwidth]{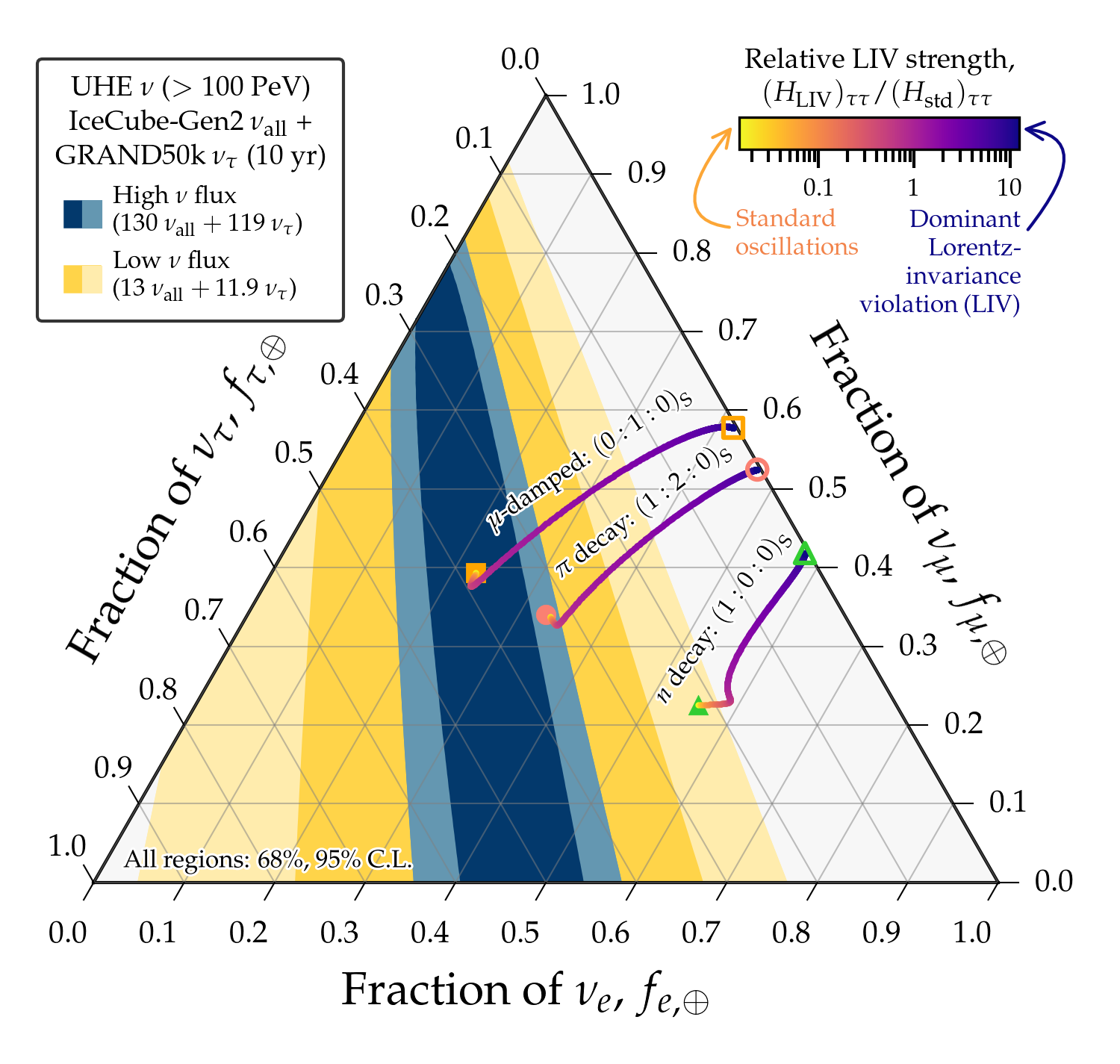}
 \includegraphics[trim={0.34cm 0 0 0},clip,width=1.03\columnwidth]{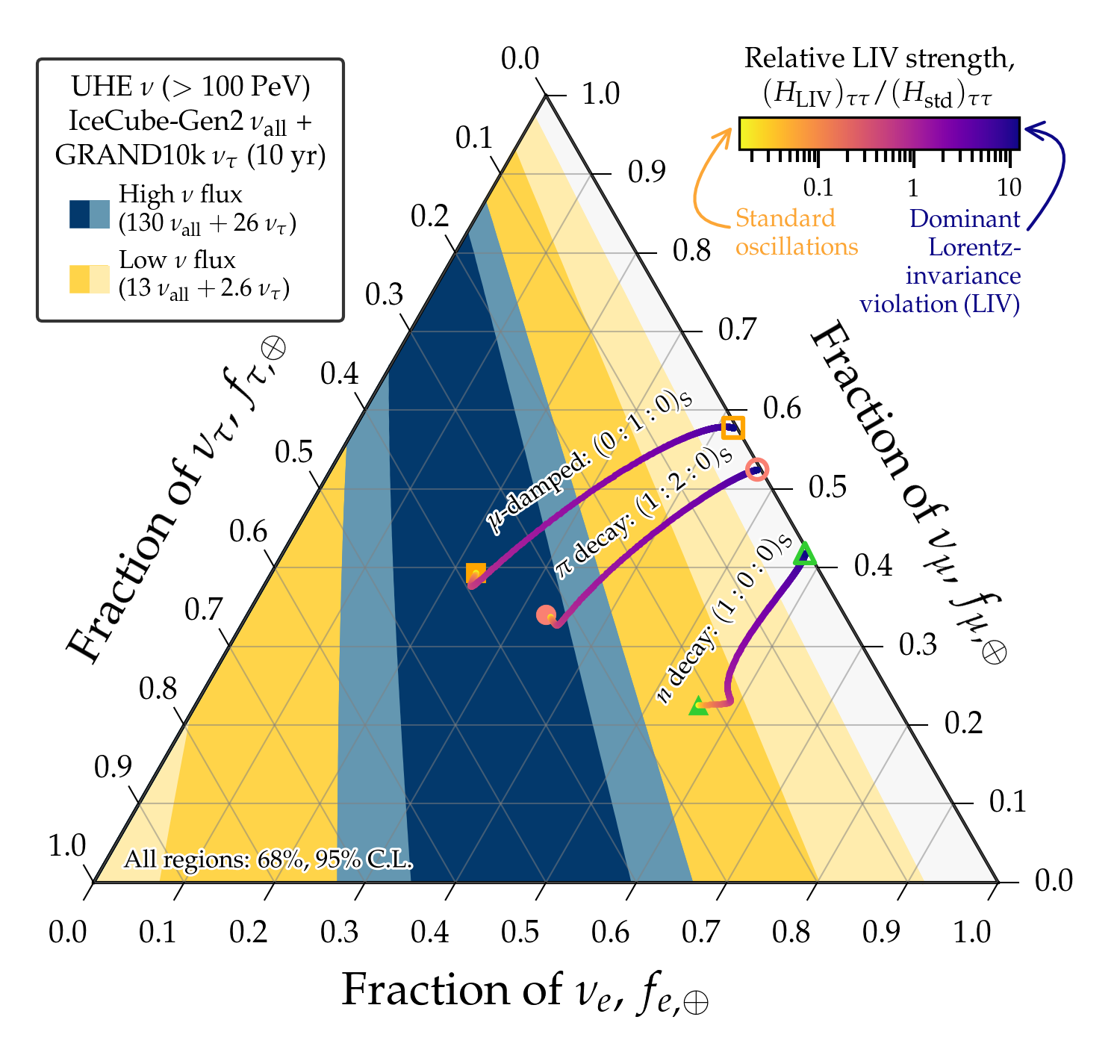}
 \caption{\textbf{\textit{Measurement of the UHE neutrino flavor composition at Earth, assuming $(0,1,0)_{\rm S}$.}}  {\it Left:} Using GRAND50k, \ie, same as \figu{ternary_grand50k} in the main text, but for $(0,1,0)_{\rm S}$.  {\it Right:} Using GRAND10k, \ie, same as \figu{ternary_grand10k}, but for $(0,1,0)_{\rm S}$.  Figure~\ref{fig:likelihood_sources_muon_damped} shows the corresponding inferred flavor composition at the sources.}
 \label{fig:ternary_muon_damped}
\end{figure*}

\begin{figure*}[t!]
 \centering
 \includegraphics[width=0.48\textwidth]{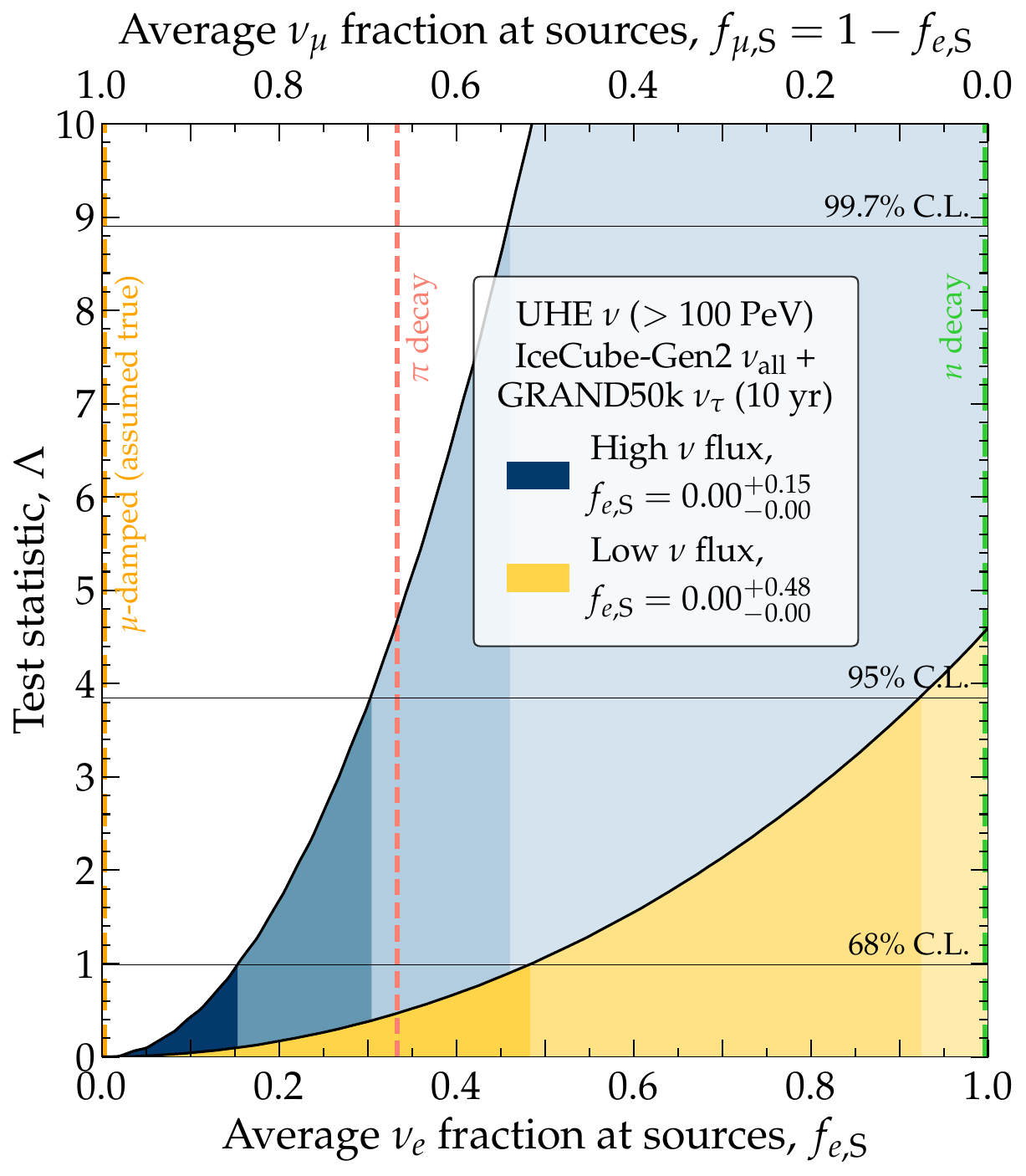}
 \includegraphics[width=0.48\textwidth]{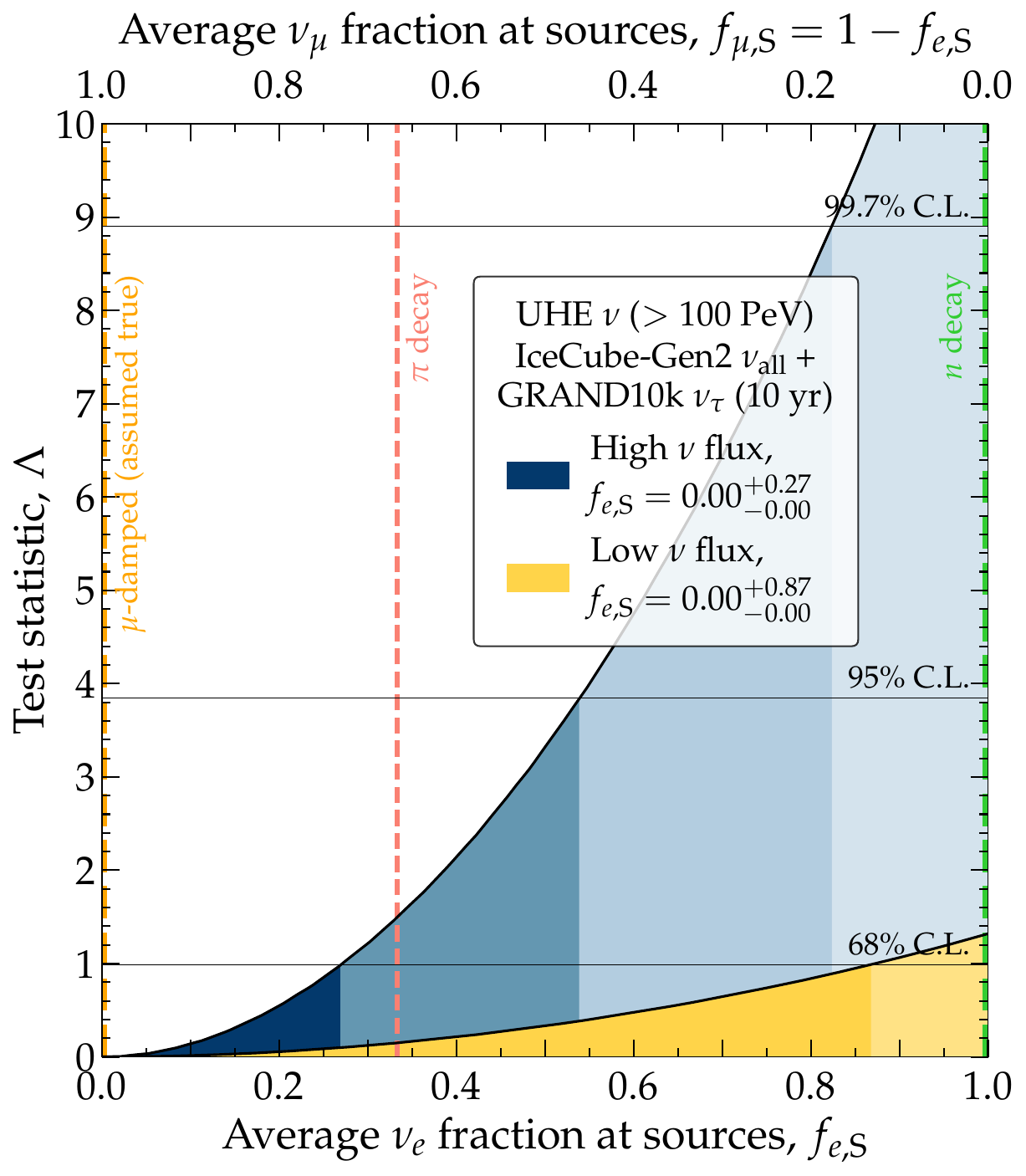}
 \caption{\textbf{\textit{Inferred flavor composition of UHE neutrinos at their sources, assuming $(0,1,0)_{\rm S}$.}}  {\it Left:} Using GRAND50k, \ie, same as \figu{likelihood_sources} in the main text, but for $(0,1,0)_{\rm S}$.  {\it Right:} Using GRAND10k, \ie, same as \figu{likelihood_sources_grand10k}, but for $(0,1,0)_{\rm S}$.}
 \label{fig:likelihood_sources_muon_damped}
\end{figure*}

To produce our main results, shown in the main text, we use flux benchmarks built assuming a flavor composition at the sources from pion decay, $\left( \frac{1}{3}, \frac{2}{3}, 0 \right)_{\rm S}$, and detection in GRAND10k.  Below we show results obtained under different assumptions:
\begin{description}
 \item[Figure~\ref{fig:ternary_grand10k}]
  Flavor composition at the Earth assuming $\left( \frac{1}{3}, \frac{2}{3}, 0 \right)_{\rm S}$ and GRAND10k.
 \item[Figure~\ref{fig:likelihood_sources_grand10k}]
  Inferred flavor composition at the sources assuming $\left( \frac{1}{3}, \frac{2}{3}, 0 \right)_{\rm S}$ and GRAND10k, \ie, computed from the results in \figu{ternary_grand10k}.
 \item[Figure~\ref{fig:ternary_muon_damped}]
  Flavor composition at the Earth assuming muon-damped flavor composition $(0, 1, 0)_{\rm S}$, and GRAND50k and GRAND10k.
 \item[Figure~\ref{fig:likelihood_sources_muon_damped}]
  Inferred flavor composition at the sources assuming $(0, 1, 0)_{\rm S}$, and GRAND50k and GRAND10k, computed from the results in \figu{ternary_muon_damped}, using the same methods as before~\cite{Bustamante:2019sdb, Song:2020nfh}.
\end{description}
The main observations on and interpretation of Figs.~\ref{fig:ternary_grand10k}--\ref{fig:likelihood_sources_muon_damped} are in the main text.  Because the flavor composition at Earth for pion-decay and muon-damped production is different, the rates of detected events are different in each case; compare the rates shown in Figs.~\ref{fig:ternary_grand50k} (in the main text) and \ref{fig:ternary_grand10k} {\it vs.}~Figs.~\ref{fig:ternary_muon_damped} and \ref{fig:likelihood_sources_muon_damped}. 

\onecolumngrid
\section{Results for other LIV operators}\label{app:LIV_operators}
\twocolumngrid

\begin{figure*}
 \includegraphics[width=\textwidth]{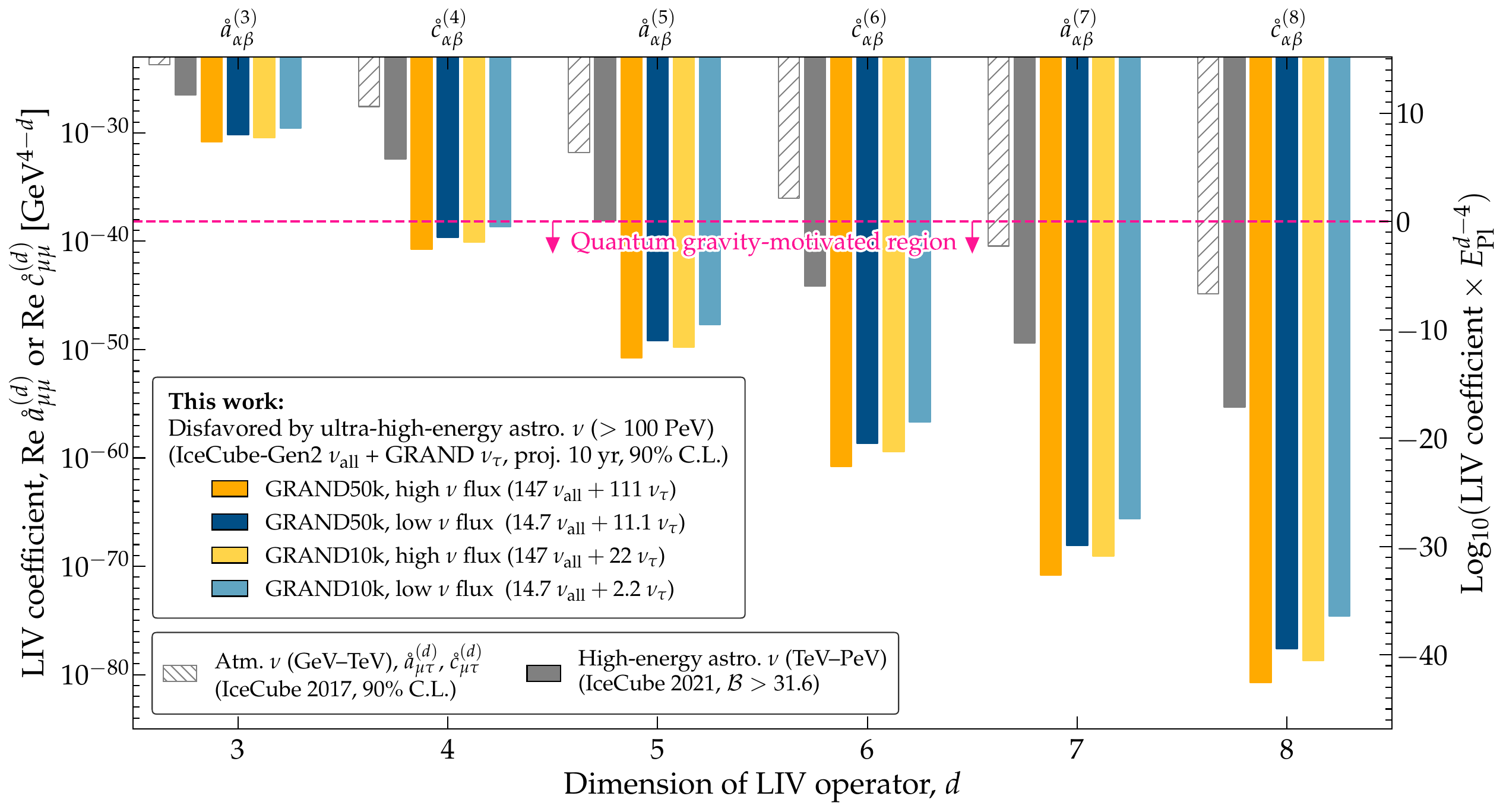}
 \caption{\textbf{\textit{Projected limits on LIV coefficients $\mathring{a}_{\mu\mu}^{(d)}$ and $\mathring{c}_{\mu\mu}^{(d)}$ from UHE neutrinos.}} Same as \figu{limits_tt_vs_dim} in the main text, but for the $\mu\mu$ LIV coefficients.  See Figs.~\ref{fig:limits_em_vs_dim} and \ref{fig:limits_et_vs_dim} for other coefficients.}
 \label{fig:limits_mm_vs_dim}
\end{figure*}

\begin{figure*}
 \includegraphics[width=\textwidth]{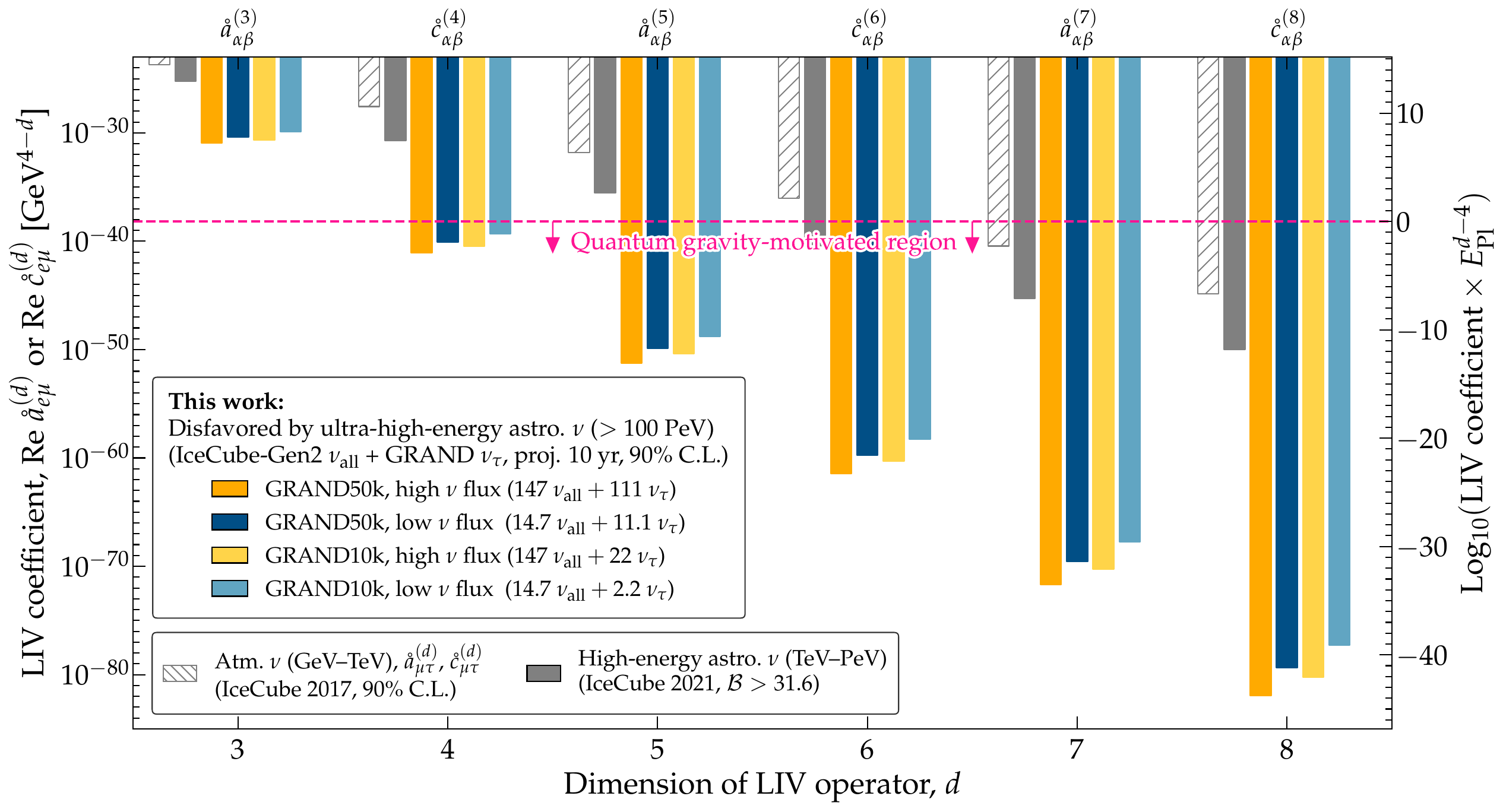}
 \caption{\textbf{\textit{Projected limits on LIV coefficients $\mathring{a}_{e\mu}^{(d)}$ and $\mathring{c}_{e\mu}^{(d)}$ from UHE neutrinos.}} Same as \figu{limits_tt_vs_dim} in the main text, but for the $e\mu$ LIV coefficients.  See Figs.~\ref{fig:limits_et_vs_dim} and \ref{fig:limits_et_vs_dim} for other coefficients.}
 \label{fig:limits_em_vs_dim}
\end{figure*}

\begin{figure*}
 \includegraphics[width=\textwidth]{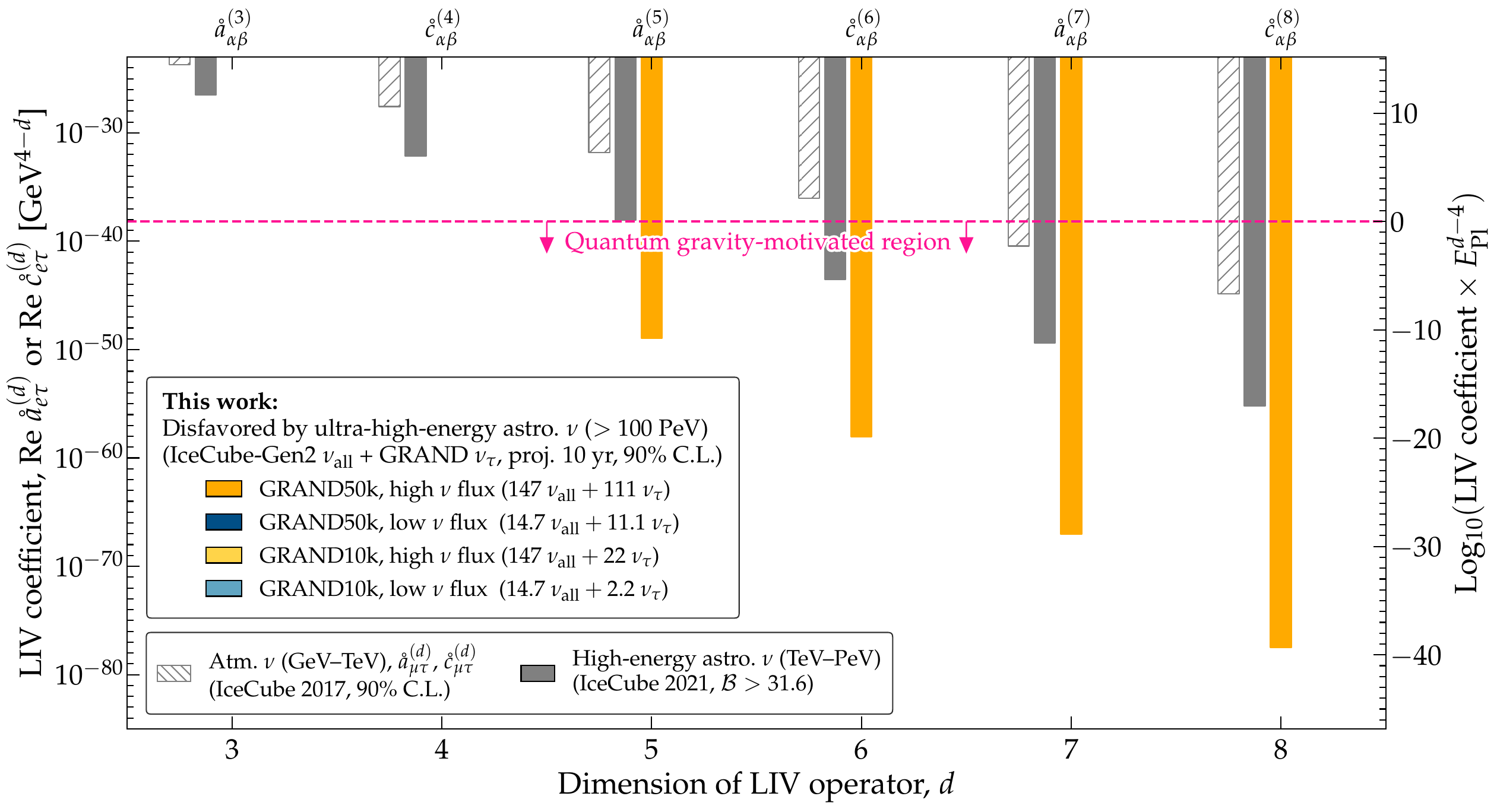}
 \caption{\textbf{\textit{Projected limits on LIV coefficients $\mathring{a}_{e\tau}^{(d)}$ and $\mathring{c}_{e\tau}^{(d)}$ from UHE neutrinos.}} Same as \figu{limits_tt_vs_dim} in the main text, but for the $e\tau$ LIV coefficients.  Limits are only achievable in the most optimistic scenario, using GRAND50k and our benchmark UHE neutrino flux.  See Figs.~\ref{fig:limits_mm_vs_dim} and \ref{fig:limits_em_vs_dim} for other coefficients.}
 \label{fig:limits_et_vs_dim}
\end{figure*}

In the main text (\figu{limits_tt_vs_dim}), we show upper limits on the CPT-odd and CPT-even LIV coefficients $\mathring{a}_{\tau\tau}^{(d)}$ and $\mathring{c}_{\tau\tau}^{(d)}$, for operators of dimension $d = 3$ to 8.  Below we show limits on other LIV coefficients:
\begin{description}
 \item[Figure~\ref{fig:limits_mm_vs_dim}]
  Limits on $\mathring{a}_{\mu\mu}^{(d)}$ and $\mathring{c}_{\mu\mu}^{(d)}$.
 \item[Figure~\ref{fig:limits_em_vs_dim}]
  Limits on $\mathring{a}_{e\mu}^{(d)}$ and $\mathring{c}_{e\mu}^{(d)}$.
 \item[Figure~\ref{fig:limits_et_vs_dim}]
  Limits on $\mathring{a}_{e\tau}^{(d)}$ and $\mathring{c}_{e\tau}^{(d)}$.
\end{description}
Like for \figu{limits_tt_vs_dim}, we show results assuming that the flavor composition at the sources is $(\frac{1}{3}, \frac{2}{3}, 0)_{\rm S}$, from pion decay. In all cases, our projected limits from UHE neutrinos improve on existing limits.  However, for $\mathring{a}_{e\tau}^{(d)}$ and $\mathring{c}_{e\tau}^{(d)}$, limits are achievable only under the most optimistic scenario, using GRAND50k and our benchmark high neutrino flux because the LIV couplings affect directly only $\nu_e$ and $\nu_\tau$, which make up the smallest contributions to the flux generated by pion decay. 

\end{document}